\documentstyle [psfig]{mn2e}
\input epsf
\def\ha{H$\alpha$}
\def\deg{\hbox{$^\circ$}}
\def\scos{SuperCOSMOS}
\def\ergcms{erg cm$^{-2}$ s$^{-1}$}
\def\hamr{$H\alpha - R$}

\title{The AAO/UKST SuperCOSMOS \ha\ Survey (SHS)}
\author [Quentin Parker et al.]
{Q.A. Parker$^{1,3}$, S. Phillipps$^{2}$, M.J. Pierce$^{2}$,
M. Hartley$^{4}$, N.C. Hambly$^{5}$, M.A. Read$^{5}$, 
\newauthor H.T. MacGillivray$^{5}$, S.B. Tritton$^{5}$,
C.P. Cass$^{4}$, R.D. Cannon$^{3}$, M. Cohen$^{6}$, J.E. Drew$^{7}$,
\newauthor D.J. Frew$^{1}$, E. Hopewell$^{7}$, S. Mader$^{8}$, D.F. Malin$^{3}$, M.R.W. Masheder$^{2}$,
D.H. Morgan$^{5}$,
\newauthor R.A.H. Morris$^{2}$, D. Russeil$^{9}$, K.S. Russell$^{4}$, R.N.F. Walker$^{2}$\\
$^{1}$ Macquarie University, Sydney, Australia,\\
$^{2}$Astrophysics Group, University of Bristol,
Tyndall Avenue, Bristol, U.K.\\
$^{3}$Anglo-Australian Observatory, Epping, New South Wales, Australia\\ 
$^{4}$UK Schmidt Telescope, Anglo-Australian Observatory, Siding Spring,New South Wales, Australia\\
$^{5}$ Institute for Astronomy, School of Physics, University of Edinburgh,
$^{6}$UC, Berkeley, USA\\
$^{7}$Imperial College, London, U.K.\\
$^{8}$Australia Telescope National Facility, Parkes, Australia, \\
$^{9}$Observatoire de Marseille, 2 Place le Verrier, Marseille, 13248 cedex 4, France}
\date{ Accepted . Received ; in original form        }
\renewcommand{\baselinestretch}{1.6}
\begin{document}
\maketitle

\begin{abstract}
The UK Schmidt Telescope (UKST) of the Anglo-Australian Observatory completed 
a narrow-band \ha\ plus [NII]~6548, 6584\AA ~survey of the Southern Galactic Plane and 
Magellanic Clouds in late 2003. The survey, which was the last UKST wide-field photographic survey, 
and the only one undertaken in a narrow band,
is now an on-line digital data product of the Wide-Field Astronomy 
Unit of the Royal Observatory Edinburgh (ROE). The
survey utilised a high specification, monolithic
\ha\ interference band-pass filter of exceptional quality. In
conjunction with the fine grained Tech-Pan film as a detector it has produced
a survey with a powerful combination of area coverage (4000 square degrees),
resolution ($\sim$1~arcsecond) and sensitivity ($\leq$5~Rayleighs), reaching a 
depth for continuum point sources of
$R\simeq20.5$. The main survey consists of 233 individual fields on a grid of centres
separated by 4\deg\  at
declinations below +2\deg\ and covers a swathe approximately
20\deg\ wide about the Southern Galactic Plane. The original survey films were
scanned by the SuperCOSMOS measuring machine at the Royal
Observatory, Edinburgh to provide the on-line digital atlas called 
the SuperCOSMOS \ha\ Survey (SHS).  
We present the background to the survey, the key survey characteristics, 
details and examples of the data product, calibration process, comparison with 
other surveys and a brief description of its potential for scientific exploitation.
\end{abstract}


\section{Introduction}

\ha\ emission from HII regions is one of the most direct optical tracers
of current star formation activity and is routinely used to measure star
formation rates in external galaxies (e.g. Kennicutt 1992). In our own galaxy, HII regions are
seen by direct UV illumination of molecular clouds from adjacent hot
stars and as highly structured shells, bubbles and sheets of emission
resulting from supernovae, planetary nebula, Wolf-Rayet stars and other stellar outflows. 
Some large-scale outflows can, in
turn, be themselves a trigger of star formation, and their morphology is
strongly influenced by the nature and density of the ISM into which they
expand. \ha\ imaging allows this to be studied in great detail in our
immediate Galactic neighbourhood and to be detected at a great distance
in external galaxies. The UV flux from hot stars also excites a more
diffuse emission from the ISM, unconnected to current star formation and
detectable over large areas of sky.

The perimeters of some large emission shells appear to enclose the locations of more recent star 
formation which may in turn generate further supernovae
and stellar winds (Dopita, Matthewson \& Ford 1985), while their morphology 
informs the processes by which star formation is propagated (e.g.
Elmegreen \& Lada 1977, Gerola \& Seiden 1978). 
Because of proximity, some of these structures can
present very large angular sizes such as Barnard's loop (probably the first 
large-scale emission structure
detected in the Galaxy) which subtends 13\deg\ (Pickering 1890)
and the Gum nebula which at 36\deg, is even larger  (Gum 1952). More distant complexes or
groups of HII regions, such as NGC 6334, can still be of the order 1\deg\ across and
yet present fine detail on arcsecond scales (Meaburn \&
White 1982). Given their interaction with their external large-scale environment
(Tenorio-Tagle \& Palous 1987) it was clear that emission-line imaging of these structures required
an efficient wide-field capability and high spatial resolution. 

On smaller scales, stellar \ha\ emission characterises the short
lived, least well understood stages of stellar evolution, i.e. those of pre- and
post-main sequence stars, planetary nebulae and close binary systems. Previous
efforts to detect emission sources have either offered modest
area coverage; e.g. the UBVI and \ha\ photometric surveys of
Sung, Chun \& Bessell (2000) or Keller et al. (2001) or, where a large-area survey
has been conducted, becomes incomplete at relatively bright magnitudes. An example is
the objective-prism survey of Stephenson \& Sanduleak (1977) which reaches only
$\sim$\,14$^{th}$ magnitude. Such surveys are highly incomplete so their
emission source catalogues provide only limited samples upon which to build our
understanding of these rarely observed phases of stellar evolution.

From the above, the importance of Galactic \ha\ line emission from both stars and nebulae is evident
and this has encouraged many surveys for HII regions in particular, 
e.g. Sharpless (1953, 1959), Gum (1955), Hase \& Shajn (1955), Bok, Bester \& Wade (1955), Johnson (1955,
1956), Rodgers, Campbell \& Whiteoak (1960), Georgelin \& Georgelin (1970) and Sivan (1974).
These earlier surveys were limited to relatively small targeted areas
or had such wide fields of view that small-scale detail was lost due to the 
low angular resolution; e.g. the survey of Sivan (1974) used 60\deg\ field diameters giving a
plate scale of $\sim6$\deg\ mm$^{-1}$.
Relatively little optical emission-line survey work had been done
in a way that combined wide-angle coverage with good sensitivity and
high resolution. These characteristics are essential to allow thorough
examination of the morphology and interaction of emission regions with
their environment on arcsecond to degree scales and to detect the large
variety of stellar emission sources to suitably faint levels.

Hence, in the mid-1990s, a number of the present authors suggested that the U.K. Schmidt Telescope (UKST) 
should be used to make a narrow band photographic \ha\ survey of the Southern Milky Way and 
Magellanic Clouds. The only previous wide-area UKST \ha\ 
material comes from the work by Meaburn and co-workers in the 1970s 
(see Davies, Elliot \& Meaburn 1976; Meaburn 1980).
They used a 100\AA~ band-pass multi-element mosaic filter and fast, but coarse grained, 098-04
emulsion. It covered some limited areas close to the Galactic Plane (Meaburn
\& White 1982), but was
mainly influential in the study of the ionised gas in the Magellanic
Clouds, showing the first evidence for ``supergiant-shells'' and other large-scale
features (Davies et al. 1976, Meaburn 1980). 
There are other recent wide-area \ha\ surveys such as that by the Virginia group in the northern 
hemisphere (VTSS -- Dennison et al.
1998) and the Mount Stromlo group in the south (Buxton et al. 1998). 
Recently, Gaustad et al. (2001) have released the full ``Southern \ha\ Sky Survey Atlas'' (SHASSA),
covering the entire southern sky. 
This imaging survey has rather coarse, 48-arcsecond pixels and strong artefacts from uncancelled 
stars in the continuum-subtracted product, but has the major
benefit of being directly calibrated in Rayleighs.
These surveys continue the tradition of deep, low
spatial resolution studies, but use CCDs which permit low flux densities of
a few tenths of a Rayleigh to be achieved. 

An alternative approach, taken
by the Marseille and Wisconsin Fabry-Perot groups in the southern and northern skies respectively
(see Russeil et al. 1997, 1998  and Haffner et al. 2003 for the WHAM -- Wisconsin \ha\ Mapper), 
was to obtain high resolution spectral (i.e. velocity) data, but again with
low spatial resolution (e.g. 1\deg\ pixels for WHAM).

A critical comparison between the WHAM, SHASSA and VTSS surveys was 
undertaken by Finkbeiner (2003) who presented a `whole-sky' \ha\ map. 
Significantly, none of these major surveys offer 
the arcsecond spatial resolution of the AAO/UKST \ha\ survey. A summary of fundamental properties
of these modern surveys is given in Table~\ref{surveys}.

\begin{table*}
\begin{center}
\caption{Summary details of various current \ha\ surveys}
\label{surveys}
\begin{tabular}{llllllll}\hline
Survey     & Coverage  &  Depth      & Resolution & Field size      & Filter    & Coverage & Reference\\
           & (sq.deg)  &  Rayleighs  & (arcsec)   &  (degrees)      & FWHM      &         \\ \hline
WHAM$^1$   & 17000     &   0.15      & 3600       &     1           & 0.25\AA  
& $\delta > -30$\deg& Haffner et al. 2003\\
SHASSA$^2$ & 17000     &  $\sim2$    & 48         &  $13\times13$   & 32\AA     & $\delta < 15$\deg & Gaustad et al. 2001\\
VTSS$^3$   & $>$1000   &  $\sim2$    & 96         &  $10\times10$   & 17\AA     & $\delta > -20$\deg (if completed) & Dennison et al.
1998\\
SHS$^4$    & 4000      &  $\leq$5    & 1-2        &  $5.5\times5.5$ & 80\AA     & $\delta > +2$\deg; $|b|\leq10-13$\deg & This paper\\
IPHAS$^5$  & 1800      &  $\sim3$    & $\sim1$    &  $0.3\times0.3$ & 95\AA     & $|b|\leq 5$\deg; northern plane & Drew et al. 2005\\
 \hline
\end{tabular}
\end{center}
Note: 1 Rayleigh = $10^{6}$/4$\pi$ photons cm$^{-2}$
s$^{-1}$ sr$^{-1} = 2.41\times10^{-7}$ \ergcms\ sr$^{-1}$ at H$\alpha$\\
\noindent{${^1}$: http://www.astro.wisc.edu/wham; ${^2}$: http://amundsen.swarthmore.edu/SHASSA;}\\ 
\noindent{${^3}$:http://www.phys.vt.edu/\~{}halpha; ${^4}$: http://www-wfau.roe.ac.uk/ss/halpha/}\\
\noindent{${^5}$:http://astro.ic.ac.uk/Research/Halpha/North/}
\end{table*}

\section{The AAO/UKST H$\alpha$ Survey}

The AAO/UKST \ha\ survey provides a 5~Rayleigh sensitivity narrow-band 
survey of Galactic emission (\ha\ plus [NII]~6548, 6584\AA) with arcsecond spatial resolution. 
Henceforth the survey will be refered to simply as the \ha\ survey though 
it is understood that this includes any [NII] emission
component that is sampled by the filter band-pass (such emission can completely 
dominate \ha\ for some PN types for example).
Approximately 4000~deg$^{2}$ of the Southern Milky Way have been covered to $|b|\sim10-13$\deg\ 
together with a separate contiguous region of 700~deg$^{2}$ in and around the Magellanic 
Clouds. Matching 3-hour \ha\ and 15-min broad-band (5900-6900\AA) short red (SR) 
exposures were taken over the 233 distinct but overlapping
fields of the Galactic Plane and 40 fields of the Magellanic Clouds. These were done on 
4\deg\ centres because of the circular aperture of the \ha\ interference filter 
which has a dielectric coating diameter of about 305mm ($\sim5.7$\deg) deposited on a standard 
$356\times356$mm red glass (RG610) substrate (refer Section 5). The overlapping
4\deg\ field centres enable full, 
contiguous coverage in \ha\ despite the circular filter aperture. 
Because of the slightly smaller effective field, 
a new Southern sky-grid of 1111, 4\deg\ field centres was created
(whose numbers should not be confused with the 893 standard 5\deg\ field centres of the UKST Southern Sky Surveys). 
A map of the survey region in a standard UKST RA/DEC
plot together with the new field numbers is available on the SHS web 
site\footnote{http://www-wfau.roe.ac.uk/sss/halpha/}. In the on-line UKST plate catalogue these fields
have a `h' prefix (e.g. h123) to avoid confusion with the ESO/SERC 5\deg\ fields.

The survey began in 1997 and took six years to complete.
This latest and final UKST photographic survey was the first large-scale, narrow band 
survey undertaken on the telescope and is the first where the sole method of dissemination 
to the community is via access to on-line digital data products.
Preliminary survey details and results were given by Parker \& Phillipps 
(1998, 2003). The present paper is intended as the definitive reference for the survey.
We describe the key characteristics of the survey, the on-line data product, 
some survey limitations, a flux calibration scheme, comparisons with other surveys  
and a brief overview of the potential for current and future scientific exploitation.

The arcsecond resolution of the AAO/UKST \ha\ survey makes it a particularly powerful tool, 
not only for investigating the detailed morphology of emission features across the widest 
range of angular scales, but also as a means of identifying large numbers of faint point-source 
\ha\ emitters, which include cataclysmic variables, T Tauri, Be and symbiotic stars, 
compact Herbig-Haro objects and unresolved planetary nebulae (PNe). Given the coincidence of
the broad CIV/HeII blend in late-type Wolf Rayet stars, these objects can also be detected.
Most other comparative surveys (Table~\ref{surveys}) are largely insensitive to 
point-source emitters as they lack 
spatial resolution, being optimised instead for the faintest levels of resolved 
and diffuse emission. 

On larger scales, the detailed spatial structure of 
the ionized ISM component traced by the new AAO/UKST \ha\ survey can provide key data for 
many studies, e.g. mapping of specific areas for detailed spectroscopic follow-up to obtain 
emission-line gas kinematics or for dynamical studies of star forming regions, with their 
implications for the energetics of the central stars.
Furthermore, comparisons with other indicators
of star formation from other wavebands should provide essential clues to 
the active mechanisms. The survey also complements the recent Galactic Plane
radio maps from MOST (Green et al. 1999), the new NIR maps from 2MASS (Jarrett et al. 2000) 
and the mid-infrared maps from the MSX satellite (Price et al. 2001). 

\begin{figure*}
\mbox{
\epsfxsize=0.9\textwidth\epsfbox{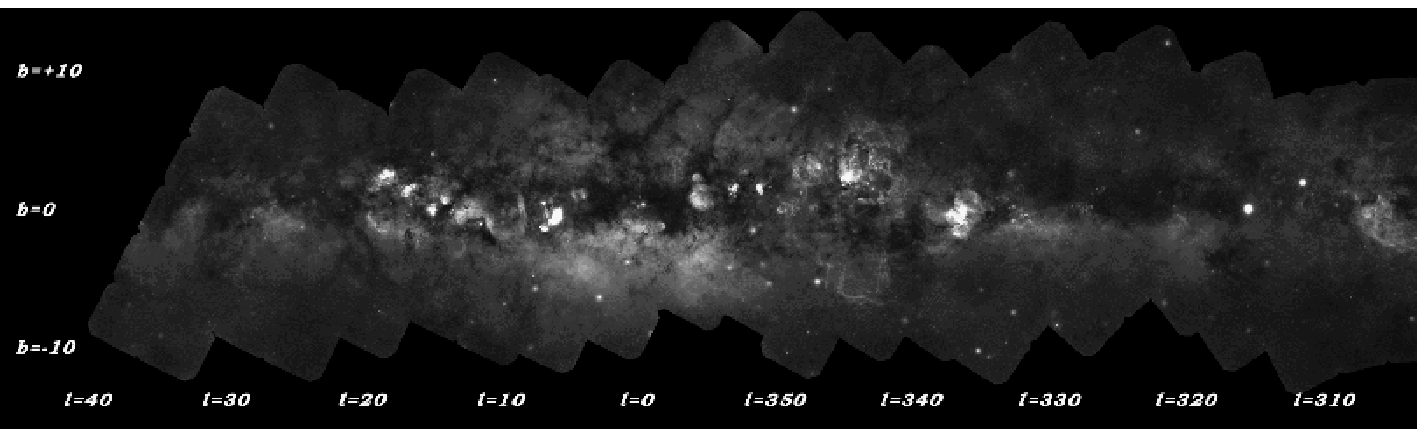}}
\mbox{
\epsfxsize=0.9\textwidth\epsfbox{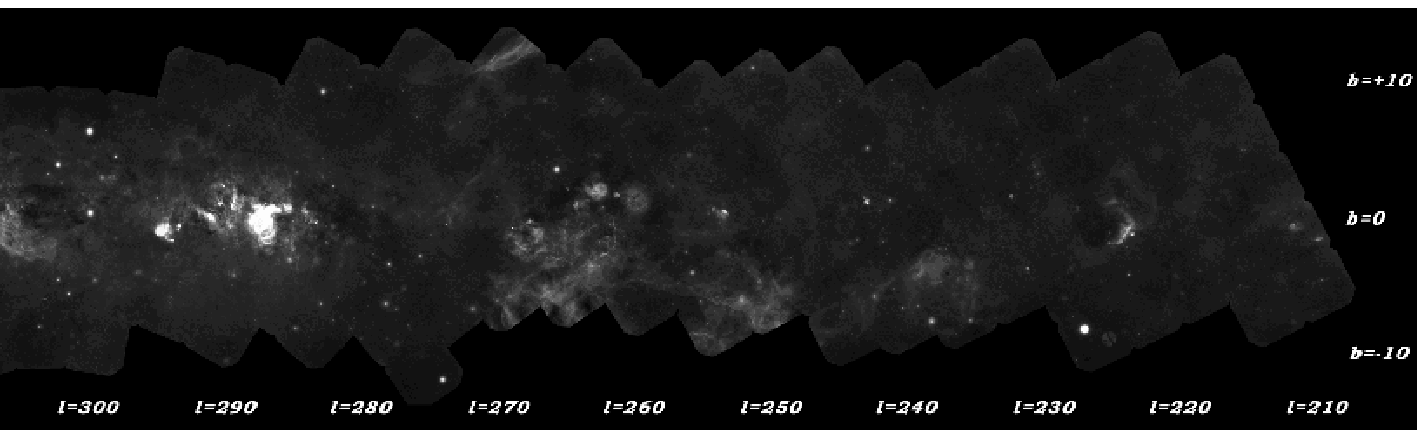}}
\caption{All 233 survey fields mosaiced together by M.Read: Top panel cover galactic 
longitude l=40 to 310 degrees, bottom panel l=300 to 210 degrees.}
\label{mosaic-map}
\end{figure*}

Figure~\ref{mosaic-map} presents two panels showing the 233 survey fields (mosaiced together 
by M.Read) to illustate the overall survey coverage.
The entire survey has been incorporated into an on-line mosaic within the freeware `Zoomify' 
environment (see http://www.zoomify.com) which enables preliminary survey visualisation and scanning. 
The lowest resolution map can be zoomed-in to a level where each pixel
represents about 12~arcseconds. This interactive map is available on-line.\footnote{
http://surveys.roe.ac.uk/ssa/hablock/hafull.html}
This map is a factor of 18 lower in resolution than the full 
0.67~arcsecond pixel survey 
data available on-line which should be used for serious scientific
work.  The success of this survey has led directly to a 
northern counterpart, currently 
underway on the 2.5m Isaac Newton Telescope on La Palma using a wide-field CCD camera and 
\ha\, R and I band imaging; the Isaac Newton telescope Photometric \ha\ Survey (IPHAS). 
This important survey is the subject of a separate paper 
(Drew et al. 2005) though a brief comparison in an
overlap region is included later in Section 12. 

\section{The detector: Technical Panchromatic film-based emulsion}

\begin{figure*}
\mbox{
\epsfxsize=0.3\textwidth\epsfbox{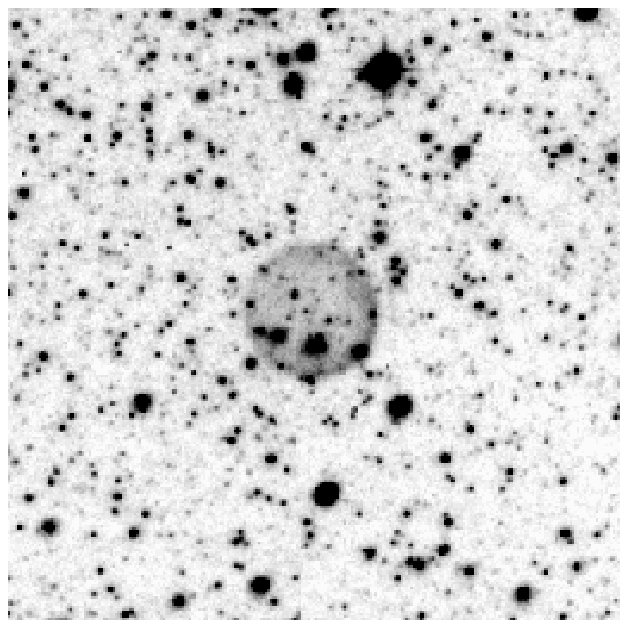}}
\mbox{
\epsfxsize=0.3\textwidth\epsfbox{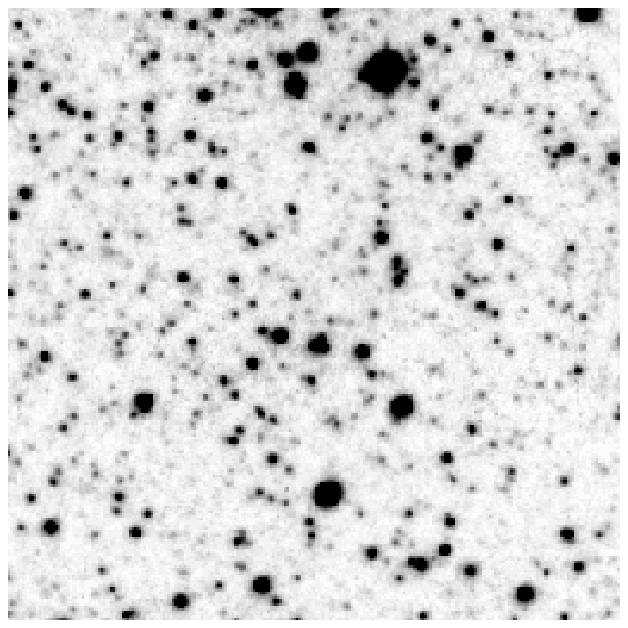}}
\mbox{
\epsfxsize=0.3\textwidth\epsfbox{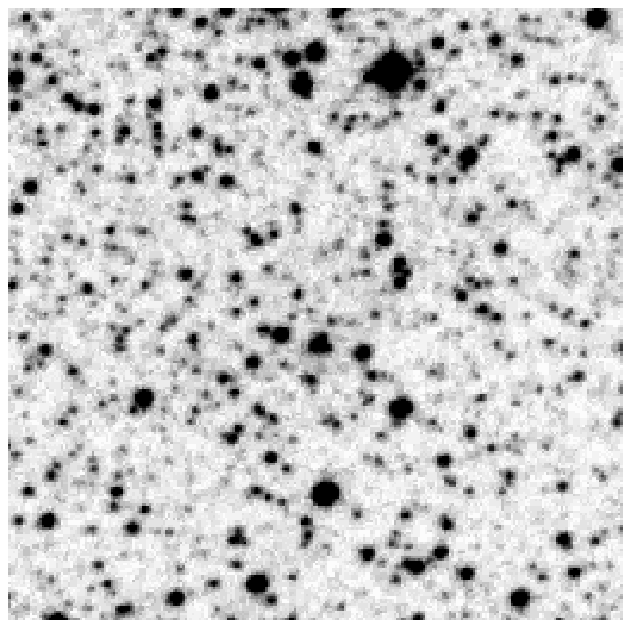}}
\caption{$3\times3$~arcminute extracts of SuperCOSMOS data around a newly 
discovered PN (PHR1706-3544) from the 3-hour \ha\ survey data
(a - left), matching 15~minute Tech-Pan SR data (b - middle) and earlier epoch 60~minute IIIa-F R-band 
data (c - right). The new PN is only visible in the \ha\ image. Note the well matched depth for point 
sources between all three exposures and the improved resolution of the Tech-Pan images compared with the
IIIa-F exposure.}
\label{HA-SR-R}
\end{figure*}

The survey was carried out using Kodak Technical Panchromatic (Tech-Pan) Estar based films (e.g. Kodak 1987). 
The superb qualities of this emulsion and its adaptation for UKST use has been described in detail by
Parker \& Malin (1999) so only a very brief summary is given here.
The Tech-Pan emulsion has remarkably high quantum efficiency for a
photographic material with hypersensitised films having a DQE
approaching 10 per cent (Phillipps \& Parker 1993). Due to its original development in 
connection with solar patrol work, it has particularly high efficiency 
around  \ha. The Tech-Pan films are also extremely
fine grained, with an inherent resolution of $\sim5\mu$m, 
leading to an excellent high-resolution imaging capability and a depth for point
sources that exceeded that achieved for the more widely used glass-based IIIa-F emulsion by about 
a magnitude for standard UKST R-band survey 1-hour exposures (e.g. Parker \& Malin 1999).
These factors, combined with the wide area coverage available
to Schmidt photographic surveys, made Tech-Pan an ideal choice for the Southern Galactic Plane \ha\ survey.  
The colour term stability of Tech-Pan compared to the IIIa- emulsions used at
the UKST is given by Morgan \& Parker (2005) where these terms are shown to be stable, reproducible, 
generally small, and similar to those previously derived for the older IIIa- 
emulsions. This gives 
confidence in the survey's photometric integrity. Over the survey life-time, photography on a Schmidt 
telescope still offered several advantages over CCD images, especially low cost and very fine 
spatial resolution and uniformity across a large physical area ($356\times356$~mm) 
giving a 40~deg$^2$ wide field of view. 
However, a key limitation is that the detector response is linear over only a narrow dynamic
range so recovering and calibrating the intensity
information needs careful treatment (see Section 11). 
In Figure~\ref{HA-SR-R} we present small, $3\times3$~arcminute regions to demonstrate the
qualitative difference between the 3-hour \ha\  and 15-min SR Tech-Pan exposures and the
standard 60-min R-band IIIa-F UKST survey data. This region includes a newly discovered planetary nebula
(PHR1706-3544) found from the \ha\ survey data as part of the Macquarie/AAO/Strasbourg \ha\ 
planetary nebulae catalogue
(Parker et al. 2003 and 2005 in preparation). Note the improved resolution of the 
Tech-Pan image, the very similar depth of the respective
exposures for point sources and the tighter point-spread function (psf) for 
the Tech-Pan compared to the IIIa-F emulsion.

\section{The narrow band \ha\ band-pass filter}

To take advantage of UKST's large field of view it
was necessary to obtain a physically large narrow band-pass filter to be
placed as close as possible to the telescope's focal plane. The issues involved with mounting 
such filters with Schmidt telescopes has been described by Meaburn (1978) and previous large 
interference filters were generally of the mosaic type (e.g. Meaburn 1980). 
Such smaller scale interference filters
are easier to manufacture and can be made to higher optical quality. However, difficulties 
associated with their mounting often lead to problems
of missing data in the filter gaps, degraded, variable resolution and lack of homogeneity over large 
survey areas, even when the optical quality of the elements themselves are excellent. This was the case for the
Meaburn mosiac filters which did not fully deliver the anticipated performance due to an unfortunate index 
mis-match in optical cement between the components which resulted in reflection ghosts (which can be got rid of 
numerically after scanning), coupled with the practical difficulty of mounting the components in a sandwich to 
eliminate optical path variations (Meaburn, private communication).

Fortunately, it proved possible for the AAO to obtain a custom-made, exceptionally large,
monolithic, thin-film interference filter from Barr
Associates in the USA which avoids the problems that can be associated with mosaic filters. 
Detailed filter specifications and characteristics 
are given by Parker \& Bland-Hawthorn (1998). The
essential features are reviewed here for completeness together with some additional modeling of the
filter profile in the converging beam when off-axis (see Pierce 2005 for further details).
An RG610 glass substrate was cut
to $356\times356$mm ($\sim6.5$\deg), the standard size of
UKST filters and coated with a multilayer, dielectric stack to give a 
3-cavity design with a clear aperture of $\sim$\,305\,mm diameter and with an effecive 
refractive index of the equivalent monolayer of 1.34. 
This circular aperture of layered coating constitutes the interference filter so the corners of
the square glass substrate do not behave as an \ha\ filter.
Nevertheless this is probably the world's largest astronomical, narrow band filter. 
At the UKST plate scale this covers an on-sky area roughly
5.7\deg ~in diameter (slightly less than the full Schmidt
field). To ensure complete and contiguous
survey coverage with the circular aperture interference filter it was necessary to 
move use 4\deg\ field centres.

The filter central wavelength was set slightly longward of
rest-frame \ha\ for two reasons, one instrumental and one
astronomical. First, the UKST has a fast, f/2.48 converging beam. This
leads to the interference filter `scanning down' in transmitted wavelengths for
off-axis beams compared to beams incident normal to the filter. Secondly
we wished to survey positive velocity gas (in our own
and nearby galaxies). Given a band pass (FWHM) of 70\AA, we chose
to centre the filter at  6590\AA ~in collimated light compared to 6563\AA ~for rest-frame H$\alpha$. 
The peak filter transmission is around 90 per cent. Measurements of the filter at the 
CSIRO National Measurement Laboratory in Sydney quantitatively confirmed the excellent 
conformity of the filter to our original specifications (see Parker \& Bland-Hawthorn 
1998). First light filter images were obtained in April 1997. 

\subsection{The filter model}
\begin{figure}
\centering  
\epsfysize=5.5cm  \epsfbox{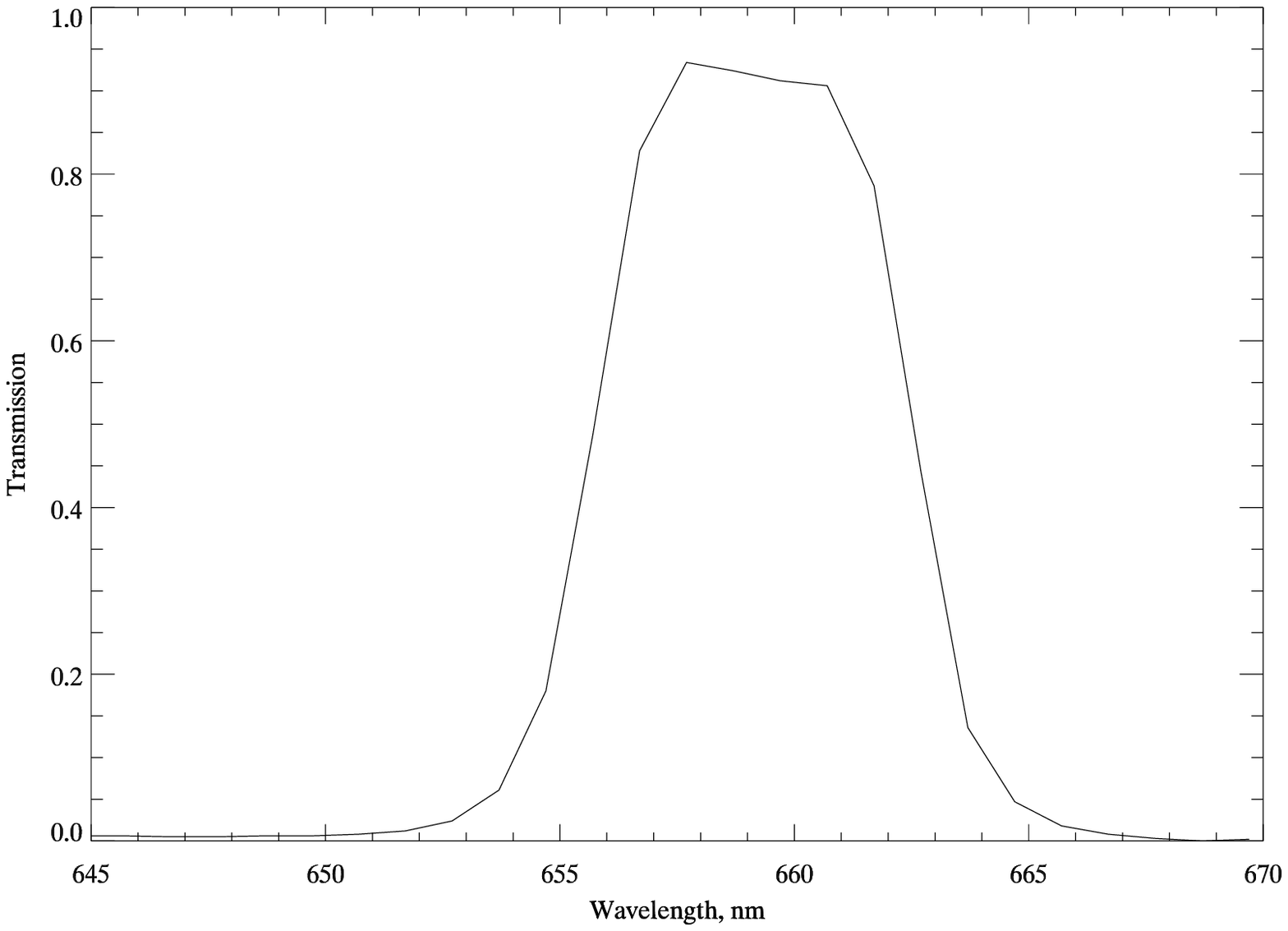}
\epsfysize=5.5cm  \epsfbox{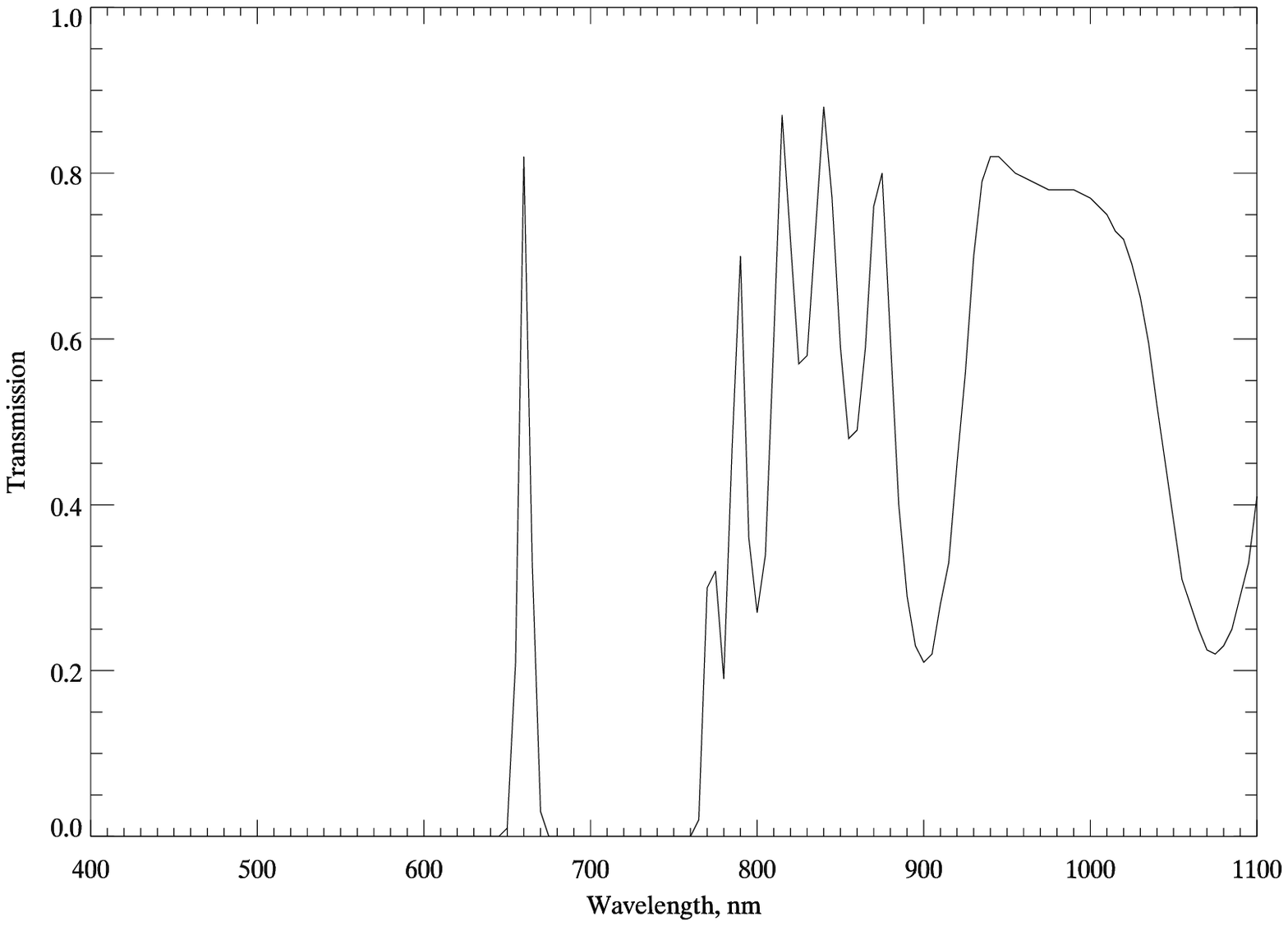}
\caption{The upper plot (a) shows a high resolution transmission scan from
central area of the \ha\ filter. The lower plot (b) shows a wide wavelength range
transmission scan from the central area of the \ha\ filter showing the isolated narrow
peak around \ha. Transmission at longer wavelengths is not recorded in the
survey data as the Tech-Pan film emulsion cut-off is at 6990\AA\ and hence 
not sensitive to light at longer wavelengths.}
\label{specscan}
\end{figure}

Figure~\ref{specscan}a-b shows two spectral scans of the filter, both
taken near the centre using light at normal
incidence. Figure~\ref{specscan}a is the result of a high resolution scan around
the \ha\ region and shows that the bandpass is well centred on
6590\AA\ and has $\sim$\,70\,\AA\ FWHM. The transmission is high across
the reasonably flat top of the bandpass, reaching over 90 per cent. Figure~\ref{specscan}b 
is based on a scan with an extended spectral
range from 4000 to 11000\AA. The CSIRO tests show that the out-of-band filter
transmission is 0.01 per cent or less up to 7600\AA.
Figure~\ref{specscan}b shows that the filter does transmit redward of
7600\,\AA\ at up to $\sim$\,85 per cent, but the survey data will be
unaffected by this as the Tech-Pan film used as detector is insensitive beyond 6990\,\AA.

While this satisfies the intended performance of the filter in light
of normal incidence, in the $f/2.48$ beam of the UKST, light from an
object in the field centre is focused into a cone and
enters the filter at a range of angles up to 11.4\deg. The
bandpass of an interference filter is blue-shifted for light entering
at an angle. This was modelled by breaking down the contributions from the light
cone into a series of concentric rings of size 1\deg\ covering the
telescope beam over a range 0.4\deg\ to 11.4\deg,
each entering the filter at a different angle. The contribution from
the central part of the cone will not be significantly
blue-shifted. The spectral shift was calculated for each ring according
to Equation~\ref{eq:blue} adapted from Elliott \& Meaburn (1976).

\begin{equation}
\label{eq:blue}
\lambda_{\theta} =	\lambda_0\,\cos(\arcsin(\sin(\theta)/\mu))
\end{equation}

\begin{figure}
\centering
\epsfysize=5.5cm  \epsfbox{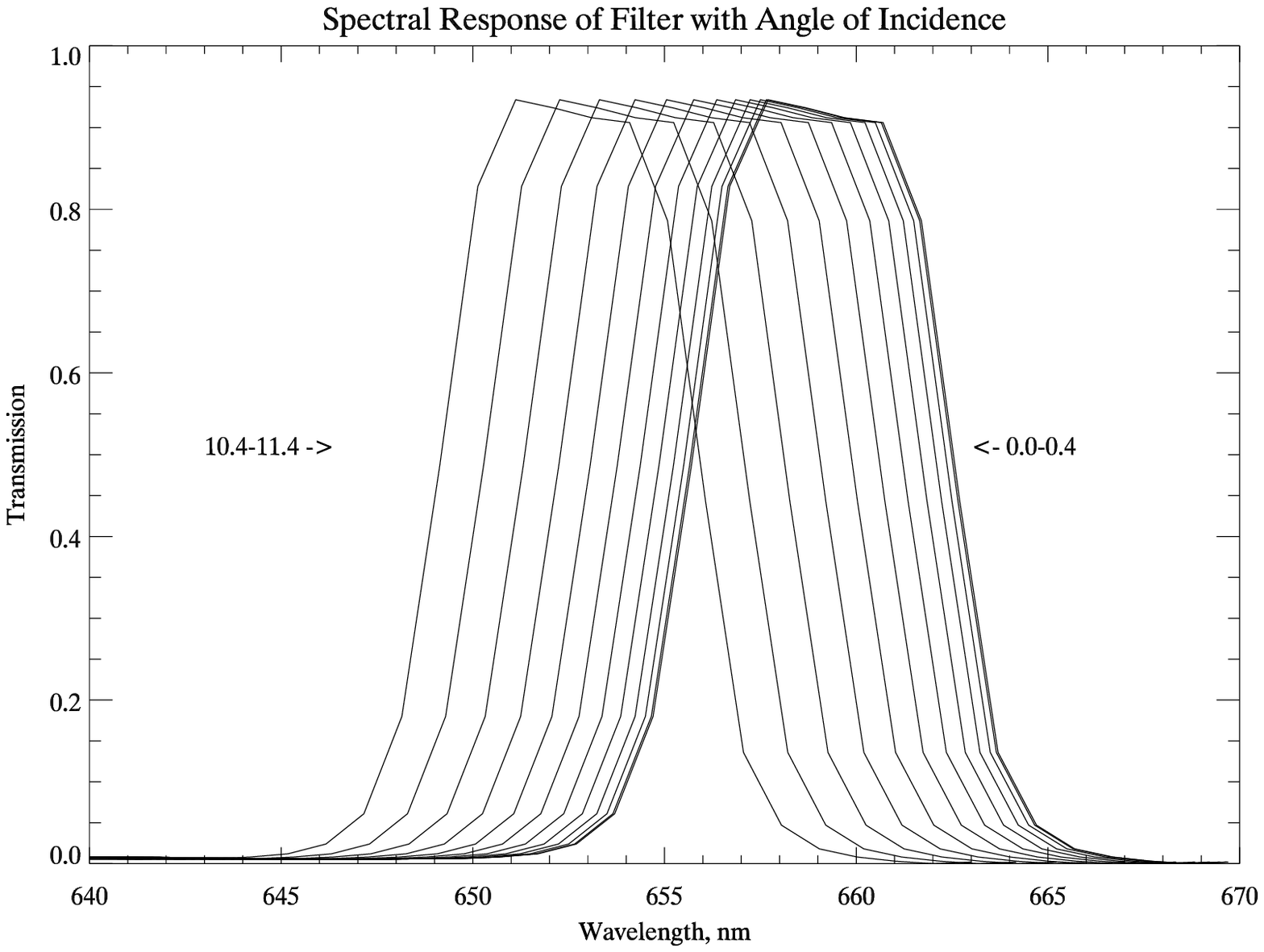}
\epsfysize=5.5cm  \epsfbox{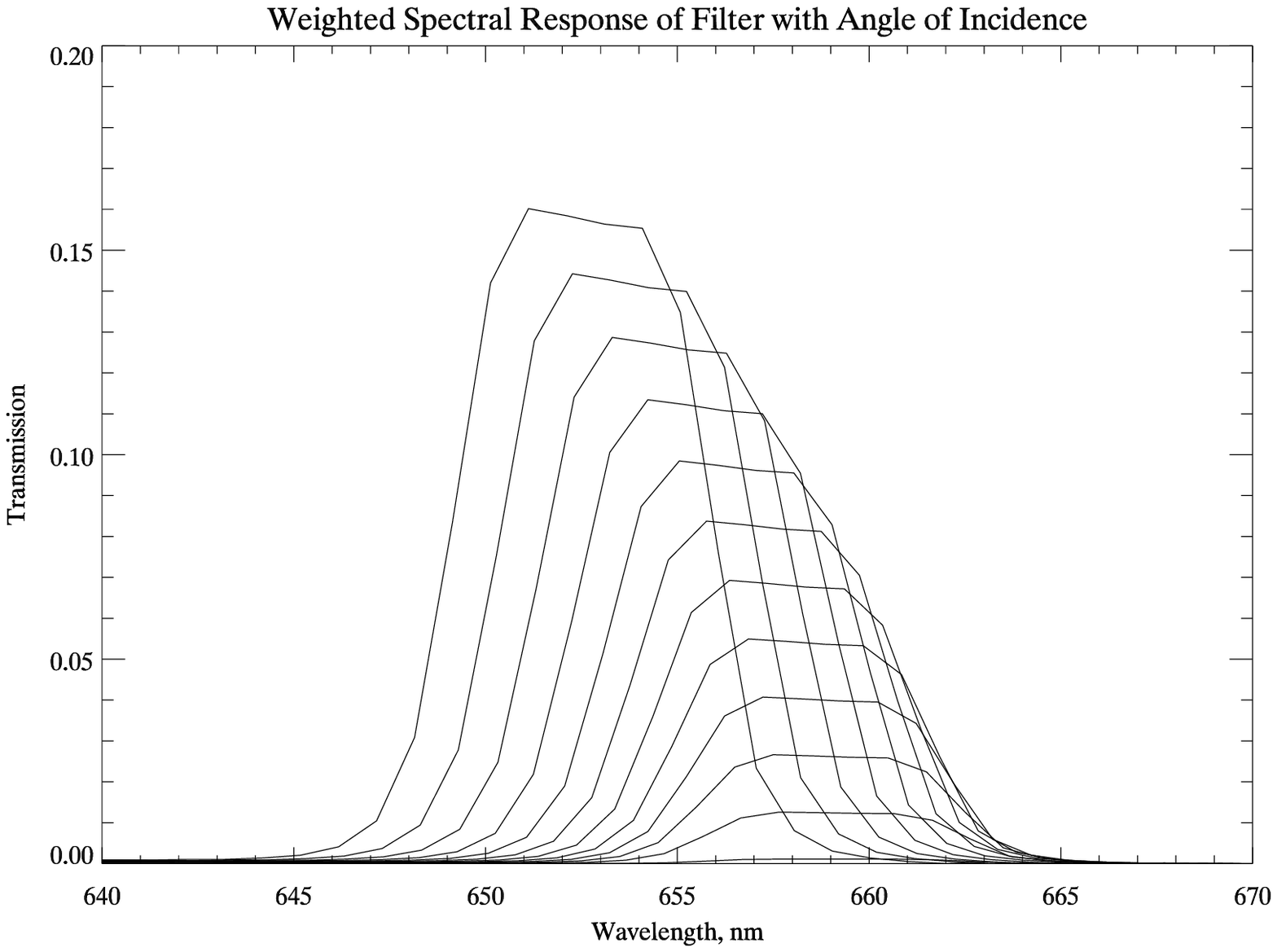}
\epsfysize=5.5cm  \epsfbox{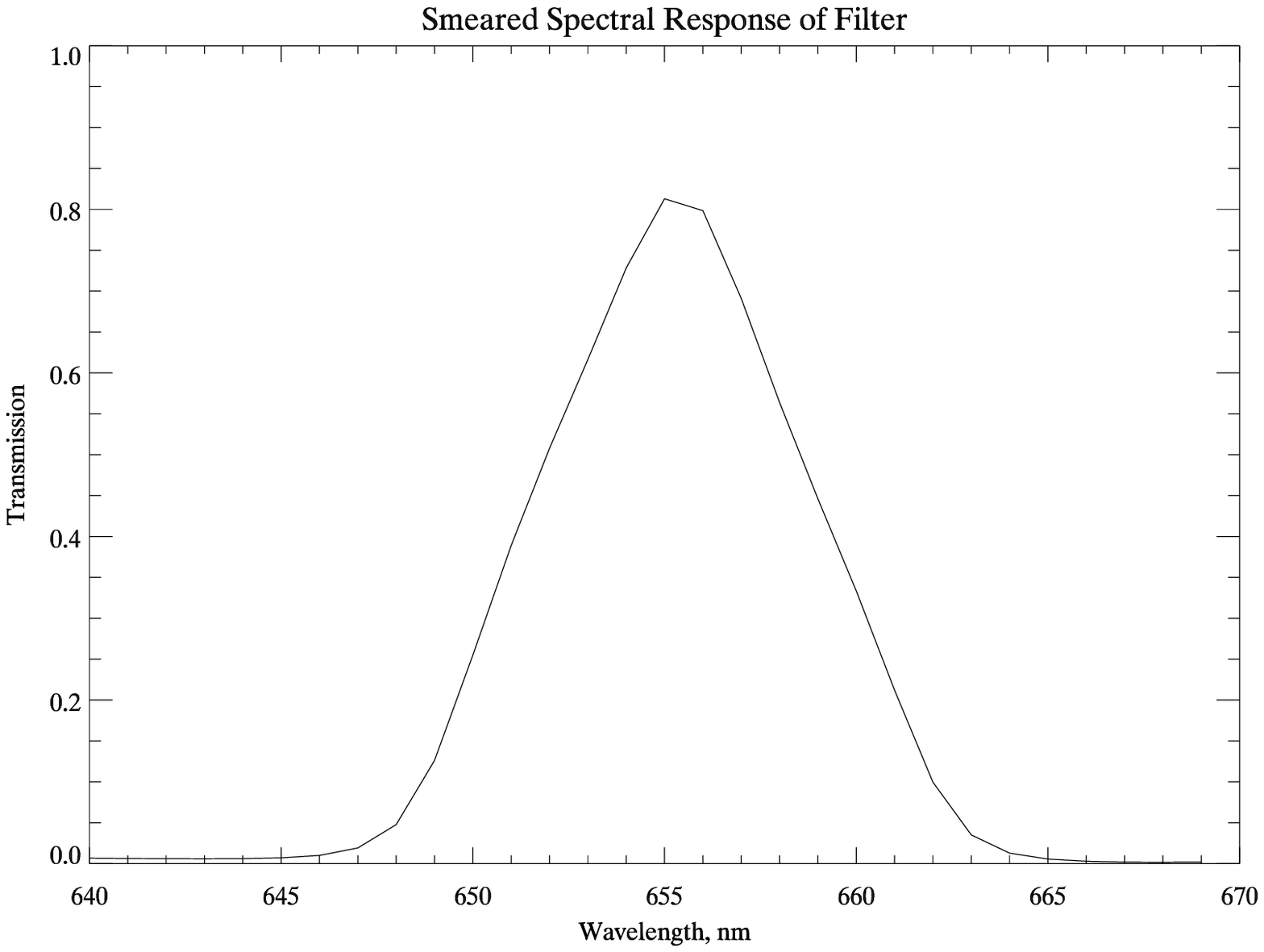}
\caption{Blue-shifting response of interference filter in converging UKST
beam as one moves out from the centre of the field. 
Top plot (a) shows the shift for each concentric ring of the beam,
the second plot (b) shows these shifted response curves weighted according
to the area of the ring. The final plot (c) shows the summed, smeared out
filter transmission curve. The central wavelength of this smeared profile is
6550\,\AA\ and the FWHM is 80\,\AA.}
\label{blueshift}
\end{figure}

Here $\lambda_0$ is the chosen central wavelength for the filter
bandpass in this case 6590\,\AA, $\lambda_{\theta}$ is the shifted central wavelength of
the filter profile based on the
angle, $\theta$, of the incident light and $\mu$ is the refractive
index. A higher refractive index will minimise the blue shifting of
the filter transmission with incident angle of light and the filter
was designed with this in mind. Tests performed by the CSIRO using light at 0\deg,
5\deg\ and 10\deg\ incidence found the effective refractive index of
all the layers combined, ie. the effective monolayer,  is $\mu$~=~1.34. This is the value 
used in Equation~\ref{eq:blue} to generate the shifts for the spectral
response of the \ha\ filter in the UKST beam. These shifts are shown
in Figure~\ref{blueshift}a. The solid lines are the shifting response
curves with the most red response curve being applicable to light of
normal incidence and the most blue response curve tracing the filter
response to light entering at the most extreme angle from the
telescope beam.

In order to combine these to generate a smeared filter response curve
which accounts for the telescope beam, each shifted bandpass is
weighted by the area of the contributing ring as a fraction of the
whole cone. The weighted response curves are shown in
Figure~\ref{blueshift}b and the resulting, summed bandpass is shown in
Figure~\ref{blueshift}c. The FWHM of this smeared bandpass is 80\,\AA,
centred on $\sim$\,6550\,\AA.

This models the transmission of the filter in the centre of a survey field. Towards the
edges of the field the shape of the cone changes and the maximum angle
of incidence is over 14\deg\ which will shift the filter response
further to the blue. The smeared out filter profile is significant as it permits
calculation of the contribution of the contaminant [NII] lines at
6548\,\AA\ and 6584\,\AA, to the flux recorded by the survey. Based on
the smeared out filter response shown in Figure~\ref{blueshift}c the
filter transmits \ha $\lambda$6563\,\AA\ at 80 per cent,
[NII]$\lambda$6548\,\AA\ at 82\% and [NII]$\lambda$6584\,\AA~at
50 per cent. Given that the [NII]$\lambda$6584\,\AA\ line is quantum
mechanically fixed to be three times as strong as the
[NII]$\lambda$6548\,\AA\ line (Osterbrock 1989), this gives a
transmission of 58 per cent for any [NII] emission compared with 80 per cent
transmission for the \ha\ line. This is especially important when 
considering planetary nebulae (PNe) because the strength of the [NII] 
lines varies with respect to the \ha\ line from PNe to PNe and will have a very
significant impact on any calibration scheme based on PNe line flux
standards if not taken into account. Of course, for general diffuse \ha\ emission, the point to 
point \ha\ to [NII] ratio is in general unknown without independent spectroscopic information, 
so we assume a [NII]/\ha\ of 0.3, typically used for the warm ionised medium 
(e.g. Bland-Hawthorn et al. 1998).

\subsection{Survey depth and quality control}

\begin{figure*}
\centering  
\epsfysize=5.5cm  \epsfbox{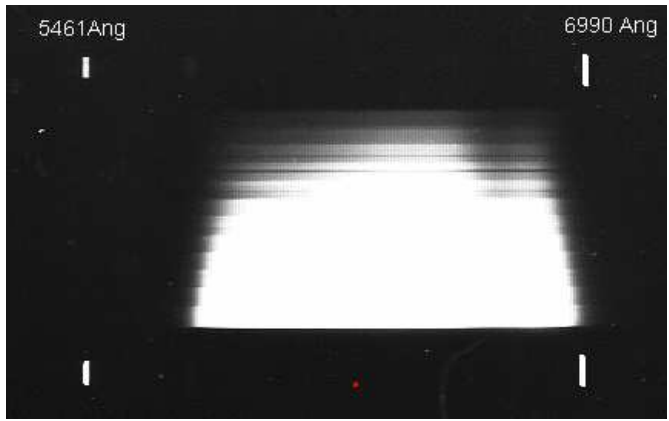}
\caption{Calibration spectrograph result of the effective SR bandpass as a function of wavelength
from the combination of the red OG590 filter and the Tech-Pan emulsion.}
\label{OG590}
\end{figure*}

The \ha\ films are not sky-limited after a 3-hour exposure, but this was chosen as a pragmatic
limit which optimises depth, image quality and survey productivity.
Field rotation and atmospheric differential refraction can adversely affect longer exposures (Watson 1984)
which are also more susceptible to short-term weather and seeing variations. 
The associated 15-minute broad-band SR exposures were taken through the OG590 
red filter. At this exposure level they are
well matched to the depth of continuum point-sources on the matching \ha\ exposure. 
For completeness we include in Figure~\ref{OG590} the effective SR bandpass as a function of wavelength 
obtained from a calibration spectrogram for the OG590 filter in combination with the Tech-Pan emulsion.

With photographic surveys, the magnitude limit for a given survey field
is not a fixed parameter but is a function of factors such as seeing, hypersensitisation 
and development of the films after exposure, emulsion batch variations and the brightness of the
night-sky. Nevertheless, it is clear from comparison with the generally deeper, 
standard UKST R-band survey data, that the approximate magnitude limit for a typical \ha\ 
survey field in an equivalent R magnitude for continuum
point sources is $\sim20.5$ (Arrowsmith \& Parker 2001). This value can be 
directly determined by examining the number magnitude counts
from the matched \ha\, SR and R band SuperCOSMOS Image Analysis Mode (IAM) data 
(see later) for a given field and determining the point where 
completeness breaks down. As an illustration we give magnitude limit 
estimates for continuum point sources in A and B grade
exposures of two \ha\ survey fields in Table~\ref{depth}.

Additionally, the use of the same emulsion for both \ha\ and SR exposures ensures an 
excellent correspondence of their image psf's when film pairs are
taken under the same observing conditions. The intention was to take the \ha\ and SR 
exposures consecutively as far possible. 
This greatly simplifies the inter-comparability of both types of exposure. Of the 233
survey fields, only 100 are in fact sequential pairs while most of the rest were
taken a few days apart. However 45 fields had a gap of one or more years between
the \ha\ and SR survey exposures because one or other of the exposures had to be 
repeated to satisfy the stringent survey quality acceptance criteria.
\begin{table}
\begin{center}
\caption{The depth of each of the four original images measured in R
equivalent magnitudes.}
\begin{tabular}{|c|c|c|c|} \hline
Exposure & Survey & Survey & Histogram num/mag peak\\
Number & field  & Grade  & (R equiv. magnitude)\\ \hline
HA18745 & h527 & A2 & 20.77 $\pm$ 0.025\\
HA17850 & h527 & BI3 & 20.42 $\pm$ 0.025\\
HA18749 & h678 & A1 & 20.62 $\pm$ 0.025\\
HA17935 & h678 & BIE4 & 20.42 $\pm$ 0.025\\
\hline
\end{tabular}
\end{center}
\label{depth}
\end{table}
Strict quality control has been applied to the survey pairs by M.Hartley and S.Tritton
according to well established criteria before any exposure is allowed to be incorporated into the 
survey. This ensures that the most uniform and homogeneous data set possible is 
created. Each exposure grade is determined by
means of a score with `0' being the best and `3' being the limit
for an exposure to be considered an `A' grade (highest quality).  The image grade is
recorded in the information and data sheets which accompany the survey
data, together with a letter code to indicate which is the most significant
contribution to the score. Long, 3-hour exposures are
prone to field rotation which can cause image trailing (denoted by T
in the image grade), poor weather can lead to curtailed exposure times
(U for underexposed). Cosmetic defects such as emulsion faults
(E), haze halos (H) and processing streaks (P) can also contribute to
a poor grade. These defects can be present in either the \ha\ or the
SR image. Where possible, any survey exposure which was not
rated A grade was repeated. Unfortunately, a few B-grades had to be accepted into the survey though
over 90 per cent were deemed survey quality, maintaining the high standards set for all UKST surveys.

\section{Astrometric accuracy of the SHS}

Astrometric calibration of survey photographic material measured on SuperCOSMOS
is discussed in Hambly et al.~(2001). The calibration procedure consists of
applying a six coefficient (linear) plate model to measured positions of
Tycho--2 catalogue reference stars, along with a radial distortion
coefficient appropriate to Schmidt optics (i.e. $\tan r / r$) and a fixed,
higher order two--dimensional correction map to account for distortion
induced by mechanical deformation of the photographic material when clamped
in the telescope plate holder to fit the spherical focal surface. 
As demonstrated in Hambly et al.~(2001c), this
yields absolute positional accuracy of typically $\pm0.2$~arcseconds for glass
plates. The SHS, on the other hand, employs film media which cannot be as
mechanically stable as glass on the largest scales. However, provided a
sufficiently dense grid of reference stars is available, it is possible
to map out the unique distortion pattern that any one film may present. 

\begin{figure*}
\epsfbox{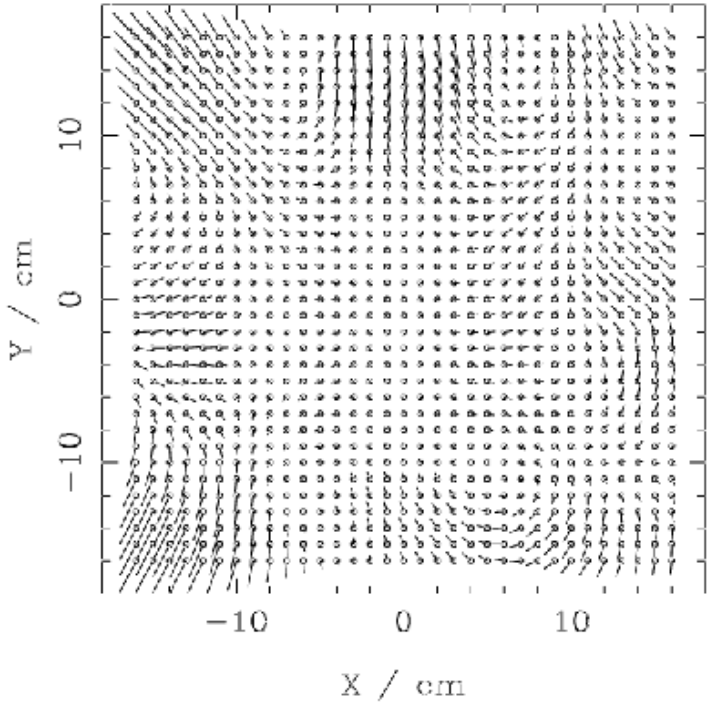}
\caption{Systematic astrometric distortion pattern of SHS H$\alpha$ survey field h67, 
reduced using a standard six coefficient linear fit plus
a radial distortion term, when compared to the UCAC catalogue. The scale 
size of the vectors is 0.5~arcsec to 1~cm. Systematic positional errors 
of more than 1~arcsec (corresponding to one tick mark on either axis) are
observed in the film data, e.g.\ in the left--hand corners.}
\label{vector}
\end{figure*}

\begin{figure*}
\epsfbox{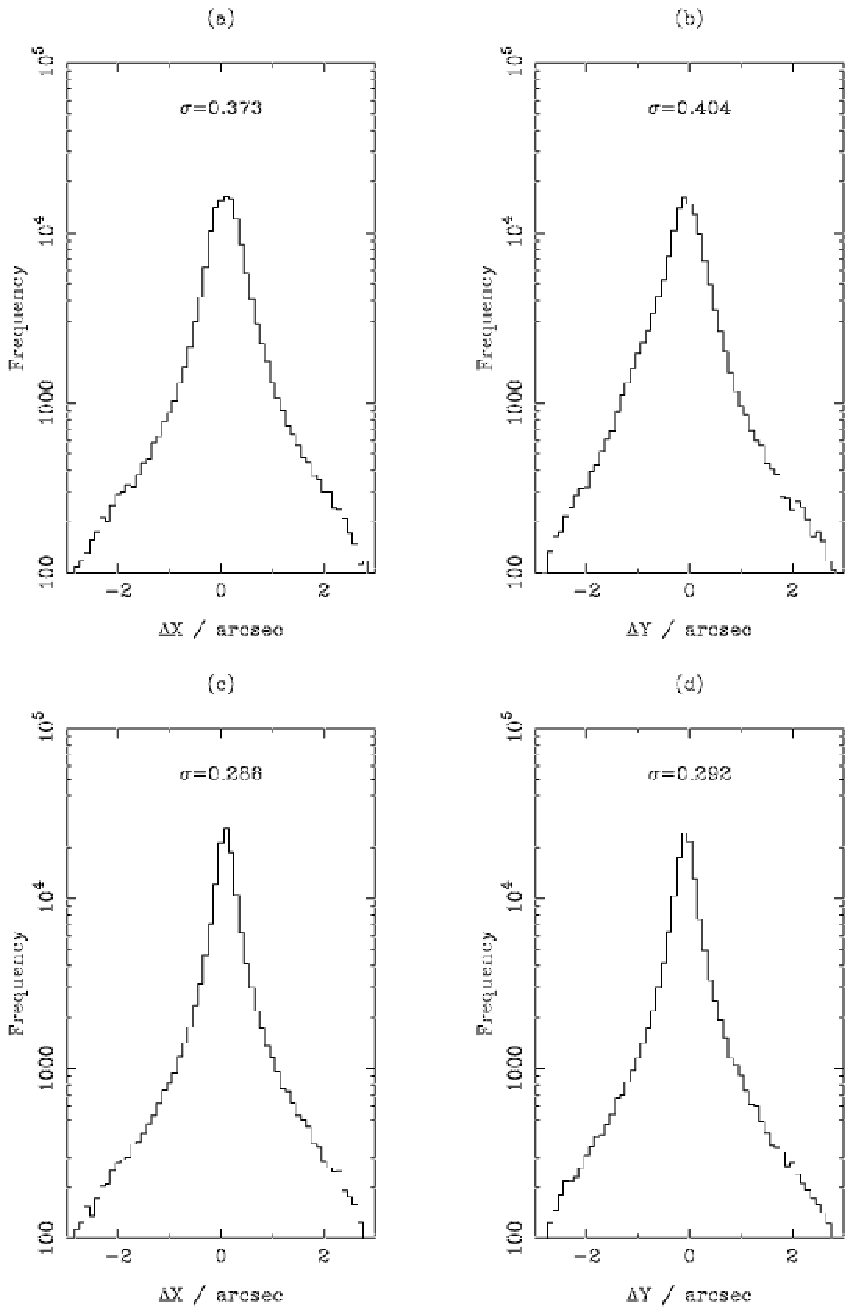}
\caption{Histograms of residuals between SHS and UCAC astrometry (see text)
for: (a,b) uncorrected positions, and (c,d) positions corrected 
during the astrometric reduction procedure using the
distortion map shown in Figure~\ref{vector}. The accuracy is quantified by a
robustly estimated RMS residual, in either co-ordinate, and shows a
$\sim30$\% improvement when the correction map is employed. From this
analysis, SHS absolute astrometry is typically accurate to 
$\sim0.3$ arcsec.}
\label{histfig}
\end{figure*}

In order to achieve the best possible astrometry for the SHS, the generic SuperCOSMOS Sky Survey
(SSS) astrometric reduction procedure was modified by replacing the averaged
distortion map with a correction stage where the individual
film distortion pattern is measured with respect to the UCAC astrometric
reference catalogue (Zacharias et al.~2004). In Figure~\ref{vector} we show the
results of comparing first-pass SHS astrometry (i.e. without correction
of any higher order systematic distortion) with the UCAC catalogue for
a single SHS film. Residuals have been averaged in 1~cm boxes and smoothed
and filtered using a scale length of 3 box widths. A systematic distortion
pattern is clearly seen, and comparing with figure~1 of Hambly et 
al.~(2001) there is no four-fold symmetry in the pattern, which is a
characteristic of mechanical deformation of rigid glass plates. Moreover,
similar plots for different films show different patterns, so a fixed
correction map cannot be applied across the entire survey film set.
Figure~\ref{histfig}(a,b) shows histograms of the residuals of individual UCAC
standards from which Figure~\ref{vector} is derived; a robustly estimated RMS
(i.e. a median of absolute deviations scaled by 1.48, to be equivalent to a
Gaussian sigma) is found to be about 0.4~arcseconds. Now, if the SHS positional
data are corrected during the astrometric reduction procedure
using the map values displayed in Figure~\ref{vector}, the RMS
drops to $\sim0.3$~arcseconds; the new histograms of individual residuals are
displayed in Figure~\ref{histfig}(c,d). The value of $\pm0.3$~arcseconds can be taken as
indicative of the typical global astrometric accuracy of the SHS in either
co-ordinate, and compares favourably with the figure quoted for the SuperCOSMOS Sky Surveys (SSS)
of $\sim0.2$~arcseconds, given the higher level of crowding of the SHS fields.

\section{The Survey SuperCOSMOS digital data}

The high speed `SuperCOSMOS' measuring machine at the Royal Observatory Edinburgh 
(e.g. Miller et al. 1992, Hambly et al. 1998) has been used to scan the 
\ha\ and SR exposure A-grade pairs at 
$10\mu$m (0.67~arcsec) resolution. 
The same general scanning and post-processing reduction 
process is employed as for 
the directly analogous SuperCOSMOS broad-band surveys of the Southern Sky (SSS)
currently on-line and outlined in detail by Hambly et al. (2001 a,b,c). 
The user interface is broadly equivalent and the main features are
summarised neatly in Figure 1 of Hambly et al. (2001a). However, due to the special nature of the
survey, some additional processing steps and \ha\ specific options have been added to 
create the on-line SuperCOSMOS \ha\ Survey (SHS) described below. 

\subsection{Basic characteristics of the on-line `SHS' \ha\ Survey} 

The Wide-Field Astronomy Unit (WFAU) of the Institute for Astronomy Edinburgh is responsible 
for maintaining the \ha\ survey data products. Both the \ha\ and SR data for
the 233 Southern Galactic Plane survey fields are available on-line
\footnote{http://www-wfau.roe.ac.uk/sss/halpha}. Unfortunately, there are no plans for the 40-field Magellanic
Cloud H$\alpha$ and SR survey pairs to also be put online.
The data products are given as FITS files (see http://heasarc.gsfc.nasa.gov/docs/heasarc/fits.html)
with comprehensive FITS header information detailing 
key photographic, photometric, astrometric and scanning parameters (e.g. Hambly et al. 2001b). 
The FITS images also have an accurate built-in World Co-ordinate System (WCS). 
This permits easy incorporation
into other software packages such as the STARLINK {\it GAIA} environment for subsequent 
visualisation, investigation, manipulation and comparison with other data. 
The entire survey data are stored on RAID disks for fast access and a comprehensive set of web-based 
documentation has been provided.
The pixel data map for each field is about 2~Gb. 
The scanned pixel data are processed through the standard SuperCOSMOS thresholded object
detection and parameterisation software (e.g. Beard, MacGillivray \& Thanisch 1990) 
to produce the associated Image Analysis Mode (IAM) data for each
field. This process determines a set of 32 image-moment parameters which provide the astrometry,
photometry and morphology of the detected objects. Full details of the
image detection and parameterisation are given in Hambly et al. (2001b). 
For the SHS survey, a selection of the 32 most important IAM parameters from the merging of the H$\alpha$,
SR and I band data for each detected image in the SHS are available and are given in Table~\ref{iamparams}.

\begin{table*}
\centering
\caption{The selected 32 IAM parameters used in the merged SHS catalogue data}
\label{iamparams}
\begin{tabular}{cllll}
\hline
  \multicolumn{1}{c}{Number} &
  \multicolumn{1}{l}{Name} &
  \multicolumn{1}{l}{Type} &
  \multicolumn{1}{l}{Description} &
  \multicolumn{1}{l}{Units}\\
\hline

1	&RA	&Double	&Celestial Right Ascension	&radians (FITS)\\
2	&DEC	&Double	&Celestial Declination	&radians (FITS)\\
3	&EPOCH	&Real	&Epoch	&year\\
4	&l2	&Real	&Galactic longitude	&dec.degrees\\
5	&b2	&Real	&Galactic latitude	&dec.degrees\\
6	&R\_Ha	&Real	&\ha\ equivalent R-mag	&mags\\
7	&SR2	&Real	&matching SR magnitude (SR2)	&mags\\
8	&SR1	&Real	&First epoch SR magnitude (SR1)	&mags\\
9	&I	&Real	&First epoch I magnitude	&mags\\
10	&AREA\_Ha	&Integer	&Total area \ha\ image	&pixels\\
11	&AREA\_SR2	&Integer	&Total area SR2 image	&pixels\\
12	&AREA\_SR1	&Integer	&Total area SR1 image	&pixels\\
13	&AREA\_I	&Integer	&Total area I image	&pixels\\
14	&ELL\_Ha	&Real	&H-alpha image ellipticity\\
15	&ELL\_SR2	&Real	&SR2 image ellipticity\\
16	&ELL\_SR1	&Real	&SR1 image ellipticity\\
17	&ELL\_I	&Real	&I image ellipticity\\
18	&PRFSTT\_Ha	&Real	&\ha\ N(0,1) profile classification statistic	&0.001 sigma\\
19	&PRFSTT\_SR2	&Real	&SR2 N(0,1) profile classification statistic	&0.001 sigma\\
20	&PRFSTT\_SR1	&Real	&SR1 N(0,1) profile classification statistic	&0.001 sigma\\
21	&PRFSTT\_I	&Real	&I N(0,1) profile classification statistic	&0.001 sigma\\
22	&BLEND\_Ha	&Integer	&\ha\ Deblending flag (0 if not deblended)\\
23	&BLEND\_SR2	&Integer	&SR2 Deblending flag (0 if not deblended)\\
24	&BLEND\_SR1	&Integer	&SR1 Deblending flag (0 if not deblended)\\
25	&BLEND\_I	&Integer	&I Deblending flag (0 if not deblended)\\
26	&QUAL\_Ha	&Integer	&\ha\ image Quality flag\\
27	&QUAL\_SR2	&Integer	&SR2 image Quality flag\\
28	&QUAL\_SR1	&Integer	&SR1 image Quality flag\\
29	&QUAL\_I	&Integer	&I image Quality flag\\
30	&PA	&Integer	&Celestial position angle (selected band)	&degrees\\
31	&CLASS	&Integer	&Classification flag (selected band)\\
32	&FIELD	&Integer	&SHS field number (4\deg\ centres)\\
\hline\end{tabular}
\end{table*}

The full resolution, $10\mu$m pixel data and associated IAM parameterised data 
for both the \ha\ and SR scanned exposures are stored on-line on a field by field basis. On the \scos\ web-site the scanned survey 
data for each field has the prefix `HAL' before the survey field number (so H$\alpha$ survey field h350 = 
HAL0350 for example, when referring to the on-line digital \scos\ data).
The SR images have been transformed to exactly match the pixel grid of the master \ha\
exposures which permits direct image blinking and comparison between the pixel 
data for each field. The general \ha\ survey data products are accessed via a web interface 
that has the same look and feel as the existing broad-band SuperCOSMOS on-line `SSS' surveys but with 
some additional functionality. The IAM data produced for each 
field can be downloaded separately if desired or assembled into seamless catalogues on-the-fly
which can cover several adjacent fields using the `Get a Catalogue' option. 
The combined IAM data is organised into a full listing of 53 image parameters or a more manageable subset 
of the most useful 32 as in Table~\ref{iamparams}. 
A set of `expert' options are also available to further select catalogue extraction parameters.
A special feature to create a difference image of each field following variable image psf 
matching techniques developed by Bond et al. (2001) also exists to permit large-scale resolved emission maps 
to be created with reduced artefacts from uncancelled stars. 
This can be computationally intensive and so
is not generally available without prior arrangement with the Wide-Field Astronomy Unit. 
For most applications simple
quotient imaging between the \ha\ and SR pixel data is sufficient due to the well matched psf's and depth.

A $16\times$ blocked-down version of each field is also available as both a GIF image 
and as a FITS file which has the WCS built in to the FITS header. These 
whole field maps can be studied to select smaller regions of interest for extraction at full resolution using the
`Get an Image' option.
The full resolution pixel data access limit is currently set at $9000$~arcmin$^2$ with 
regions downloaded as FITS files (also with WCS) and both the SR and \ha\ data for the same 
region can be downloaded simultaneously. Areas for extraction can be chosen via equatorial 
(J2000 or B1950) or galactic {\it (l,b)} 
co-ordinates in a variable $m\times n$ arcminute rectangular region format. A clickable map of the 
current fields on-line enables individual field details to be displayed prior
to viewing the blocked full field image.
A batch mode enables large numbers of thumb-nail images to be extracted around objects 
of interest with the option to return \ha\ and/or SR postscript plots of the extracted images.  
An option to apply a `Flat-Field' to the \ha\ pixel data in intensity space is included to permit 
correction of the non-uniformities in the measured exposures arising from the 
excellent but slightly varying \ha\ filter transmission profile. 
This has been shown to work effectively and is described in Section~9.
A radius of 153~mm ($\sim2.85$ degrees) from each survey field centre has been adopted as the region with 
good data ($<15$ per cent correction factors). The `good data' radius from each scanned \ha\
field centre has been used in creating a confidence map which is incorporated into the extracted FITS 
image as an additional FITS extension (extension [3], e.g test-image.fits[3]). This can be used to flag areas
of the extracted image that might not be quite as good as others. Currently this has values of 100 for 
regions extracted interior to this radius and 0 for regions outside. 

\subsection{Incorporation of the SSS `$I$' band data}

\begin{figure}
\centering  
\epsfysize=5.0cm  \epsfbox{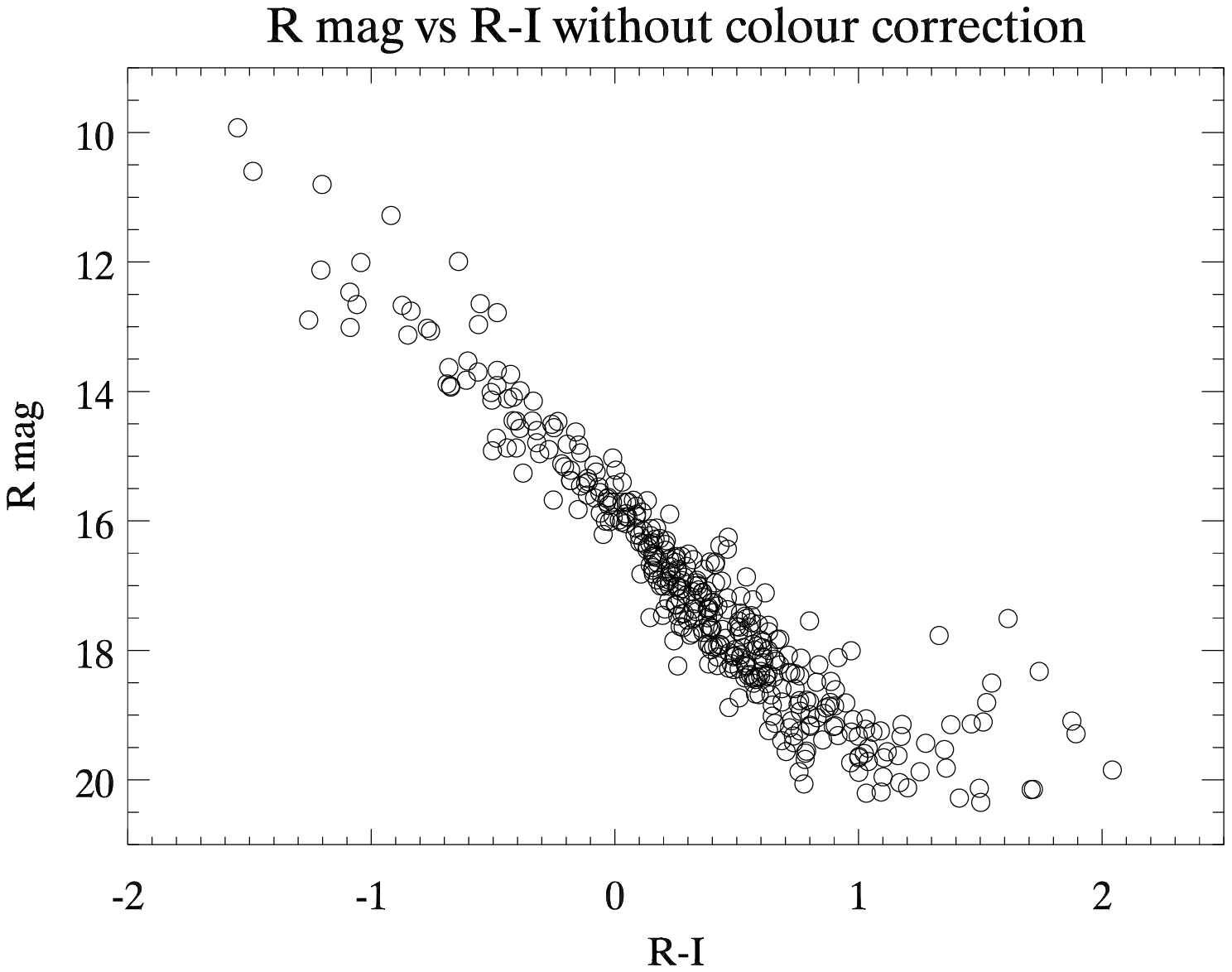}
\epsfysize=5.0cm  \epsfbox{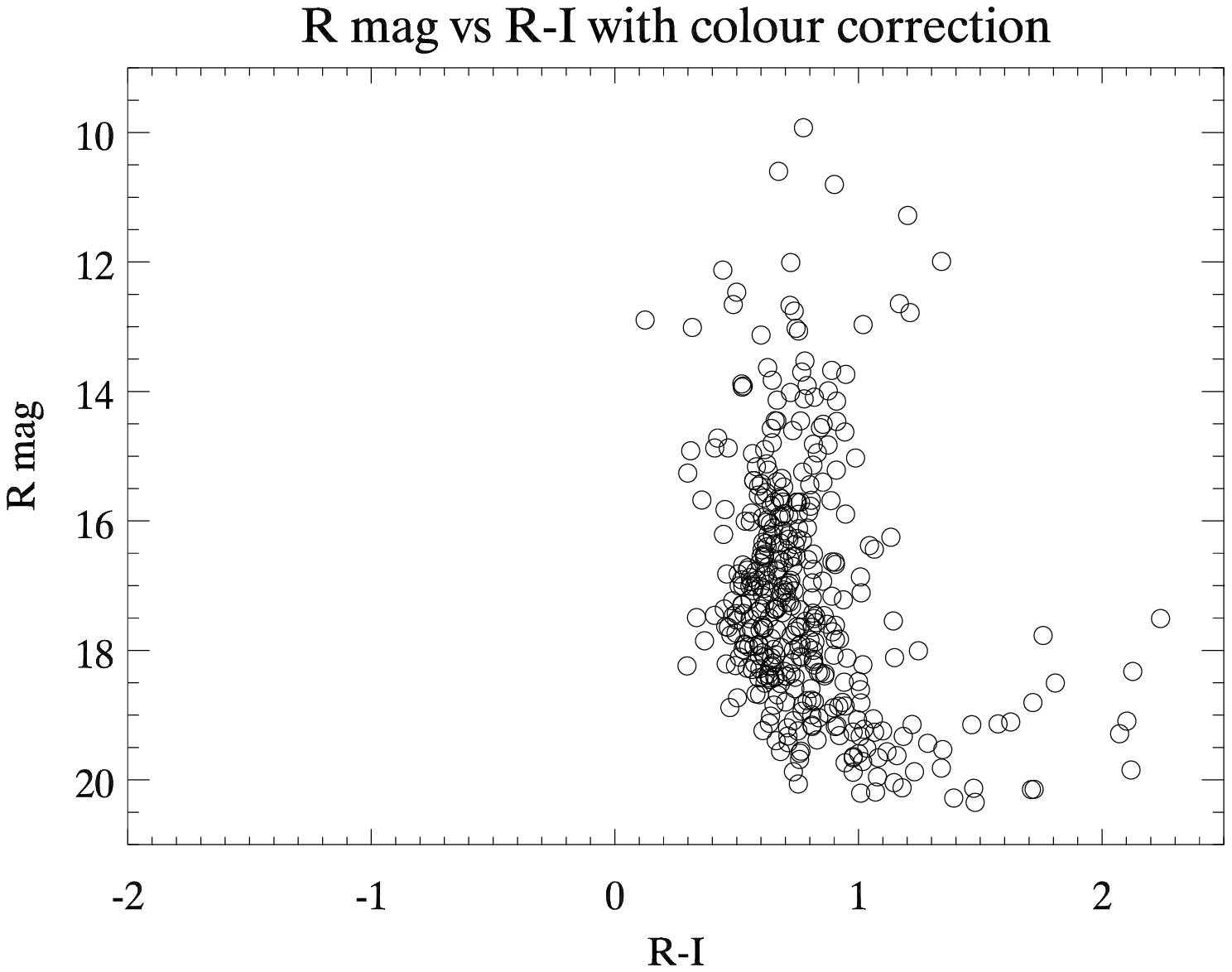}
\caption{Plots of SR magnitude versus $R-I$, from a 10~arcminute square region
in Monoceros, before colour correction for
positional and magnitude dependent errors (a - upper ) and after (b - lower).}
\label{colcorr}
\end{figure}

The IAM catalogue downloaded directly via the `Get a Catalogue' 
option or as incorporated in the FITS table extension to the
dowloaded pixel data via the `Get an Image' option, contains information 
not just from the \ha\ and the SR images but also from
the SERC-I (near infrared) survey which has been carefully matched in with the SHS data. 
The I-band data is particularly useful when searching for point-source \ha\ 
emitters as it can help to eliminate contamination from late-type stars. 
However there are issues that any user should be aware of when combining the I-band data
with the SHS magnitudes. When calibrating UKST data, positional
and magnitude-dependent systematic errors are present as a result of variations in
emulsion sensitivity and vignetting towards the image corners (Hambly et al. 2001b).
The I-band survey was taken on standard ESO/SERC
5\deg\ field centres, so the vignetting will have a different effect on
the two sets of photometry at a given survey location. Furthermore, the I-band data are calibrated to
relatively few standards. These differences are evident when
looking at a plot of SR magnitude versus $R-I$ colour derived from the
SR and paired I-band photometry. Figure~\ref{colcorr} shows two colour
magnitude diagrams (CMD) for stars taken from a 10~arcminute region centred
on the middle of SHS field h1109 in Monoceros where the low galactic
reddening of $E(B-V)$\,=\,0.24 (Schlegel, Finkbeiner \& Davis 1998) should leave the
$R-I$ colour roughly constant for much of the observed magnitude range.
Figure~\ref{colcorr}a shows the raw result, where a large,
unphysical variation of 3.5 magnitudes is seen in the $R-I$ colour
from the brightest to the faintest objects. This is removed as a first-order correction from the
survey data by selecting a master colour, in this case the SR, and
correcting the I-band across all survey fields. Note however that at fainter limits one in fact
expects redder $R-I$ colours as such stars are likely to be further away and prone to be more dust reddened 
or intrinsically fainter and therefore more likely to be late types. Hence 
some modest slope is expected. Figure~\ref{colcorr}b
shows the CMD for the same patch of sky after the colour correction
has been applied, giving a result in better agreement with
expectation. The data can be downloaded in corrected or uncorrected
form via an option in the ``Expert'' parameters of the SHS
website. It is important to ensure that the I-band correction is not applied inappropriately,
i.e. in a field of intrinsically high reddening, as such a correction will remove genuine
features from the data. It should be used with caution.

\section{SHS point source photometry}
A significant advantage of the SHS data over its rivals is the ability to
detect point sources which have been photometrically calibrated
to CCD standards (e.g. Boyle et al. 1995, Croom et al 1999).
With measurements of isophotal magnitude and object classifications,
it is possible to apply a photometric calibration to the \ha\ and SR
films by comparing the SuperCOSMOS raw magnitudes of stars from the Tycho-2
Catalogue (Hog et al. 2000) and the Guide Star Photometric Catalogue
(Lasker et al. 1988). These in turn are checked against
photometric standards derived from the CCD observations given by
Croom et al. (1999) and Boyle et al. (1995).
The narrow-band \ha\ images are calibrated
to an `R-equivalent' scale. The 3 hour \ha\ and 15 minute SR exposures 
are matched so that both reach similar
depths of m$_R$\,$\simeq$\,20.5 for point sources. Where an object is detected in one
band but not in the other a default value of 99.999 is given in the
catalogue data for the magnitude in the missing bandpass. Positional
and magnitude dependent errors are seen in the raw photometric data,
created by varying transmission profile and diffraction effects
through the thick (5.5\,mm) \ha\ filter, but these are corrected for in
the data available through the SHS website by comparison with the SR
data. Photometric consistency is achieved by using the overlap regions
between fields to match zero-points across the survey. These corrected
magnitudes provide a means of selecting point-source emitters. The variations
in measured IAM stellar parameters as a function of field position arising from the variable psf from
field rotation, vignetting etc, especially at large radii from the field centres, requires that such
selection is performed over limited 1-degree areas. In this way stars with an
emission line at \ha\ will show an enhanced \ha\ magnitude compared
with the SR magnitude. At the bright end of the magnitude distribution, severe photographic and \scos\ 
saturation effects come in to play, limiting stellar
photometry to R of about 11-12 in both the \ha\ and SR pass-bands.

\section{Spurious images in the SHS}
Spurious images appear from time to time in all photographic images scanned by SuperCOSMOS. 
They have a variety of forms and causes and are present in images extracted from the SSS as well as SHS.
They can sometimes be picked up by examination of the pixel images directly, though they are often missed, 
and can also appear as spurious detections in the IAM data. They have a variety of
sizes and shapes and may be in or out of focus depending on whether the contaminating source is on the 
emulsion surface or on the platten used by SuperCOSMOS to sandwich the film flat for scanning. 
Here, we differentiate between spurious images in the emulsion itself caused by processing defects, 
emulsion flaws and static marks, and those caused by foreign objects 
on the surface of the emulsion or on the back of the film. Holes and scratches in the emulsion 
surface can also give rise to spurious images. Satellite trails and transient phenomenon also give rise
to real developed images which may have no counterpart in other survey bands of the same region. 
We do not consider these here. 

\subsection{Basic causes}
The SuperCOSMOS facility is situated in a class-100 clean room and each film is pressure air-cleaned prior 
to scanning. However, despite best efforts, particles that may already have been present on the
emulsion before shipment to \scos\, manifest themselves as spurious images. 
The biggest cause is fine particulate dust (20--100$\mu$m). Unfortunately the Estar film base 
of the Tech-Pan emulsion is prone to static charge build-up. 

\subsection{Recognising spurious images}
\begin{figure*}
\mbox{
\epsfxsize=0.3\textwidth\epsfbox{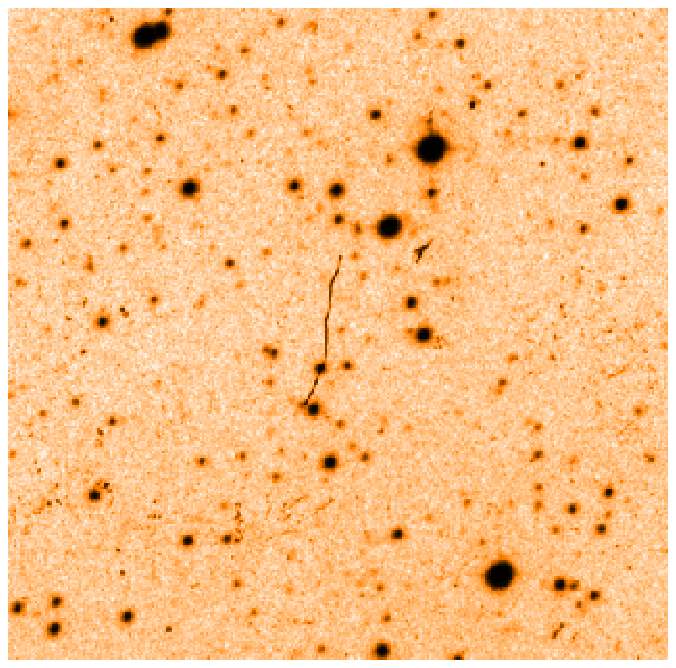}}
\mbox{
\epsfxsize=0.3\textwidth\epsfbox{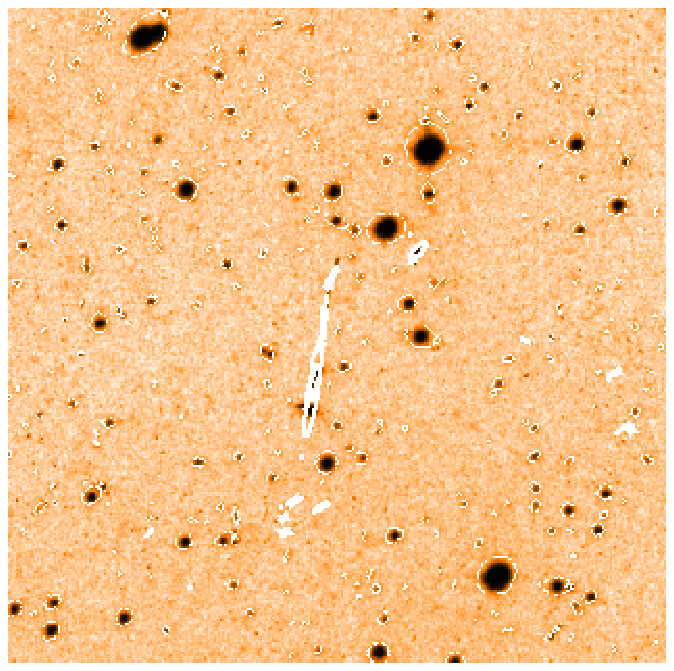}}
\caption{$5\times5$~arcminute extracts of SuperCOSMOS \ha\ data from SHS survey field h273
highlighting a contaminating spurious image together with a matching image with the IAM data overlaid and with
all the spurious images in the frame highlighted.}
\label{crud}
\end{figure*}

The SuperCOSMOS scanning system is highly specular so detritus present on the emulsion surface that is often 
invisible when viewed under diffuse illumination conditions (such as on a light table) is
revealed in sharp relief in the SuperCOSMOS data. 
The number of artefacts seen in the SHS data is somewhat worse than on other glass plate based
surveys of the SSS. Fortunately, having matched exposures in two bands makes identification of
such artefacts more straightforward. For example, since the \ha\ and SR exposures are registered on the same pixel
grid, quotient imaging can reveal the locations of spurious images. 
A $5\times5$ arcminute region extracted from h273, a field with a particularly high number of 
spurious images, is shown in Figure~\ref{crud}. 

We can take advantage of the fact that the pixel image properties of spurious images are usually quite 
distinct from real astronomical images, often having a sharpness below that possible from the 
combination of
telescope optics and seeing disk. Their shapes are often highly irregular and non-symmetrical such 
that they would not fit any normal psf. This makes them amenable to Fourier filtering.
Objects that have no counterpart in the other band are potential spurious image candidates though 
variable objects, novae and the effects of de-blending complicate the issue significantly. Various IAM 
parameters such as the profile statistic, ellipticity etc. may also aid in identification. Furthemore, spurious 
IAM objects arising from de-blending overlaying contaminating fibres or hairs usually have very high 
ellipticities
which may help in isolating likely candidates. Storkey et al. (2004) discuss techniques for recognising 
and eliminating spurious objects in the on-line SuperCOSMOS surveys. As yet this procedure has not been 
applied to the SHS data. 


\section{Flat-Fielding of the survey data}

\begin{figure}
\centering  
\epsfysize=6.5cm  \epsfbox{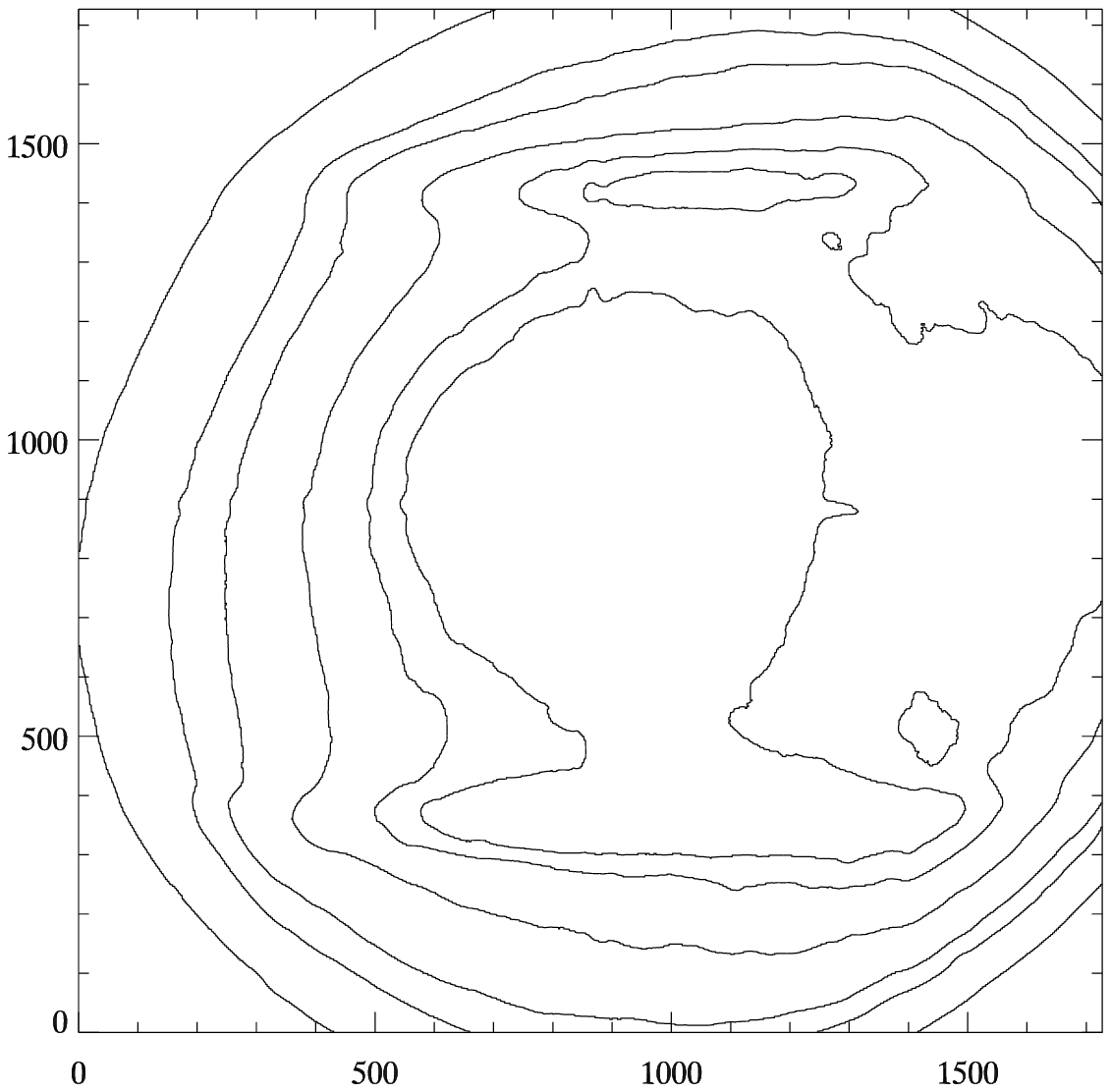}
\epsfysize=5.5cm  \epsfbox{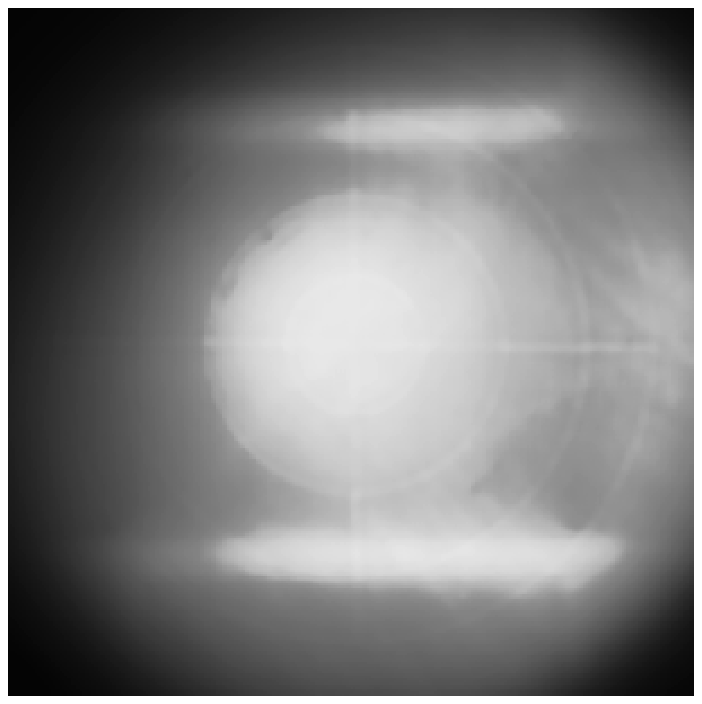}
\caption{ (a - top) Contour plot of the narrowband \ha\
filter transmission over the SuperCOSMOS scanned area of $5.156^{o}$ 
($1728\times1728$pixel). 
The contours are plotted at 85, 90, 92, 95, 97 and 98 per cent. 
The bottom image (b) is a histogram equalisation of the same flat-field data which reveals the
filter artefacts present at the 0.1 per cent level}
\label{flatfield}
\end{figure}

For any interference filter of the size used here, low-level 
non-uniformities exist which lead to residual non-physical background variations in
the exposed images. In order to establish the magnitude of 
such effects, three flat-field exposures were taken
with the filter subject to uniform illumination. The flat-field images 
permitted evaluation of the combined
effects of filter transmission in the fast, $f/2.48$ Schmidt beam and telescope
vignetting (see UKST Unit handbook, Tritton, 1983). 
The flat-field images were exposed to place them
on the linear portion of the film's characteristic curve and
were averaged to give the filter/telescope transmission profile shown
in Figure~\ref{flatfield}a-b.

Figure~\ref{flatfield}a shows transmission contours 
at 85, 90, 92, 95, 97 and 98 per cent of maximum transmission in the central region. 
The response is seen to be asymmetric, with the 97 per cent contour
extending beyond the edge of the 5.16\deg\ field on the right, which
corresponds to the west of the survey fields. In the east the transmission 
decreases more rapidly, reaching 85 per cent at the eastern edge. Towards the filter 
corners the transmission drops further. However, the 4\deg\ overlapping centres 
(Section 4) and the asymmetric nature of the response allows the selection 
of \ha\ data requiring flat-field correction of less than 15 per cent for any 
given area of sky, provided that adjacent fields are available. Most data 
will require much smaller flat-field corrections. The effect of 
flat-field corrections as large as 15 per cent
on pixel data is considered in the survey calibration section. 
Figure~\ref{flatfield}b is a histogram equalization of the actual flat-field pixel map
which reveals the extremely low level artefacts present at the 0.1 per cent level which are invisible
in a linear rendition.
The flat-field correction has been stored as a transmission array 
with maximum value unity, so it is applied by dividing survey image 
values in intensity space by the relevant correction array elements. This is available as the default
option on the SHS website. The correction breaks down towards the corners of the scanned 
image and in regions outside of the clear aperture because the density of the 
exposures at the edges of the circular aperture is too low to lie on the straight 
line portion of the characteristic curve leading
to an over-correction and also because the \ha\ filter transmission is 
becoming increasingly skewed in these extreme regions. Raw transmission or photographic density values and
generically calibrated intensity values without flat-field correction can also be requested
on the download form. The IAM data is obtained 
from the raw \scos\ scans without flat-field correction.

\subsection{Specific filter features}
Despite the superb quality of the filter, low-level, large-scale variations 
in transmitted flux can be seen in the \ha\ survey images under 
certain exposure conditions. In particular there
are two parallel bands of slightly enhanced transmission 
(leading to elevated photographic density) going E-W in the north and south of the filter. 
These bands are only 1--3 per cent higher in intensity than the
surrounding regions. A series of low-level artefacts which are not obvious in the contour plot because
they are at a level of $<$ 1 per cent can just be discerned in the filter transmission image
in Figure~\ref{flatfield}a-b. They can also
just be seen in the contour plot as the spike in the 98 per cent contour just
right of centre. The observed shape of these artefacts mimics the series of shallow concentric 
grooves scored into the surface of the mandrel to enable the Tech-Pan film
to be sucked under light vacuum to the curved focal surface of the plate-holder to ensure proper focus.
They are thought to arise from the backscattering off these 
grooves of light that has passed right through the film. 
Again, it is gratifying that these artefacts, present at the $<0.1$ per cent level, 
are effectively removed by application of the flat-field.

\begin{figure}
\centering  
\epsfysize=7.0cm  \epsfbox{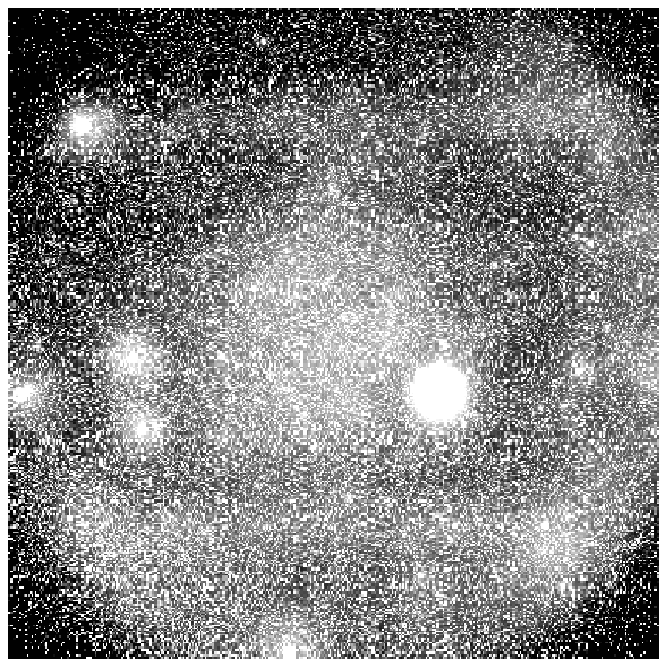}
\epsfysize=7.0cm  \epsfbox{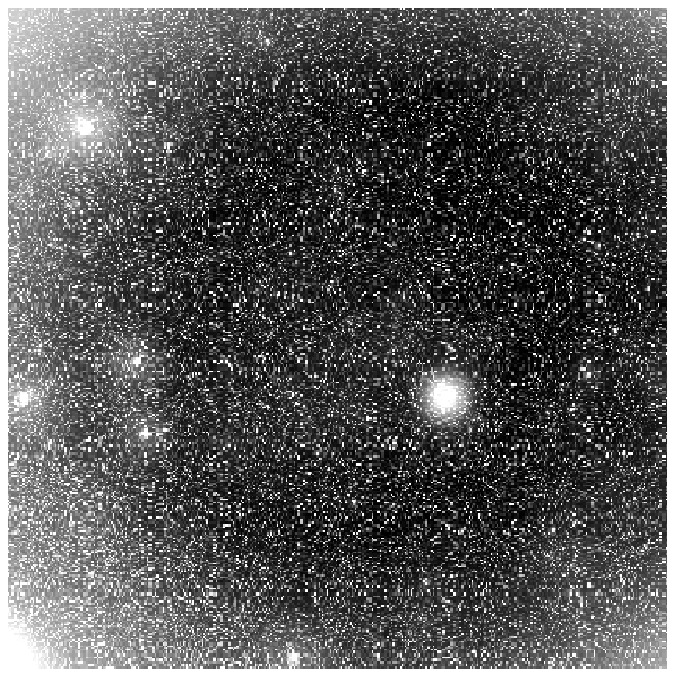}
\caption{Two 16$\times$ blocked down, 5.156\deg\ SHS \ha\ images of field
h410. The top image (a) is the raw intensity data without flat
field correction. The bright areas trace not Galactic emission but the
filter transmission profile seen in Figure~\ref{flatfield}. The bottom image (b)
has had the flat-field correction applied which
successfully removes the filter artefacts from within the circular aperture.}
\label{HAL0410}
\end{figure}

\subsection{Application of the flat-field and correction validity}

Field h410, which sits away from the Galactic Plane
on the extreme edge of the survey at
($l, b$) 330.2\deg, +10.28\deg\, has been chosen to test the vailidty of the
flat-field correction as it contains a very low-level
isotropic background of Galactic line emission. In the survey image
this will be moderated by the filter response. Figure~\ref{HAL0410}a-b
shows two 16$\times$ blocked down images of survey field h410. The
top image Figure~\ref{HAL0410}a is the raw \ha\ data, before the application of the
flat-field correction. The structure evident on the field as lighter
areas is not Galactic emission but matches the filter transmission
profile. The two horizontal bars are present and the image is less
exposed towards the corners where the recorded intensity is lower and
the star density also falls. The bright star just to the right of
centre is $\epsilon$ Lupi.

Application of the flat-field correction results in Figure~\ref{HAL0410}b 
which should have a flat background wash
of emission across it. The structure from the filter is no longer
evident and the bright areas are now in the corners, where the larger
flat-field correction over-corrects the SuperCOSMOS intensity counts.  
This will not adversely affect the majority of the
pixel data available on the SHS website. Data can always be taken from the
best area of filter response and no flat-field correction larger
than 15 per cent is necessary for any of the survey data that overlap a
neighbouring field. Data from the edge of the survey, where no
adjacent field exists, have been made available and may require a
correction greater than 15\%. Areas affected in this way are flagged
in the third extension table which accompanies the downloaded FITS image.

\section{Geocoronal \ha\ emission}

Geocoronal \ha\ emission is caused by fluorescence after solar
Lyman~$\beta$ excitation of atomic hydrogen in the exosphere. Because
imaging surveys lack velocity resolution for the emission they record,
the geocoronal contribution will be present in all of them but
indistinguishable from bona fide Galactic \ha\ emission. Fortunately, one
modern \ha\ survey, WHAM (e.g. Haffner et al. 2003), 
offers very good velocity resolution
($\sim$12~km/s) and is able to separate the atmospheric
emission from the Galactic emission and measure the
intensity. Nossal et al. (2001) report on \ha\ observations carried out
by the WHAM instrument in 1997, the same time as the SHS imaging was
starting at the UKST. They find that the geocoronal emission intensity
depends on how much the earth shades the line of sight from sunlight. 
Their resulting plot of geocoronal \ha\ emission as a function of earth
shadow height shows that at heights greater than 6000~km
only a very low level
$\sim$2~R wash of geocoronal \ha\ emission is present.
Based on the observational information available in the headers of the
SHS images, it is possible to calculate the shadow heights for any
field. For a random sample of 6 SHS fields the shadow height for the
whole three hour observation and across the five-degree field of view
was found to be greater than 6000~km, so low-level geocoronal emission
is not problematic, as most Galactic Plane fields covered by the SHS will contain significantly
stronger emission.

\section{Application of an absolute Calibration to the \ha\ Survey data}

The AAO/UKST \ha\ survey data needs an absolute
intensity calibration if the full scientific value of its
sensitivity to faint, diffuse emission is to be realized. The intensity
calibration must provide a reliable means of transforming the pixel
intensity values from SuperCOSMOS scans of the 
\ha\ images into meaningful intensity units such as
Rayleighs
which is consistent from field to field. We show
that continuum emission can be successfully removed from the \ha\
images by scaling and subtracting the SR continuum image. 
Unlike CCD data, which enjoys a linear response
over a wide range of emission strength, photographic data can be very
difficult to calibrate because the response of the
emulsion and \scos\ scanner is linear only over a relatively small dynamic
range. Variations in sensitivity and background occur from exposure to
exposure, especially when the \ha\ and SR pairs were taken on different nights, phases of the moon
etc. Despite this, we show that
the survey data have been well exposed to capture Galactic emission on
the linear part of the characteristic curve 
and can be calibrated by means of comparison with the
complementary, accurately intensity calibrated  SHASSA survey (Gaustad et al. 2001). This process
does not form part of the current SHS release but can be undertaken by the user as required.

\subsection{Image comparison with SHASSA}

The Southern \ha\ Sky Survey Atlas by Gaustad et al. (2001)
provides wide-field narrow-band CCD \ha\ images of the southern sky
below $\delta$~=~+15\deg\ taken with a robotic imaging camera sited at
Cerro Tololo Inter-American Observatory (CTIO). The camera used
a small, fast, f/1.6 Canon lens which gave a very large (13\deg)
field of view and a spatial resolution of 48~arcseconds. Each SHASSA field
was imaged through a narrow-band interference filter of width 32\,\AA\
centred at 6563\,\AA\,  as well as a continuum filter with two bands of
61\,\AA\ at 6440\,\AA\ and 6770\,\AA, on either side the \ha\
line. The SHASSA website\footnote{http://amundsen.swarthmore.edu/SHASSA/} makes available the
raw \ha\ and continuum images as well as 48~arcsecond and 4~arcminute resolution
continuum subtracted, intensity calibrated data.
\begin{figure*}
\mbox{
\epsfxsize=0.47\textwidth\epsfbox{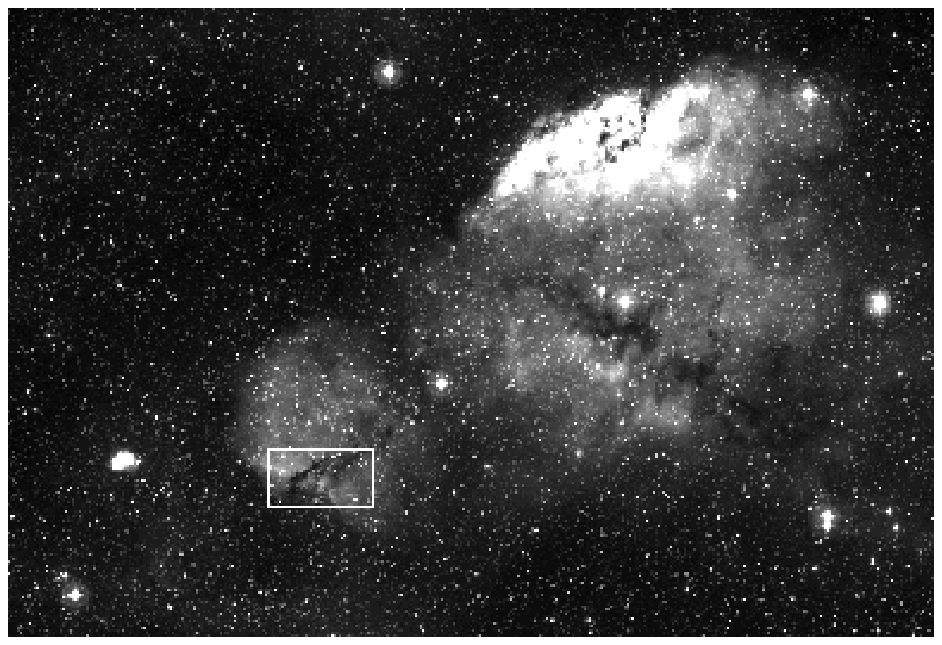}}
\mbox{
\epsfxsize=0.45\textwidth\epsfbox{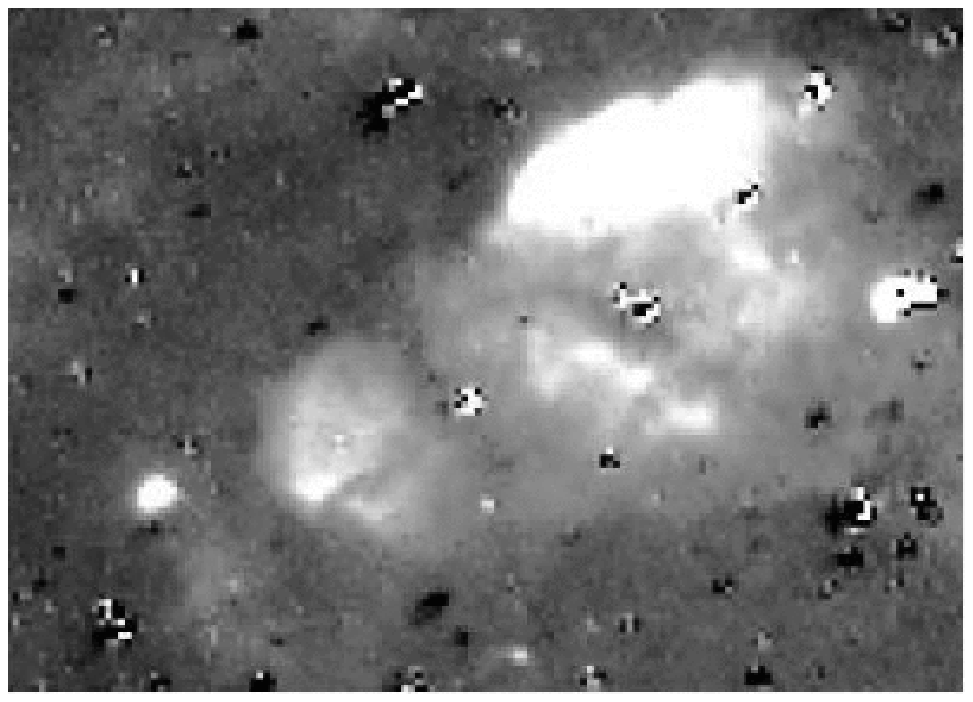}}
\mbox{
\epsfxsize=0.6\textwidth\epsfbox{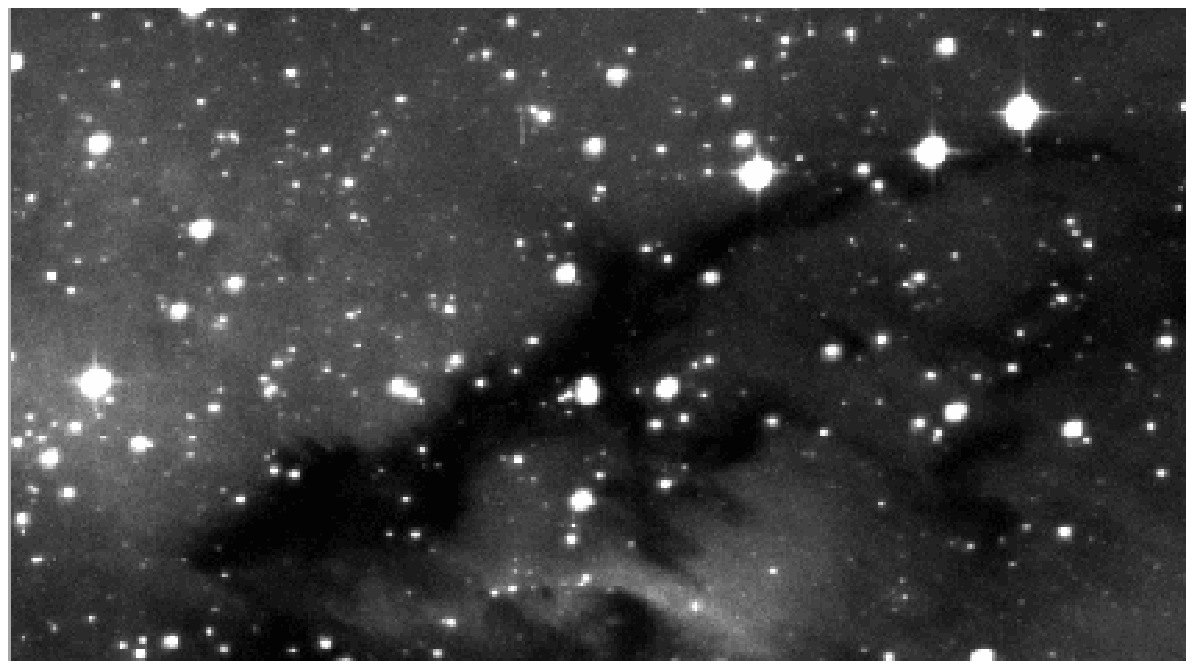}}
\caption{An $105\times75$~arcminute region centred on well known HII region 
RCW~19 (GUM~10) at $\alpha$,$\delta$ 
$= 08^h16^m17^s$, -35\deg57'58'' (J2000) extracted from the SHS 
survey (top left) and the equivalent SHASSA region (top right). A full 
resolution $8.3\times4.6$~arcminute region from the SHS survey 
data is shown below centred on the southern HII region component at
$08^h19^m$, -36\deg14' and indicated in the lower resolution SHS image with a rectangle.}
\label{shassa-shs}
\end{figure*}
The SHS data is superior in terms of resolution while the general sensitivity of both surveys
appears qualitatively similar for large-scale emission features. For example, in Figure~\ref{shassa-shs} 
we present SHS and SHASSA images of the HII region RCW~19.

\subsection{The SHASSA intensity calibration}

The SHASSA intensity calibration was derived from the
planetary nebula spectrophotometric standards of
Dopita \& Hua (1997) after the continuum images had been scaled and
subtracted from the \ha\ frames. Aperture photometry for eighteen of
the bright PNe standards was measured from the SHASSA images and used
to calculate the calibration factor for the whole survey. A difficulty
in applying PNe line fluxes to \ha\ narrow-band imaging is the
proximity of the two [NII]~$\lambda\lambda$6548,~6584 lines which are
included in the flanks of the SHASSA \ha\ filter bandpass. These vary in
strength relative to \ha\ between PNe and could significantly affect the
result. Calculating the transmission properties of the interference
filter to these lines is complicated by the blue-shifting of the
bandpass with incident angle, an effect which must be treated
carefully in the SHS data as the filter sits in the fast f/2.48 beam
of the UKST. This problem is not as severe for the SHASSA data because
in this case the filter sits in front of the camera lens, leaving only
the effects of the very large field of view. These effects are
considered in Section~4 of Gaustad et al. (2001) and carefully accounted
for in their calibration.

To allow a more detailed comparison, aperture photometry for 87 PNe with a range of surface 
brightness and integrated flux and with an independent measure of H$\alpha$ flux, 
was carried out on SHASSA images. Published spectroscopic data were used to deconvolve the 
contribution from the [N II] lines passed by the SHASSA filter.  The results agree with 
published data to $\Delta$~F(H$\alpha$)~=~--0.01dex, $\sigma$ = 0.05 for SHASSA minus 
literature fluxes (Frew 2005, in preparation; cf. Pierce et al. 2004).  
Since the PNe literature fluxes have associated errors, the SHASSA calibration  is better than $\pm10$ per cent
across the whole survey, in agreement with the nominal error supplied by Gaustad et al. (2001).

An additional uncertainty is introduced to the zero-point of the SHASSA
intensity calibration from the contribution of geocoronal emission. 
Gaustad et al. (2001) estimate this by comparison with overlapping WHAM
data points and interpolating where there are none. Our check of the
intensity calibration against independent flux measures of planetary nebulae
indicates that the geocoronal contribution to the SHASSA \ha\ images
has been successfully removed. Finkbeiner (2003) also showed there is no significant offset 
between WHAM and SHASSA data. So we conclude that the SHASSA data has been well 
calibrated to a zero-point consistent with independent measurements and
therefore have confidence in its use as a baseline calibration for the SHS data.

\subsection{Continuum Subtraction of the SHS}

Diffuse emission recorded through the narrow-band \ha\ filter on the Tech-Pan films
will be a combination of Galactic \ha\ line emission, continuum emission, night-sky 
auroral lines and geocoronal emission. Ideally all of these components would need to be
disentangled to extract just the Galactic \ha\ emission. In practice the geocoronal and auroral
emission is considered as a low ($\sim2$~R) level but temporally varying uniform wash which 
simply elevates the general background on each exposure to a slightly varying degree.

The matching SR images provide a measure of the continuum
component and, properly scaled, can be used to produce continuum
subtracted \ha\ images. Although the \ha\ and SR exposures are generally exposed to attain 
the same depth for continuum point sources, the nature of photography and the vagaries of the observing
conditions (e.g. if the exposure pairs were not contemporaneous and taken 
in different moon phases or if the seeing changed) 
mean that the depth and image quality between the \ha\ and SR exposures can and does vary. 
Hence it is necessary to determine a continuum subtraction scaling factor between \ha\ and SR on a 
field by field basis. This factor must be precisely determined for the
continuum subtraction to be effective. For high-dynamic range CCD exposures, the standard 
method for determining the appropriate scaling factor to subtract continuum from narrow-band 
is to compare aperture photometry for stars on both images whose exposures are normally 
interleaved on short timescales. 

Unfortunately, this does not work well with the \ha\ SR film exposure pairs (Pierce 2005),
often leading to under or over-subtraction of the continuum. This arises
due to varying backgrounds on the film exposures caused by; low-level emulsion sensitivity variations 
between films (especially if they come from different hypersensitised batches), inherent chemical fog
variations in the emulsion, processing variations and true sky background variations arising for the 
reasons given above. These varying backgrounds result in the same magnitude stars saturating at 
different levels on different exposures as their Gaussian point spread functions are superimposed on
top of any diffuse emission and elevated background which can severely truncate their peaks. 
The limited dynamic range of
\scos\ also acts as a further low ceiling above the background, which leaves
little room for these bright stellar Gaussians making it hard to effectively utilise stellar photometry to
determine the correct scaling factors.

\begin{figure}
\centering  
\mbox{\epsfxsize=0.35\textwidth \epsfbox{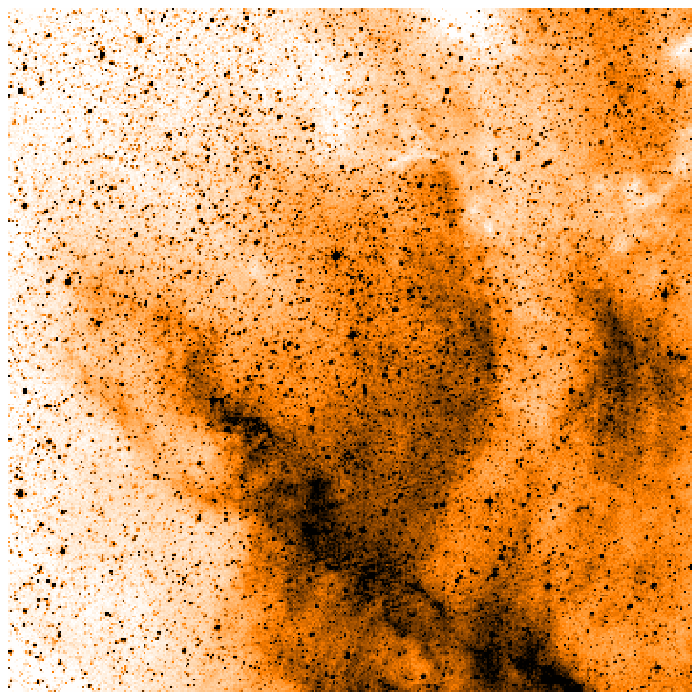}}
\mbox{\epsfxsize=0.35\textwidth \epsfbox{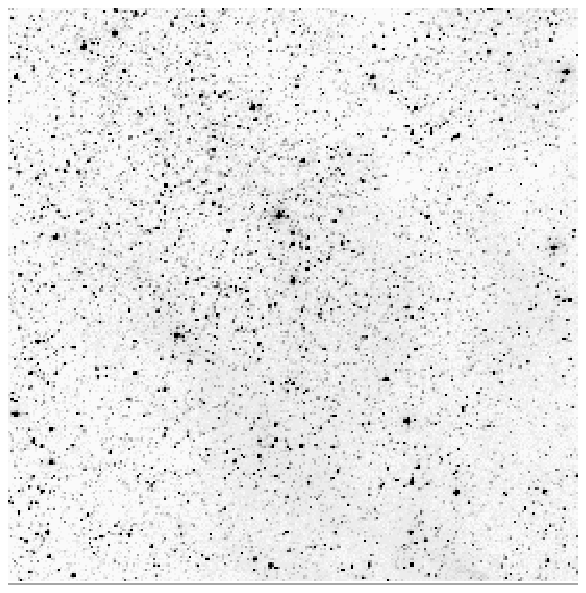}}
\mbox{\epsfxsize=0.35\textwidth \epsfbox{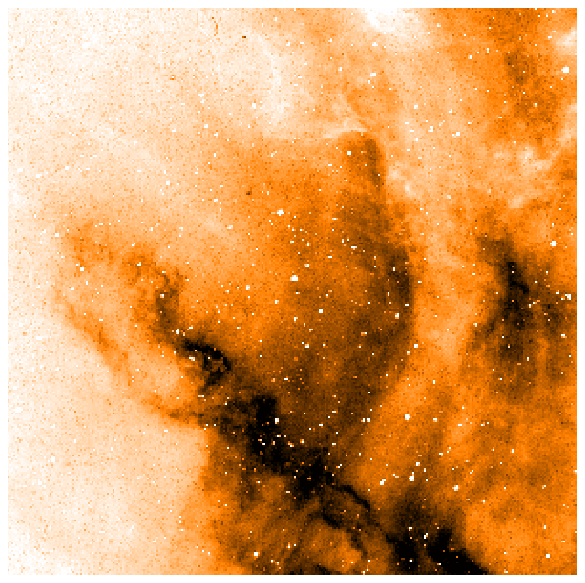}}
\caption{30~arcminute region at $16^h47^m,-49^o 00'$
(J2000) from near the centre of SHS field h350 with bright areas of emission shown darkest. The top image
is the raw \ha\ image, the middle the matching SR while the bottom
image is a continuum-subtracted image of the same area. The subtraction, based on a
comparison with SHASSA data, has worked well, leaving only
minor residuals around stars on the bright diffuse emission.}
\label{1647m4900_2}
\end{figure}

Fortunately, we are able to use the existing SHASSA data to 
provide a well-determined scale with an independently confirmed
zero-point to compare with and calibrate the SHS images. Even exposure pairs taken years
apart can be successfully continuum subtracted. A detailed investigation of the SHS calibration 
process has been undertaken by Pierce (2005) but the
essential aspects of this process and its application are given here.
For example, Figure~\ref{1647m4900_2} shows three images of a 30~arcminute region
taken from field h350, which shows strong, varying Galactic \ha\ emission. The top image 
is the \ha\ image downloaded from the SHS website, the middle image is its SR counterpart while the
bottom image has had the SR `continuum' image scaled and subtracted
via a comparison with SHASSA data. In general this subtraction is very
good, removing most of the stellar images and the diffuse continuum.
Only the stars which sit in the strongest emission have been
over-subtracted and appear as white spots in the image.

For a given area
of sky, pixel data from each survey can be downloaded and, after
matching for spatial resolution, the \scos\ intensity counts can be
compared directly with the Rayleigh values in the SHASSA data. A plot
of continuum-subtracted SHASSA pixels against equivalent SHS pixels
should return a linear relation with a common zero point if the
reduction and intensity calibration have been properly carried
out. Comparing incorrectly continuum-subtracted SHS data with SHASSA
data results in an offset between the two surveys. A range of values
for the scaling factor can be applied to the SHS data until the
zero-point of the continuum-subtracted UKST survey images best matches
the zero-point of the equivalent SHASSA data, indicating the
appropriate value to use. A calibration based on the SHASSA data will 
provide an advantage for the SHS over the CTIO survey as it can be applied 
to the full resolution pixel data. This offers the chance to determine 
intensities for emission structures not resolved by SHASSA such as the 
new sample of extended PNe discovered from the SHS data 
(e.g. Parker et al. 2003, 2005, Pierce et al. 2004, 
Frew \& Parker, in preparation and see Section 14.1).

Each scanned SHS survey field, at the full 0.67~arcsecond resolution,
contains over 2~Gb of data so it was not practical to
download and compare all the pixel data for each field. Instead,
most of the emission variation on a given survey field can be
sampled using carefully selected 30~arcminute regions. For fifteen fields,
several 30~arcminute areas were downloaded to sample the complete dynamic
range of emission present. Once the scaling reliability over a range of flux levels and
central aperture locations was established, the best single,
30~arcminute region was chosen from the whole frame for the rest of the 233 survey fields to provide the
base calibration for each field. The \ha\ filter flat-field correction
was applied to remove the low-level non-uniformities in transmission across the
narrow-band filter.

Data requiring a flat-field correction of up to 15 per cent have been shown to be
suitable for inclusion in the survey, though in most
cases data returned from the SHS website require less correction
than this. For the regions considered here, the pixel data only required
flat-field correction $\leq3$ per cent in most cases and, in general, no more
than 8 per cent. For
two fields, h350 and h1109 regions well away from the best
area of the filter were also selected to confirm that the pixel data behaves as expected
when larger flat-field corrections are required. The results are discussed below.

\subsection{The SHS Calibration process}
A detailed description of the SHS survey calibration process based on zero-pointing
each SHS field to SHASSA is given in the thesis of Pierce (2005). The essentials
of this scheme are described here. Each SHS field is completely covered
by just one 13\deg\ SHASSA field and data from the best area of SHASSA filter
response was chosen for comparison with the SHS images. 
For direct comparison, the \ha\ and SR SHS data at
0.67~arcsecond was re-binned to match the 48~arcsecond SHASSA pixels  
using the {\sc IDL} routine {\sc Hrebin}\footnote{Interactive Data Language: http://www.rsinc.com}. 
This returns the mean value of the 72\,$\times$\,72 full
resolution SHS pixels that constitute a single SHASSA 48~arcsecond pixel, so the
calibration factor determined from the comparison plot applies to the
full resolution SHS data. 
At this coarse resolution the SR data was scaled and subtracted from the
\ha\ image. Because the correct scaling factor was not yet known, a range of scaling
factors from 0.4 to 2.0 was applied so that the best value could be
selected by matching the SHASSA zero-point.
The equivalent area of SHASSA data was selected, aligned and trimmed
to match using the {\sc IDL, Hastrom} routine. 
These images were then compared directly, pixel by
pixel, with the re-binned, continuum-subtracted SHS data to give a plot
of SHASSA values in Rayleighs versus SuperCOSMOS counts per re-binned pixel. The
linear portion of the resulting comparison was then fitted to
determine the number of \scos\ counts per 0.67~arcsecond pixel per Rayleigh.
Averaging \scos\ data in this way only works properly if all the \scos\ 
pixels are on the linear portion of the characteristic curve at the faint end and unsaturated at
the bright end. Once the \scos\ dynamic range is exceeded pixel saturation 
arises and the SHASSA to \scos\ linear relation breaks down as the \scos\ flux becomes increasingly 
underestimated (e.g. Phillipps \& Parker, 1993) as seen in Figure~\ref{HAL0350calib}.

\begin{figure}
\centering  
\epsfysize=7.0cm  \epsfbox{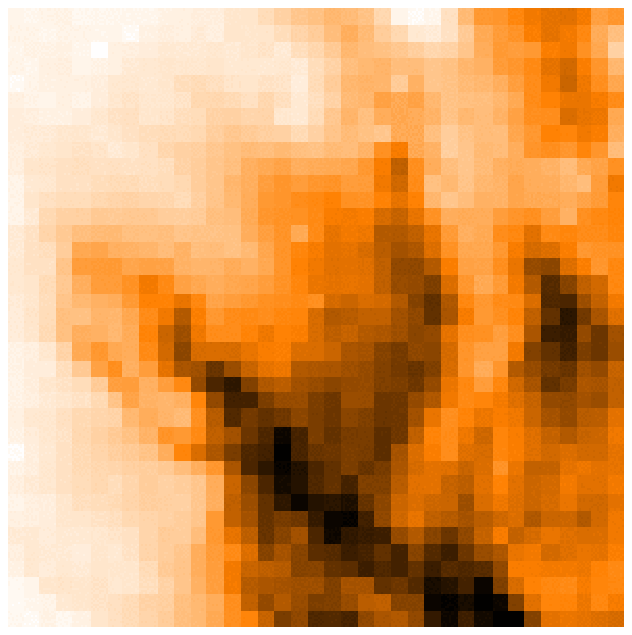}
\epsfysize=6.5cm  \epsfbox{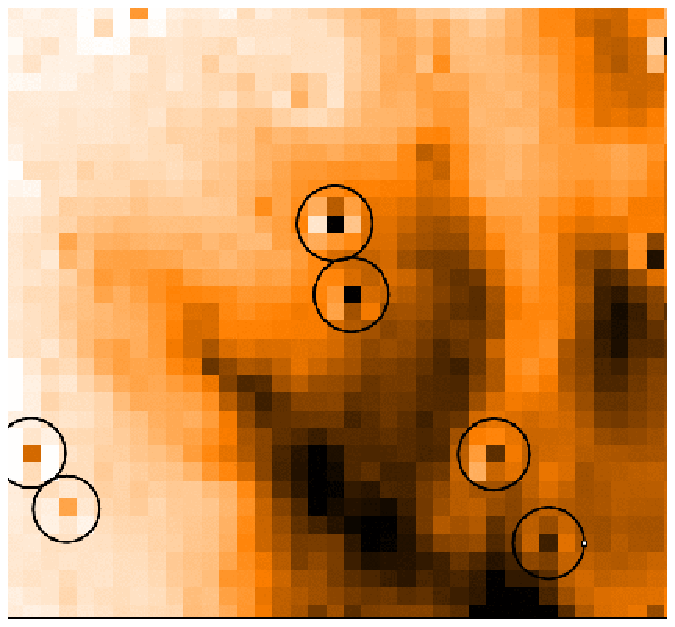}
\caption{Two \ha\ images of a  30$'$ region of SHS field h350
centred on $16^h 47^m -49\deg\ 00$'' (J2000) as in Figure~\ref{1647m4900_2}. 
The top image is the SHS image blocked down to match the 48~arcsecond resolution of the SHASSA data
and continuum subtracted. The bottom image is the equivalent
SHASSA image, trimmed and aligned ready for comparison. Areas of uncancelled bright stars are highlighted}
\label{1647m4900_48}
\end{figure}

\begin{figure} 
\begin{center} 
\epsfysize=0.35\textwidth \epsfbox{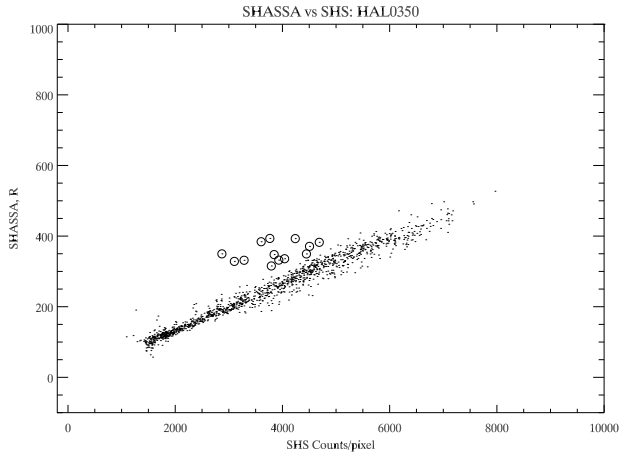}
\caption{The comparison plot of SHS and SHASSA data from the area
shown in Figure~\ref{1647m4900_48}. Each point on the plot is a the
re-binned pixel value from the SHS data against the Rayleigh value form
the SHASSA data. Note the linear relation and the large dynamic range
covered by this comparison. The outliers due to uncancelled bright stars in the SHASSA data are highlighted.}
\label{1647m4900_calib}
\end{center}
\end{figure}

The adopted process was followed for the same 30~arcminute region from SHS field
h350 as shown in Figure~\ref{1647m4900_2}. The upper image in 
Figure~\ref{1647m4900_48} shows 
the SHS image with the continuum-subtracted at 48~arcsecond resolution. The lower
image is the trimmed and aligned SHASSA data. Bright
stars on the continuum subtracted SHASSA images leave
significant residuals while in the SHS data any stellar residuals are
barely visible and then only for the very brightest stars.
The pixel-to-pixel comparison plot for this area is shown in
Figure~\ref{1647m4900_calib}. Each point on this plot is the pixel
value from the SuperCOSMOS scan against the Rayleigh value from the
SHASSA data. Immediately, it can be seen that the SHS data follows a
tight linear relation with the SHASSA values over a range of several
hundred Rayleighs. The outliers from this distribution at around 4000
SuperCOSMOS counts correspond to the SHASSA stellar residuals
noted from Figure~\ref{1647m4900_48}.


\begin{figure}
\centering  
\mbox{\epsfxsize=0.4\textwidth  \epsfbox{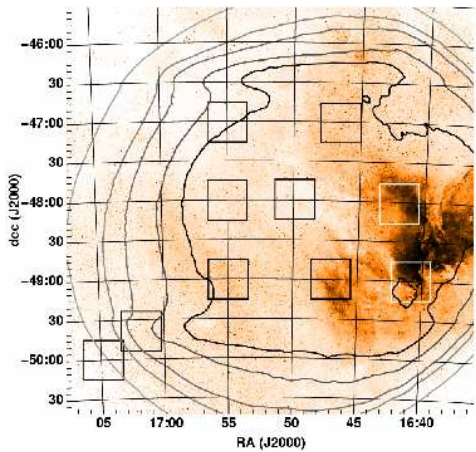}}
\caption{Blocked down image of survey field h350 with overlaid
contours at 85, 90, 92, 95 and 97 per cent. 
The 30~arcminute areas used to make the comparison plot shown in
Figure~\ref{HAL0350calib} are framed by the squares which are drawn
in black or white to best contrast with their background. 
The bright HII region to the right is NGC~6188, excited by the young open cluster NGC~6193, 
nucleus of Ara OB 1.}
\label{HAL0350areas}
\end{figure}

\begin{figure}
\centering  
\mbox{\epsfxsize=0.45\textwidth \epsfbox{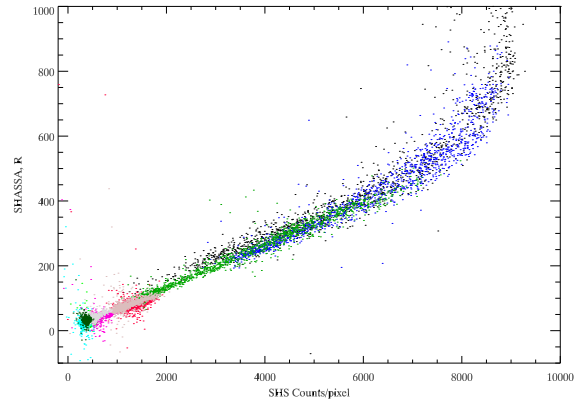}}
\caption{The comparison plot for all ten 30~arcminute areas highlighted in
Figure~\ref{HAL0350areas}. All align on the one trend, implying no
variation in sensitivity across the SHS survey field. The bright
photographic saturation is seen from $\sim$\,500\,Rayleighs. The 
grey points at the faint end of the plot are taken from the area of
poor filter response in the SE corner of h350 and align well with
the rest of the data, indicating that, in areas of bright emission,
flat-field corrections as large as 20 per cent still return reasonable
intensity values.}
\label{HAL0350calib}
\end{figure}

A further nine 30~arcminute areas were examined from field h350 to cover
the whole dynamic range of diffuse emission evident in the field. 
Figure~\ref{HAL0350areas}
shows the 16$\times$ blocked down \ha\ image for this field with
contours of filter response overlaid and boxes indicating the areas
used. The large area of strong emission in the west of this field is
caused by the UV flux from OB association Ara~1A~A
(Mel'Nik \& Efremov 1995). In all but two cases the areas were selected from
the best area of the filter response, the exception being the two
areas extracted from the SE corner which were included to test the
effect of using pixel data from the edge of the SHS survey fields
where the flat-field correction is larger.

Figure~\ref{HAL0350calib} shows the resulting SHASSA versus SHS comparison
plot with differently shaded points belonging to different 30~arcminute
areas. Almost the full range of the SHS data is shown, with the pixel values
showing a good linear relation to the SHASSA data from the faintest
emission on the field at $\sim$\,20\,Rayleighs right up to $\sim$\,500\,Rayleighs. 
The curve in the trend beyond 500\,Rayleighs is due to saturation effects with the
SHS data. These points were discarded when making the calibration fit. 

The reciprocal of the slope of the linear part of the relation provides a
calibration factor of 15.1~counts/pixel/Rayleigh to convert the SuperCOSMOS
intensity counts to Rayleighs for this field. An estimate of the error
in this calibration is possible, based on the vertical scatter of
points about this trend ,as a given \scos\ value matches a Rayleigh
value in this vertical distribution. In this field the 1$\sigma$
scatter is 21\,Rayleighs. 

The data taken from the two areas in the SE corner of the SHS
survey field required flat-field correction of up to 20 per cent. They 
behave very well when compared with the data requiring less
correction, neatly overlaying the main trend, with
no change in slope or increase in scatter. This justifies use of the
data out to the 15 per cent flat-field contour where the emulsion records
strong emission. 

\subsection{SHS sensitivity limit}
\begin{figure}
\centering  
\epsfysize=7.0cm  \epsfbox{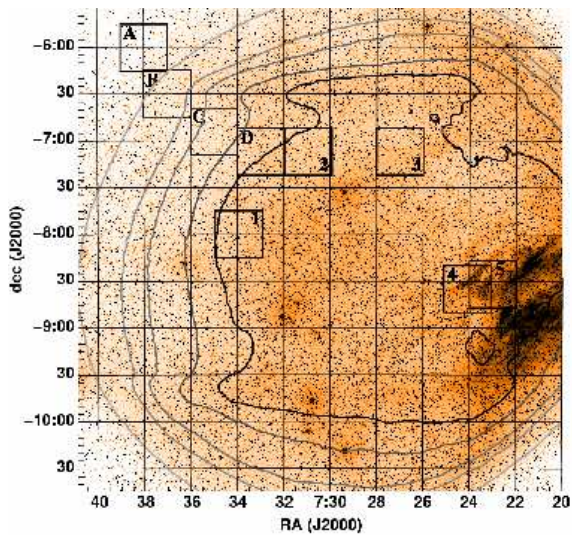}
\caption{Blocked down SHS \ha\ image of field h1109 with contours of
filter response at 85, 90, 92, 95 and 97 per cent overlaid. Two groups
of 30$'$ areas are marked by squares and labelled to indicate group
membership.}
\label{HAL1109areas}
\end{figure}

While the results from field h350 have shown that the survey has
been well tuned to the detection of diffuse emission, as well as
giving an approximate limit to the point at which the photographic
emulsion saturates, the faint limit has not yet been constrained
because there is no really faint emission on this field. At the
faintest point of 20\,Rayleighs there is no sign of either survey struggling
for sensitivity so a field exhibiting fainter emission is required to probe the SHS 
faint sensitivity limit. 
According to the SHASSA data an area of the Southern Galactic plane in Monoceros harbours
diffuse emission that reaches a level as faint as $\sim$\,2\,Rayleighs, which
is ideal to test the faint limit of the SHS
data. Figure~\ref{HAL1109areas} shows the 16 times blocked down \ha\
image of SHS field h1109, which covers this area of sky, with
contours of filter response and the areas selected shown as with
Figure~\ref{HAL0350areas}. On this field, nine 30~arcminute regions have been
examined as two groups: one, labelled 1--5 in
Figure~\ref{HAL1109areas}, from the area of best filter response and
the other, labelled A--D, probing the combined effects of extremely low
levels of emission and decreased filter transmission.

\begin{figure}
\centering  
\mbox{\epsfysize=0.36\textwidth  \epsfbox{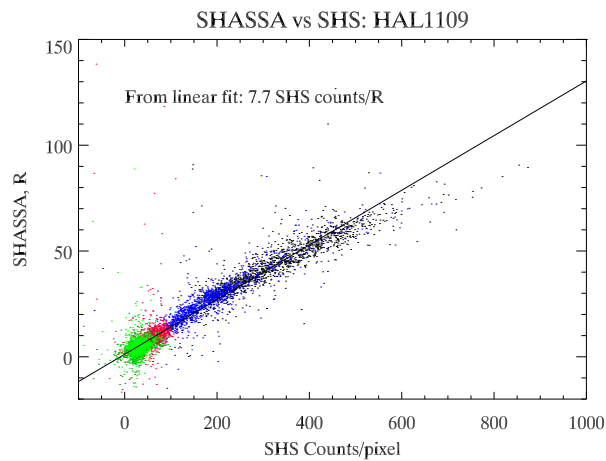}}
\caption{Calibration plot of the numbered 30~arcminute areas from the best
area of filter response on field h1109. Points from each area are
plotted as a different colour. Note the SHS data match the SHASSA data
down to the faintest intensity at $\sim$\,2\,Rayleighs. The linear fit is also plotted.}
\label{HAL1109calib}
\end{figure}

The comparison plot from the first group is shown in
Figure~\ref{HAL1109calib}. Once again, there is a clearly defined
relation between the two, although on this faint field the upper
saturation limit is not reached. It is immediately obvious that the
SHS data can match the SHASSA data right down to the faintest level of
emission, although the linear response of the SHS data is difficult to determine 
at this low level. The SHS data are therefore detecting emission structure
as faint as 2\,Rayleighs on this field, although this sensitivity is tempered
by the scatter evident in the plot and in the examination of the areas
from the area of poor filter transmission discussed below.
The linear fit returns a value of $7.7\pm0.1$ 
\scos\ counts/pixel/Rayleigh for this particular field and provides a
reasonable calibration as the 1$\sigma$ scatter to the fit is just
6.2\,Rayleighs. Note the factor of two difference in the slope of the calibration curve for this low emission
level field h1109 (no emission measure greater than about 80 Rayleighs seen in the area considered) 
compared to that obtained for high emission field h350 in Figure~\ref{HAL0350calib}, 
which returned 15.1 counts/pixel/Rayleigh, which is closer to that generally obtained for most fields. 
This serves to emphasise the need for individual field calibration due to the
variation in \scos\ pixel intensities on a given field, arising primarily from variable fog-level, 
sky background and resulting \scos\ and emulsion saturation.


\subsection{Fifteen Fields Studied in Depth}
\begin{figure}
\begin{center} 
\mbox{\epsfysize=0.33\textwidth \epsfbox{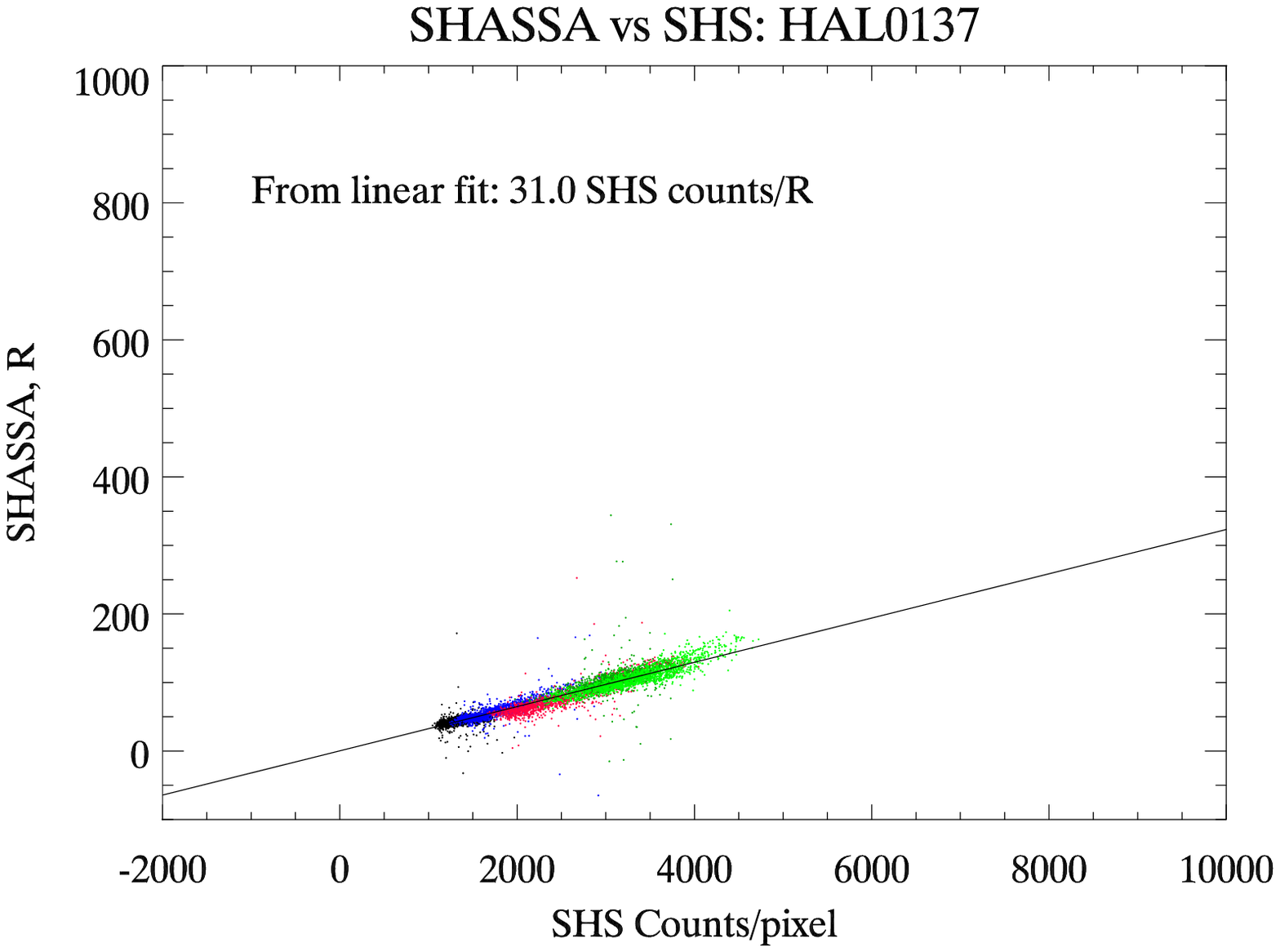}}
\mbox{\epsfysize=0.33\textwidth \epsfbox{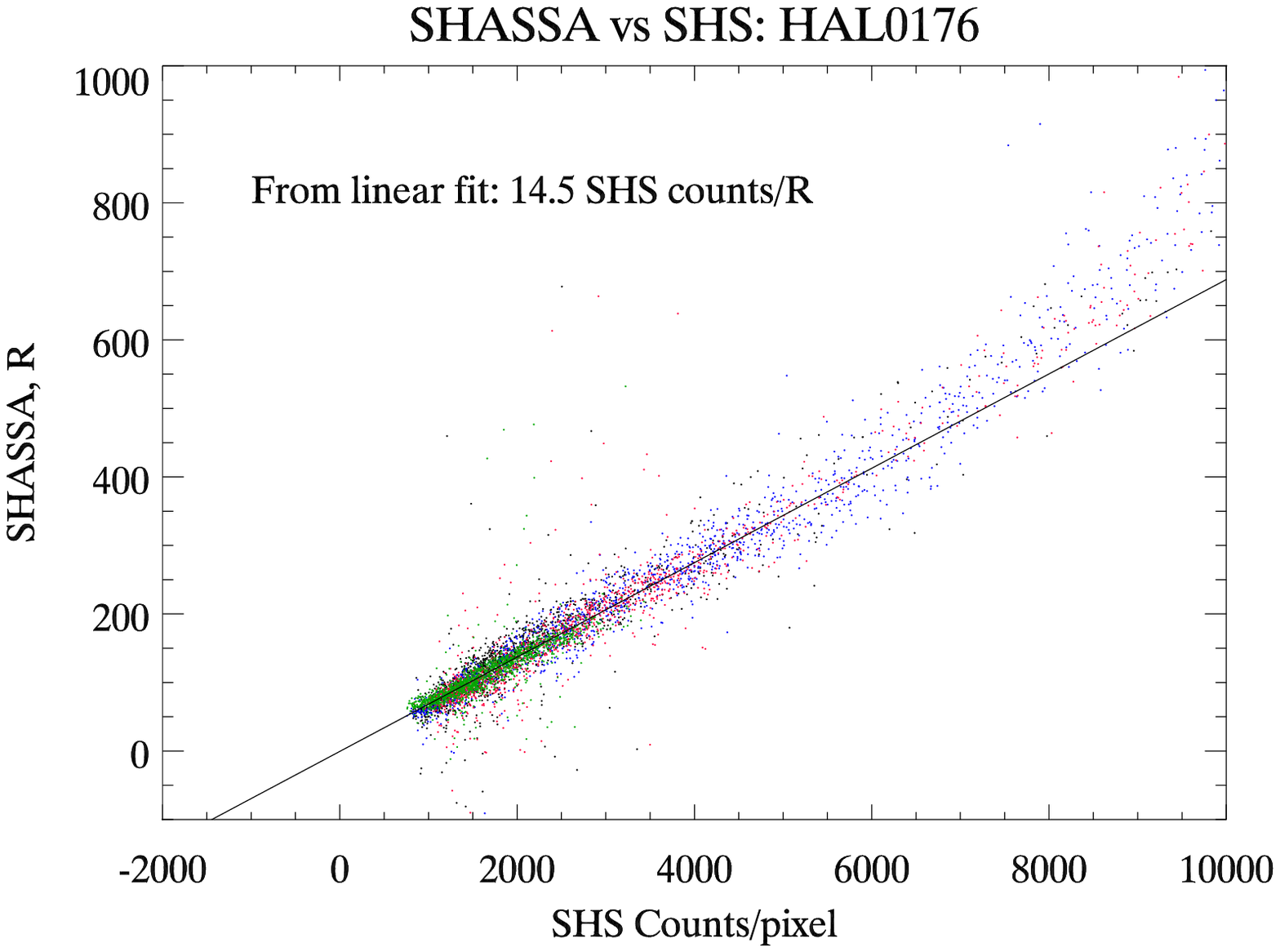}}
\mbox{\epsfysize=0.33\textwidth \epsfbox{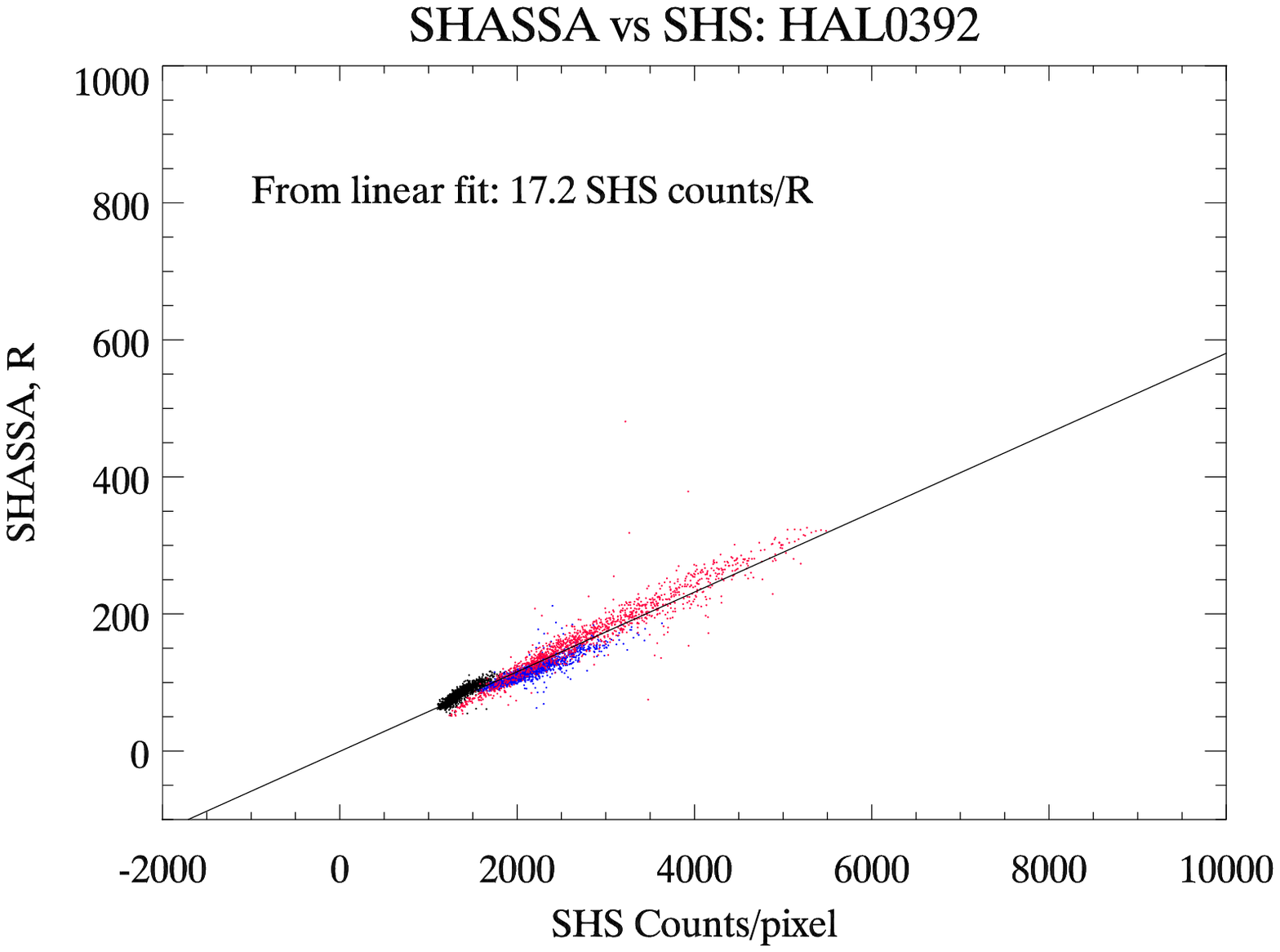}}
\mbox{\epsfysize=0.33\textwidth \epsfbox{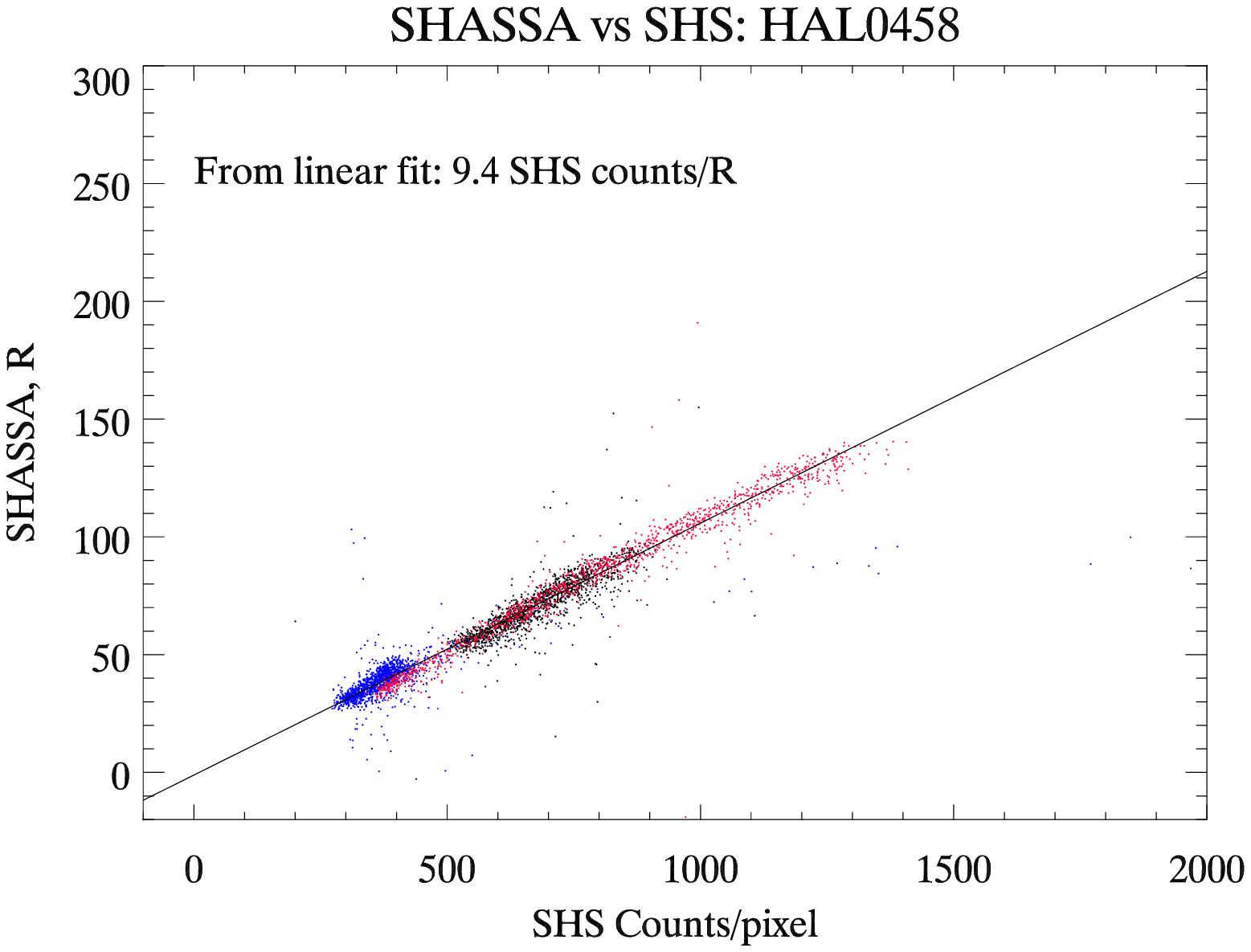}}
\caption{Comparison plots for four of the fifteen SHS fields which had several
30~arcminute areas tested against SHASSA data. These fields cover varying
dynamic ranges so the axes for different fields do not always cover
the same dynamic range.}
\label{fourcomp}
\end{center}
\end{figure}

A total of fifteen fields from a wide variety of 
Galactic environments were studied using
several 30~arcminute regions in each in order to build up a global picture of the
survey behaviour. The comparison plots for 4 of these are shown in
Figure~\ref{fourcomp} with the linear fit overlaid.
In each case a clear, essentially linear relation can be seen between the SHS and SHASSA
data. Generally, different 30~arcminute regions
follow a single trend which indicates little variation in 
emulsion sensitivity across the large SHS images. 
Three of the fields examined, h175, h350
and h555 show evidence for saturation at the bright end and fix the
bright limit at $\sim$\,500 to $\sim$\,600\,Rayleighs, while none of the
fields appear to reach the background fog level in the areas that were compared.

\subsection{Calibration of the entire SHS}
From the analysis of the fifteen fields considered above, it is clear
that each field requires individual calibration and that
one, well selected, 30~arcminute area of SHS pixel data can be compared with
SHASSA to give a working estimate of the calibration factor and a satisfactory result in most cases.
On this basis, the 30~arcminute areas of pixel data which covered the greatest
dynamic range of emission were downloaded from
each of the 233 SHS fields, re-binned and compared with the equivalent
area of SHASSA data. The results for each field can be found on the SHS web site. For
each field the position of the area used, the computed scaling factor
for continuum subtraction, the linear fit and, where appropriate, the
coefficients from a third-degree polynomial fit are given. The
1$\sigma$ vertical scatter about the linear fit is also quoted to
offer an estimate of the error in the calibration.
Of the 233 survey fields, 76 are relatively featureless and exhibit
little emission. These are difficult to fit and in 9 cases the fit
failed completely. Forty three of these show evidence of the low-level
photographic fog. Of the remaining 157 fields, 122 are well constrained by a 
linear fit. For the other 35 fields the fit can be improved with a low-order polynomial
relation. Where this is the case the coefficients are included in the table on the SHS website.

\subsection{Calibration check of SHS field overlap regions}

There is generous overlap between survey fields because of the circular aperture of the
filter which allows field-to-field consistency check of the calibration. For six overlap regions
between eight fields, a 30~arcminute area was carefully selected from the best
possible compromise of filter response between two fields, never
requiring flat-field correction greater than 15 per cent. The
calibration factor calculated from the field centre in the best region
of filter response was applied to these pixel data from the edge. For
seven of the eight fields the calibration factor determined
from the linear fit to the data was used.
Calibrated and aligned
data from the two overlapping SHS fields were plotted pixel by pixel at 0.67~arcsecond,
10~arcsecond and 48~arcsecond resolution. If the independently determined
calibration applied to each field is consistent, the resulting plot of
Rayleighs from one field against Rayleighs from its neighbour should
yield a linear relation with slope of unity and no offset.

\begin{figure}
\centering  
\epsfysize=5.5cm  \epsfbox{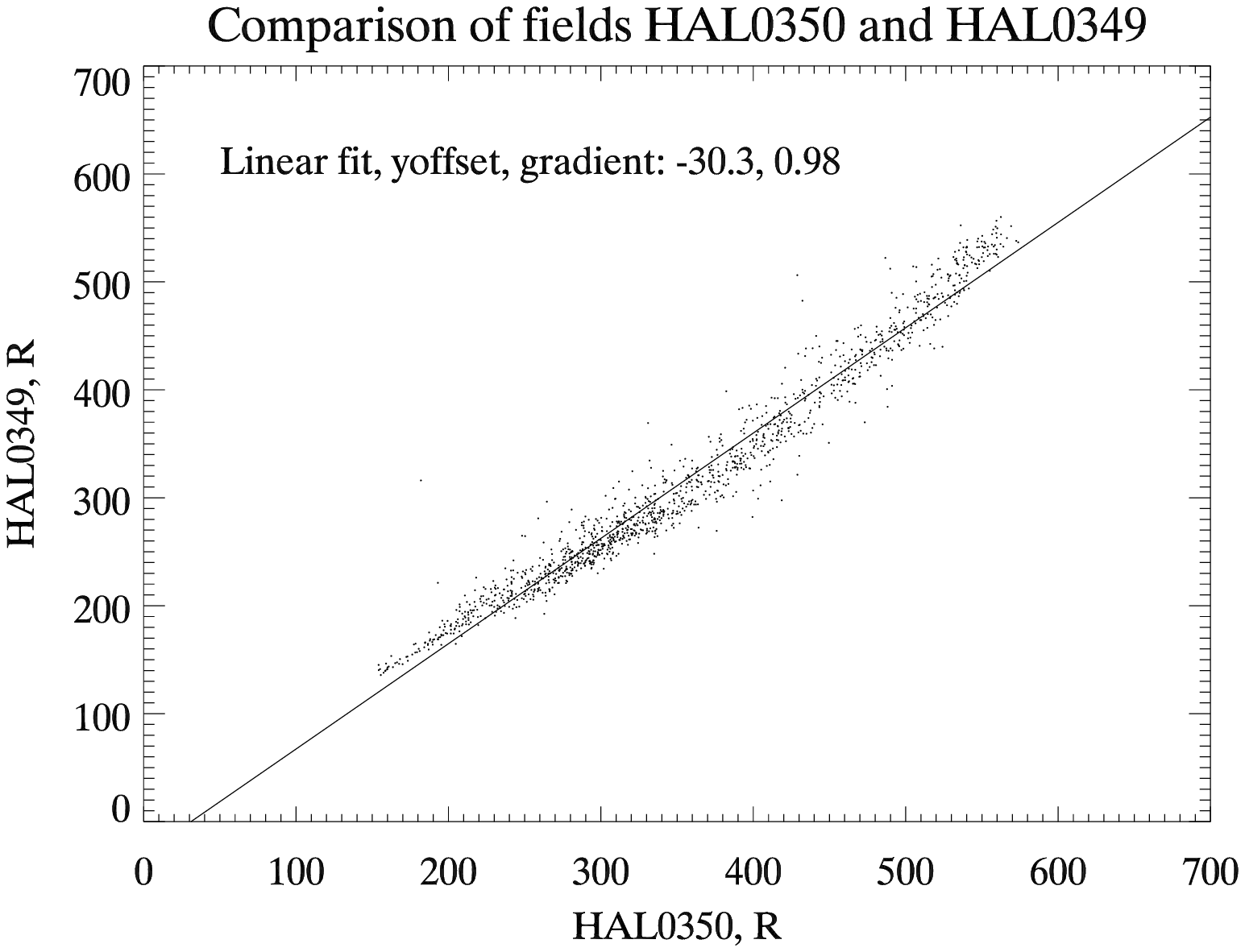}
\epsfysize=5.5cm  \epsfbox{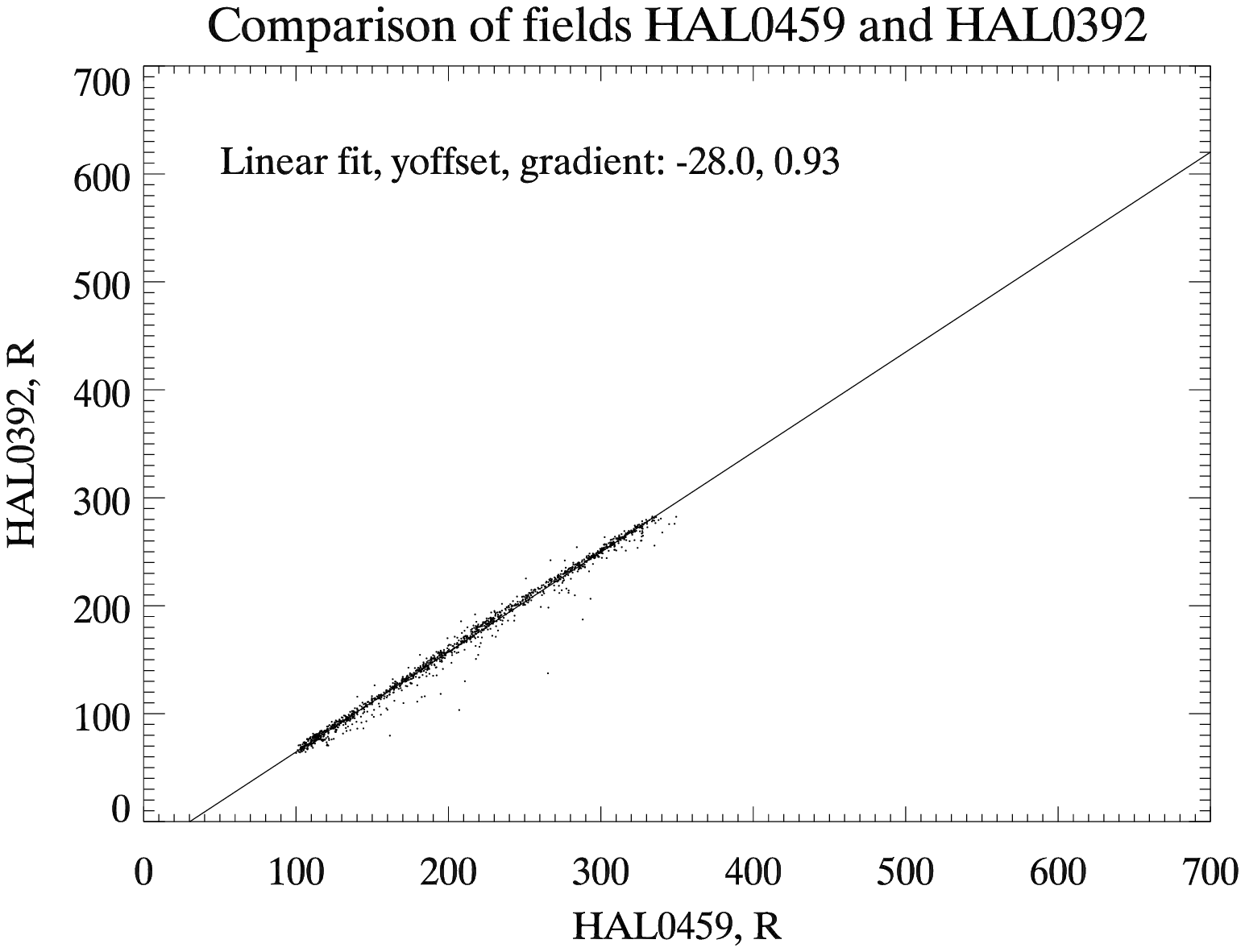}
\epsfysize=5.5cm  \epsfbox{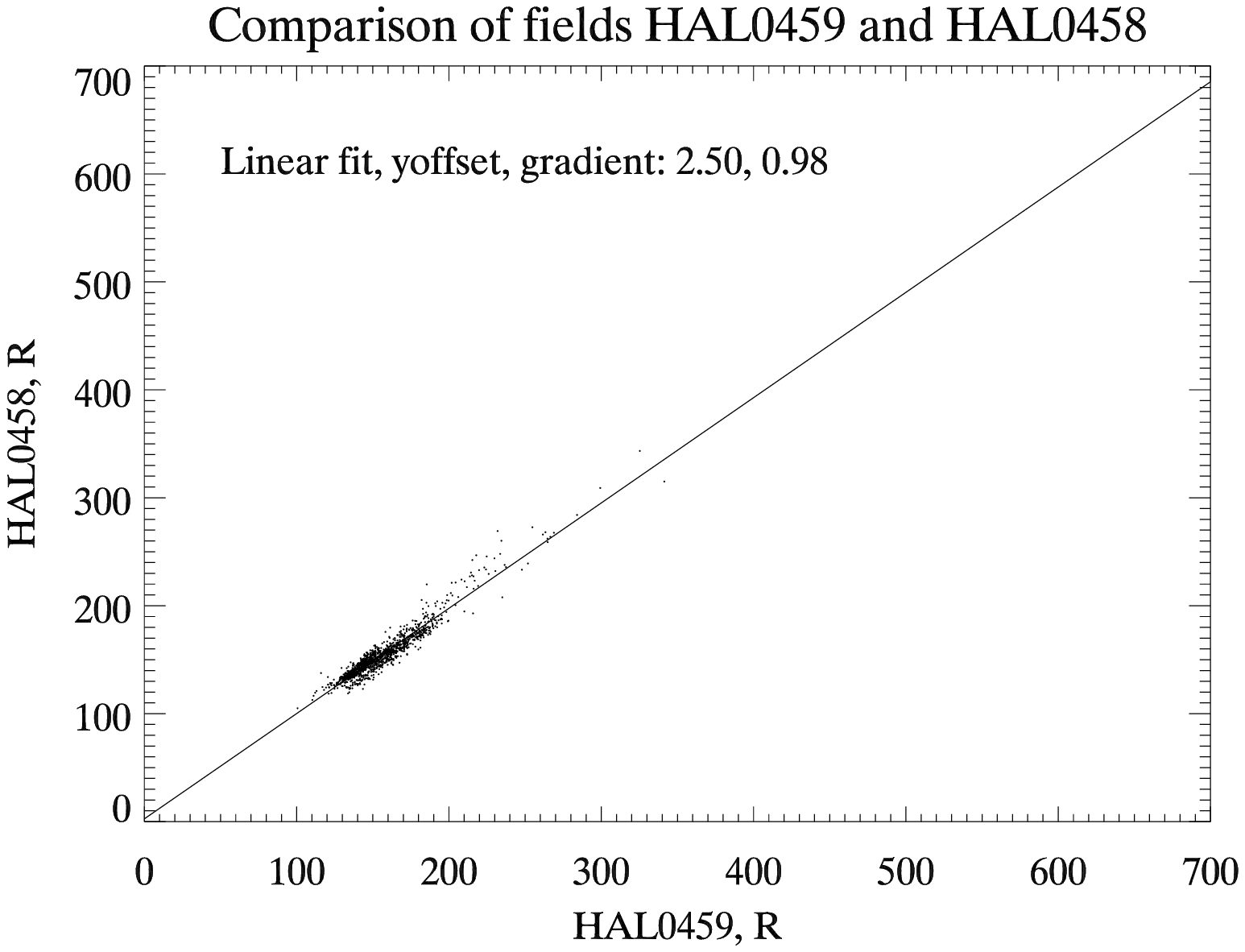}
\epsfysize=5.5cm  \epsfbox{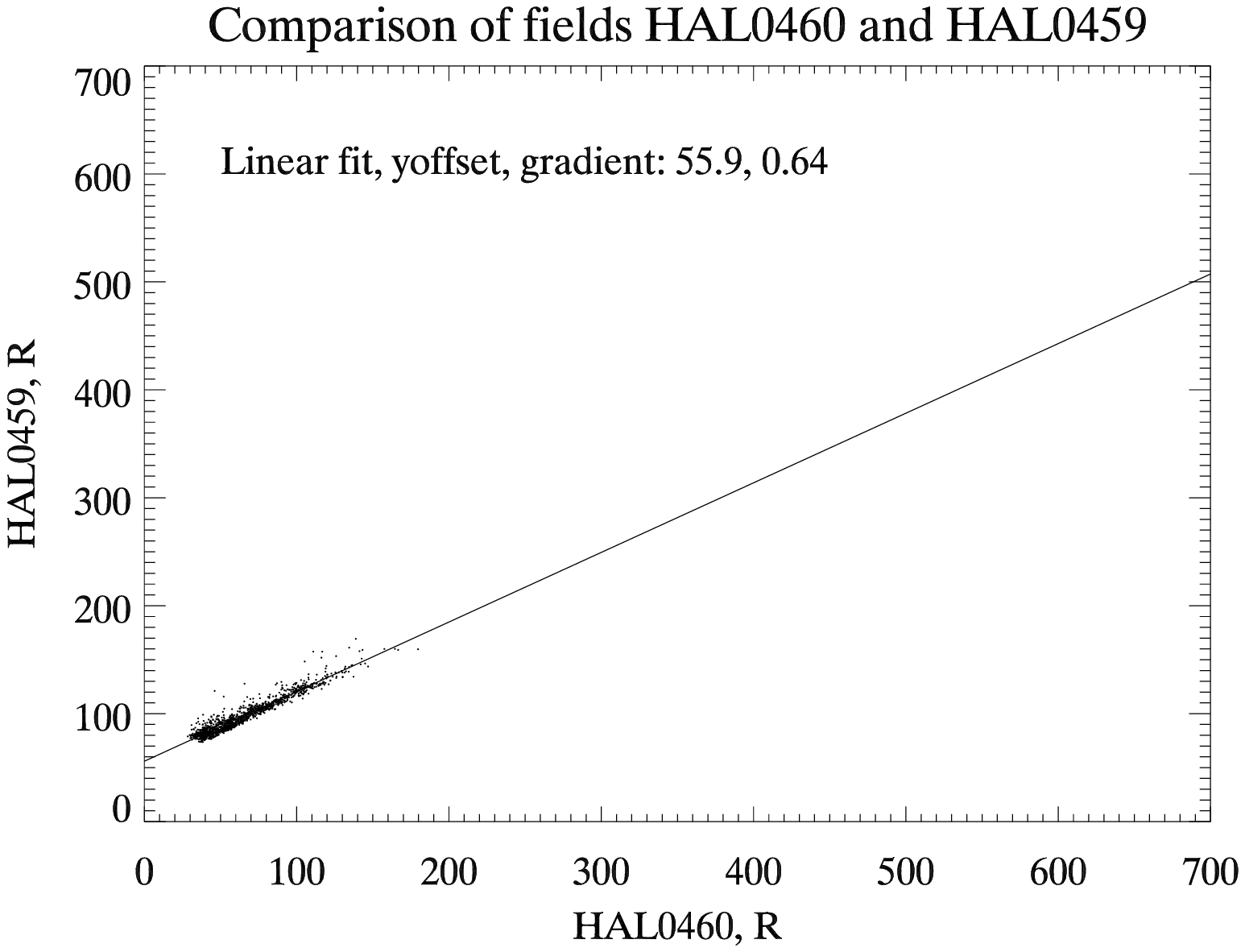}
\caption{Comparisons between adjacent fields which have been
calibrated according to a linear fit with SHASSA data.}
\label{shsoverlap}
\end{figure}

From the six fields examined in this way the results give good agreement.
Four examples 
are shown in Figure~\ref{shsoverlap} from comparisons at 48~arcecond
resolution. Here the ordinate and abscissa values are in Rayleighs
with each axis labelled with its field. The fits are quoted on the
plots and in Table~\ref{overlap}. Two comparisons, 459-392 and 459-458,
agree to better than 10 per cent, three more, 391-329, 458-391 and 460-459 also
return linear results but agree to just 14, 29 and 43 per cent
respectively. The bottom 3 comparisons are for overlapping fields so the
good match implies that one could construct seamless, large pixel mosaics.
For the  h350 and h349 the linear relation is not so well behaved.
The slighly curved trend between these two fields probably results from
the bright saturation evident in the pixel data of the overlap region
used (note the higher Rayleigh limits for this comparison).

\begin{table*}
\centering
\caption{Fits obtained from the comparison of \scos\ pixel data in overlap regions between adjacent survey
fields using linear calibration factors}
\label{overlap}
\begin{tabular}{@{}cccccc@{}}
\\
\hline
\\
Fields	&Position	&Slope	&Slope10	&Slope48	&Calib1/Calib2\\
\hline
350 and 349	&16 38 --48 00	&0.88$^a$ &0.95$^a$	&0.97$^a$	&1.03	\\
391 and 329	&08 25 --46 30	&1.15	&1.13		&1.14		&0.62	\\
458 and 391	&08 22 --42 15	&0.49	&0.54		&0.57		&1.44	\\
459 and 392	&08 48 --42 15	&0.86	&0.90		&0.93		&1.17	\\
459 and 458	&08 36 --40 30	&0.90	&0.91		&0.89		&0.93	\\
460 and 459	&08 56 --41 15	&0.60	&0.71		&0.71		&1.22	\\
\hline
\multicolumn{6}{l}{$^{a}$ This result is from a linear fit whereas a polynomial might be more appropriate.} \\
\multicolumn{6}{l}{Refer to top panel in comparison plots given in Figure~\ref{shsoverlap}}\\
\end{tabular} 
\end{table*}

\section{Comparison with IPHAS}

\begin{figure}
\centering  
\epsfysize=5.5cm  \epsfbox{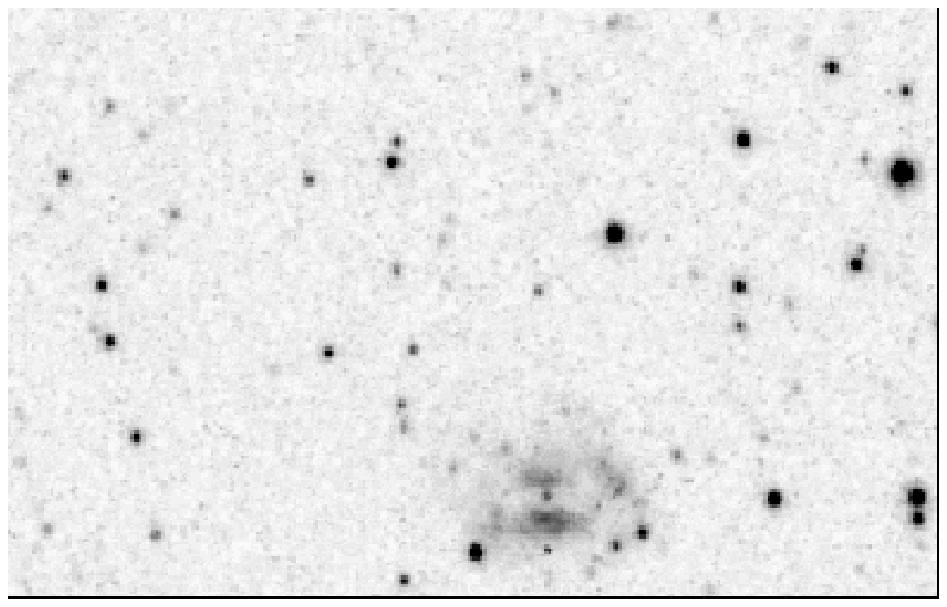}
\epsfysize=5.5cm  \epsfbox{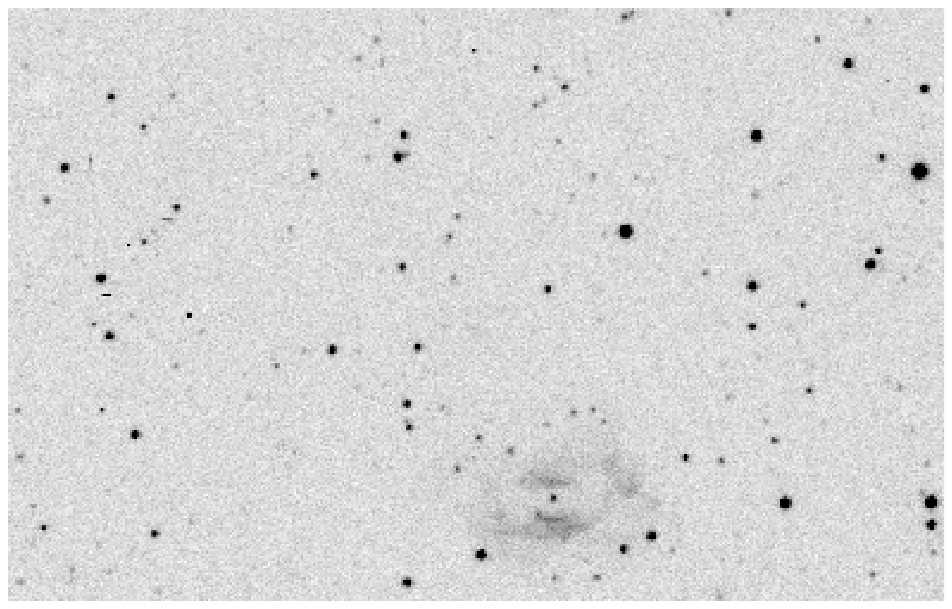}
\caption{A $3.3\times2$ arcminute comparison region between SHS (top) and IPHAS (bottom) data centred
on $18^h47^m42.6^s, +01^o33'04''$ including the 
new planetary nebula PHR1847+0132.}
\label{shs-iphas}
\end{figure}

The SHS and IPHAS \ha\ surveys have areas of overlap 
at low declinations which permit a direct 
comparison to be made between the two complementary surveys. Figure~\ref{shs-iphas} shows
a $3.3\times 2$~arcminute region centred on $18^h47^m42.6^s, +01^o33'04''$
which includes the newly discovered planetary nebula 
PHR1847+0132, taken from a
slightly shallow SHS survey field h1332 (exposure number HA18088, 
survey grade A2, but exposure time cut short to 168mins cf. 180 normally).
The data have been carefully matched in terms of co-ordinate projection but not otherwise 
processed. It is clear the two surveys achieve
similar depth for diffuse emission but that the IPHAS survey goes deeper 
for point-sources due to its better resolution. 
Further details of the IPHAS survey are given by Drew et al. (2005).

\section{Colour-colour photometric plots}

\begin{figure}
\centering  
\epsfysize=5.0cm  \epsfbox{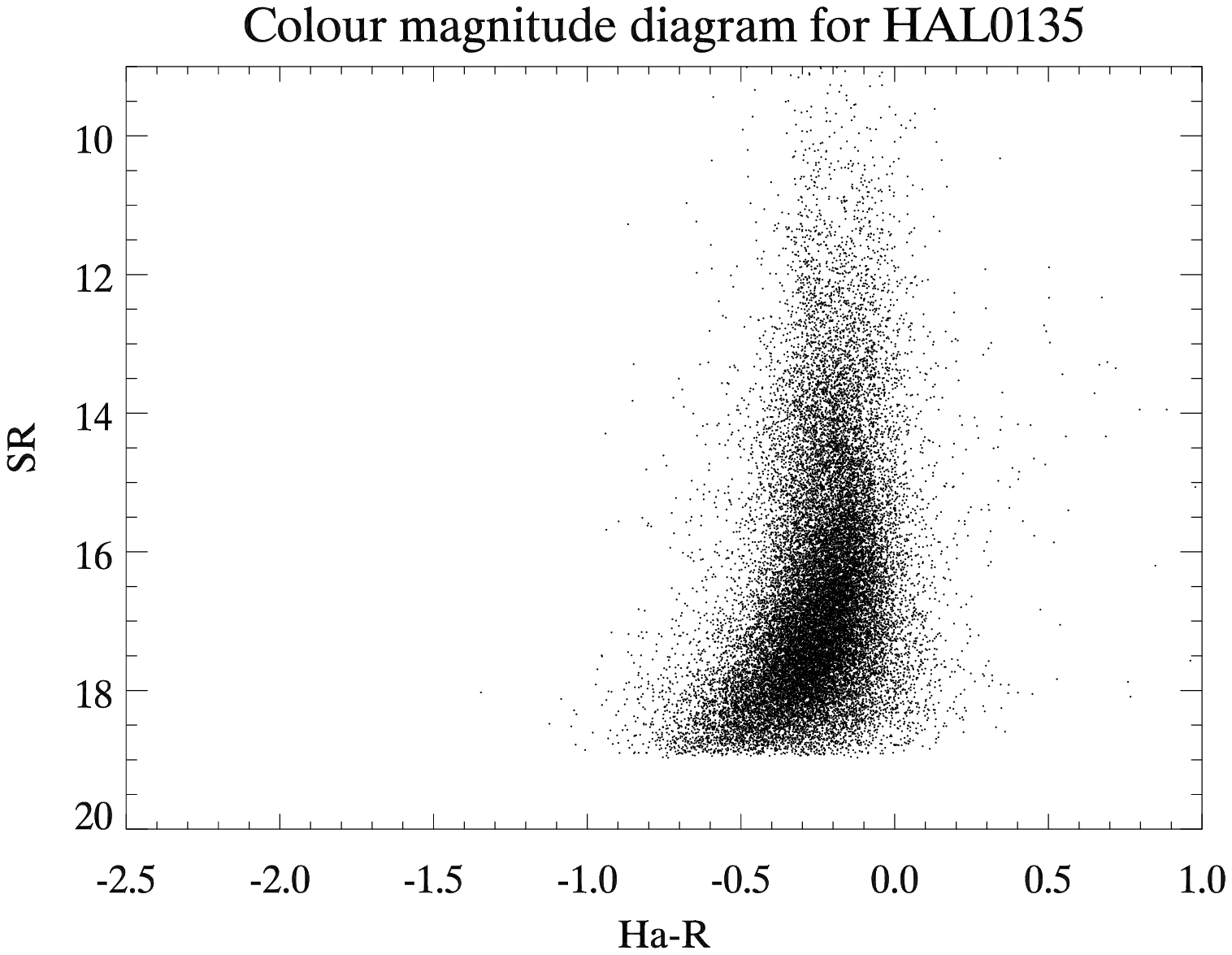}
\epsfysize=5.0cm  \epsfbox{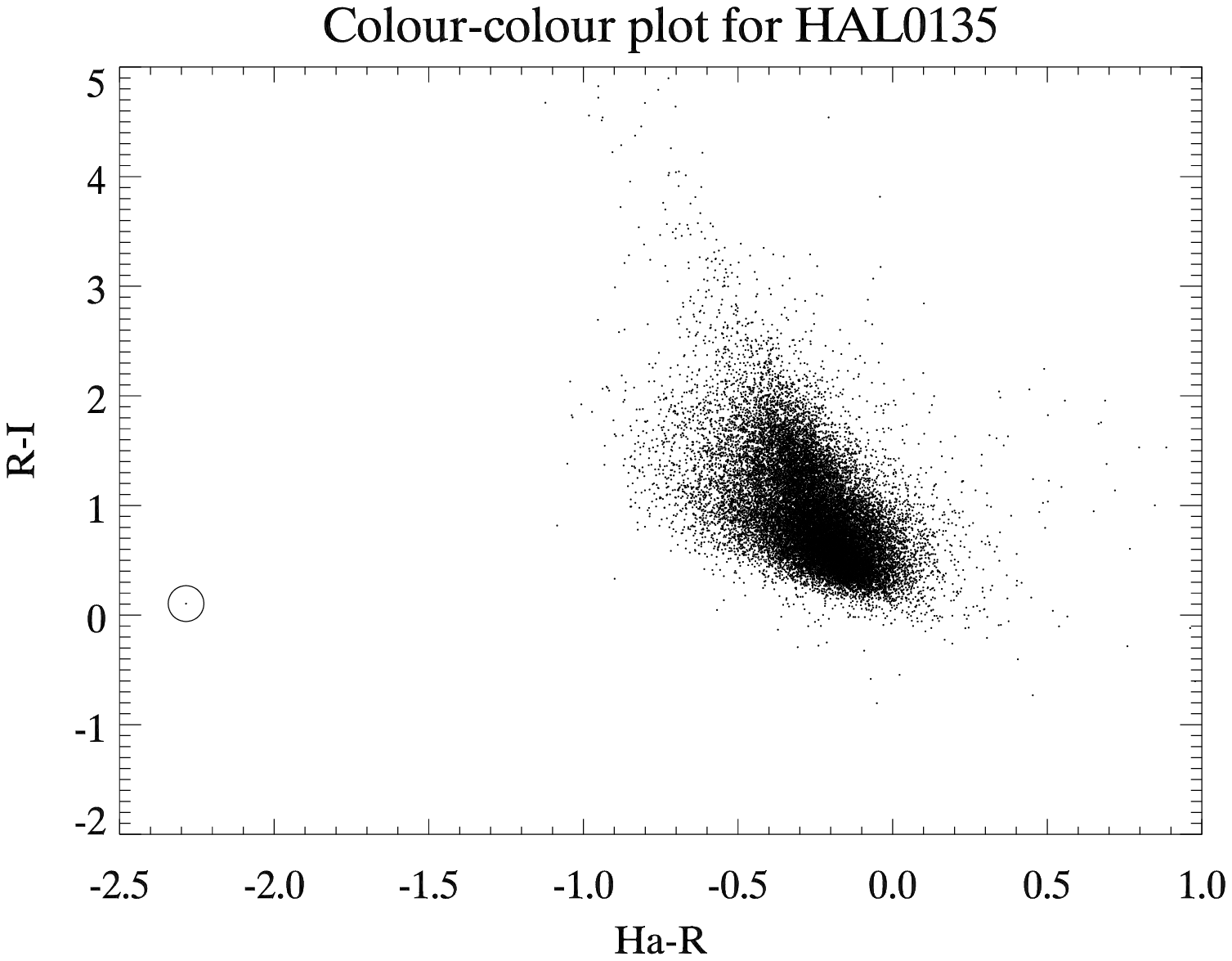}
\caption{The colour magnitude diagram of SR vs \hamr\ and the
colour-colour plot of $R-I$ vs \hamr\ for stars from an area centred
on the middle of SHS field h135. On average the stars sit at
roughly neutral values of \hamr\,=\,--0.2 and $R-I$\,=\,0.15. The
reddening in this field of up to E(B-V)\,=\,4.7\,mag is evident in the
colour-colour plot as a smearing of the stellar locus towards higher
values of $R-I$ and \hamr. The circled point indicates the location of a Planetary Nebulae.}
\label{cmdandcc135plots}
\end{figure}

The SR, \ha\ `R-equivalent' and $I$ magnitudes can be combined to
provide \hamr\ and $R-I$ colours for the objects detected by the IAM
software. The narrowband \ha\ photometry should be sensitive to
point-source emitters, so the SHS stellar photometry is of
particular interest. A colour-magnitude diagram (CMD) of SR
magnitude versus \hamr\ can be constructed from the survey data to trace
the average values of \hamr\ for normal stars by brightness and help
identify emitters by highlighting objects which stand apart from
this. Figure~\ref{cmdandcc135plots}a shows a CMD constructed from a
1\deg\ region from the centre of survey field h135. Most stars can
be seen to congregate around a stellar locus running vertically in the
diagram at \hamr\ $\sim -0.2$ and with an increasing spread
towards fainter magnitudes. This is due to the census including a more
complete population of objects at fainter magnitudes and increasing photon errors. Outliers can be
seen either side of the distribution and, if present, emission line
objects will sit to the left of the main stellar locus. 

A colour-colour plot of $R-I$ vs \hamr\ provides more information
about the stellar population along a given line of
sight. Figure~\ref{cmdandcc135plots}b presents such a plot for the
same area of sky as described above and uses the $I$ band magnitudes
which have been corrected to the SR photometry. In this plot the
stellar locus is centred at \hamr\ $\sim -0.2$ and $R-I$\, ~$=0.15$. The area
covered suffers high reddening of up to $E(B-V) = 4.7$ mag
(Schlegel, Finkbeiner, \& Davis, 1998) and this is evident in the stretching of the
stellar locus towards larger values of $R-I$ and, significantly,
\hamr\, making them appear to be emission line stars. These reddened
non-emitters can be identified in the colour-colour plot of $R-I$ versus
\hamr\ and excluded from studies of potential point-source emitters (Pierce 2005).
An additional complication is the potential contamination from late type stars. 
A TiO opacity minimum near 6536\AA ~enables the continuum to
be attained, producing the peak, compared to the TiO band heads on either side of the \ha\ filter.
Further to the red, the Tech-Pan emulsion sensitivity
cuts off at 6990\AA. Such objects can thus appear as apparent \ha\ emitters when compared to the
matching SR photometry unless the complementary I band photometry is included, as such late type stars will be
brighter in this band than \ha\ emitters. However, colour-colour plots created in 1-degree sub-regions,
avoids the smearing effects on the photometry due to small positional shifts in the stellar
locus across a \ha\ survey field. These have proved very effective in identifying point-source emission candidates
and has been successfully employed to
provide targets for follow-up multi-object spectroscopy with 6dF and 2dF at the AAO 
(e.g. Hopewell et al. 2005).

\section{Scientific Exploitation}
The SHS on-line atlas was released in stages starting in 2002 with the complete survey made available in 2003.
A variety of programmes are underway to exploit the scientific potential of this new 
resource. Several illustrative project examples are briefly mentioned below.

\subsection{Planetary Nebulae}
The largest project arising from the AAO/UKST \ha\ survey has been 
the Macquarie/AAO/Strasbourg \ha\ planetary nebula project
(MASH; Parker et al. 2003, 2005 in preparation) which has uncovered about 1000 
new Galactic planetary nebulae (PNe), 
nearly doubling the sample accrued from all sources over the last 100 years. Related projects 
concern identification of a significant new PNe population in the Galactic Bulge
(e.g. Peyaud, Parker \& Acker 2003) and the discovery of an important 
sample of Wolf-Rayet central stars of PNe 
(e.g. Morgan, Parker \& Russeil 2001; Parker \& Morgan 2003) including the 
detection of the only [WN] PN central star in the galaxy 
(Morgan, Parker \& Cohen 2003).
A possible new phase of PNe evolution has also been reported around a strongly 
masing  OH-IR star (Cohen, Parker \& Chapman 2005). A very large PN in an early stage of interaction
with the ISM has also been discovered (Pierce et al. 2004) as well as two, very large bipolar PNe previously
mis-identified as HII regions 
(Frew, Parker \& Russeil 2005). A new sample of large ($>$4~arcminute), highly
evolved, low surface brightness PNe have also been found from examination of 
16\,$\times$ blocked down FITS images of the entire 233 fields of SHS survey. These blocked-down images
effectively enhance large angular size low surface brightness features (e.g. Pierce et al. 2004; Frew
\& Parker 2005, in preparation).  All these
discoveries are being investigated with follow-up spectroscopy to
determine the fundamental parameters of this significant new sample. 

\subsection{HII regions and regions of star formation}
This type of study is currently ripe for exploitation with little work currently 
undertaken. Mader et al. (1999) report the discovery of a significant new population of Herbig-Haro objects
from an SHS extension field in Orion and a large wind-blown bubble with secondary
star-forming regions around its periphery by Cohen et al. (2002). 
Numerous very faint HII regions have also been discovered as a 
by-product of the MASH survey (Parker et al. in preparation). 
Many large features are evident including bubbles which trigger
star-formation and induce velocity departures of the associated HII regions.
Using data from the SHS and additional kinematic information from Fabry-Perot observations
(Georgelin et al. 2000), the filamentary \ha\ counterparts and triggered 
HII regions for the HI shell centered at 
290.1+0.2 (Rizzo and Arnal, 1998) have been revealed.   
The high resolution of the data permits precise description of the morphology
and extent of such HII regions. This is vital information in determining the location of
the exciting stars which can be inferred via orientation of the observed rims and dust ``elephant 
trunk'' with respect to the HII region as a whole. Furthermore, the
\ha\ counterpart and visible extension can be directly compared to radio HII regions.
This is essential information needed to determine the distance of HII
regions in the framework of the study of the large scale structure of our
Galaxy (Russeil et al., 2005).
 
\subsection{Supernova Remnants}
Several programmes searching for the optical counterparts of supernova remnants (SNRs) in the SHS
have already been undertaken. Walker, Zealey \& Parker (2001)
report finding new filamentary shell structures
traced by \ha\ emission that are likely associated with Galactic SNRs. A more recent project is
underway to uncover SNR candidates across the entire SHS from careful 
scrutiny of both the blocked-down FITS images and the original survey
films, with several new SNRs already confirmed (Stupar, private communication). 
Searches for new optical \ha\ counterparts around the 
known Galactic SNR overlapping the SHS is also underway. One new Galactic SNR 
discovered serendipitously via the MASH programme has already
been reported (Parker, Frew \& Stupar 2004). A significant increase in the 
known population of optically detected Galactic SNRs is promised.

\subsection{Point-source emitters}
One area of more recent study is the search for point-source
emitters and the subsequent follow-up spectroscopy of candidates
identified from the SHS \ha\ and SR photometry. Drew et al (2004)
report the discovery, via SHS photometry, of only the 4$^{th}$
known massive WO star in the Milky Way Galaxy identified as part of a general programme of candidate
point-source follow-up. Additionally,  Hopewell et al. (2005)
present five new WC9 stars discovered from the SHS data in a similar fashion. Pierce
(2005) and Pierce et al. (2005, in preparation) demonstrate the power of the SHS to reveal significant new
populations of \ha\ emitters via a particular study in the Vela molecular ridge, 
especially when combined with $I$ band and 2MASS photometry. These preliminary projects have been
finding, for the magnitude range explored most
thoroughly ($12 < R < 16.5$), that 10-20 per cent of candidates 
are confirmed as emission line objects via follow-up spectroscopy. Their \ha\ equivalent widths 
usually exceed 20\AA. More recent work by Pierce (2005) indicates that the situation can be improved
by weeding out M stars more thoroughly using the 2MASS data.

\section{Conclusions}
The AAO/UKST \ha\ survey as scanned by SuperCOSMOS is now complete and on-line as the SHS atlas. 
It represents a powerful tool for the study of the ionized gas content of our 
galaxy on a range of spatial scales from 
arcsecond to tens of degrees. The distribution and structure of the ionised gas result from 
a wide range of astrophysically interesting phenomena.
The astrometric and photometric properties have been described and shown to be well behaved and adequate for
most purposes. Importantly,
despite difficulties associated with photographic data and the scanning
process, comparison with the independently calibrated SHASSA images has shown
that the SHS survey faithfully records diffuse Galactic
emission over a wide range of intensities from $\sim$\,5\,Rayleighs to
500\,Rayleighs. Emission down to $\sim$\,2\,Rayleighs has been detected on one field,
h1109. A calibration scheme for all 233 survey
fields has been generated, based on comparison of a carefully selected,
30~arcminute region from each field with the equivalent area of intensity
calibrated SHASSA \ha\ image. If the limitations of the data are respected in terms of dynamic range, 
reliable flux estimates are possible.
The survey is clearly appropriate for studies of
individual \ha\ emitting objects including point-sources as well as being suitable for the study of the
ionized interstellar medium in general. A variety of projects exploiting this resource are already underway and 
many exciting discoveries have already been made. 
The community is invited to consider use of this valuable survey when undertaking any study of the 
Southern Galactic Plane. 

\section*{Acknowledgements}
The authors gratefully acknowledge the support of the AAO board, the Wide-Field Astronomy Unit at
the University of Edinburgh, the Wide-Field Astronomy Panel (UK), the Particle-Physics and Astronomy Research
Council and the AAO directors Russell Cannon and Brian Boyle and UKST astronomers-in-charge Ann Savage and
Fred Watson for making the SHS survey possible. This paper used comparison data from SHASSA which was 
produced with support from the National Science Foundation. MC thanks NASA for supporting his
participation in the SHS through LTSA grant NAG5-7936 with UC Berkeley. MJP thanks PPARC for provision of a
PhD studentship. We also thank the referee John Meaburn for valuable comments on this paper.

\section*{References}
Arrowsmith P., \& Parker Q.A., 2001, ROE internal report\\
Beard S.M., MacGillivray H.T., Thanisch P.F., 1990, MNRAS, 247, 311\\
Bland-Hawthorn J.B., Veilleux S., Cecil G.N., Putman M.E., Gibson B.K., Maloney P.R., 1998, MNRAS, 299, 611\\
Bok B.J., Bester, M.J., Wade, C.M., 1955, Daedalus, 86, 9\\
Bond I.A. et al., 2001, MNRAS, 327, 868\\
Boyle B.J., Shanks T., Croom S.M., 1995, MNRAS, 276, 33\\
Buxton M. et al., 1998, PASA, 15, 24\\
Cohen M., Green A., Parker Q.A., Mader S., Cannon R.D. 2002, MNRAS, 336, 736\\
Cohen M., Parker, Q.A., Chapman, J., 2005, MNRAS, 357, 1189\\
Croom S.M., Ratcliffe A., Parker, Q.A., Shanks T., Boyle, B.J., Smith R.J., 1999, MNRAS, 306, 592 \\
Davies R.D., Elliott K.H., Meaburn J., 1976, Mem. RAS, 81, 89\\
Dennison B., et al., 1998, PASA, 15, 147\\
Dopita M.A., Hua C.T., 1997, ApJS, 108, 515\\
Dopita M.A., Mathewson D.S., Ford V.L., 1985, ApJ, 297, 599\\
Drew J., Barlow M.J., Unruh Y.C., Parker Q.A., Wesson R., Pierce M.J., Masheder M.R.W., 
Phillipps S., 2004, MNRAS, 351, 206\\
Drew J. et al., 2005, MNRAS, submitted\\
Elliot K.H., Meaburn J., 1976, Ap\&SS, 39, 437\\
Elmegreen B., Lada C.J., 1977, ApJ, 214, 725\\
Finkbeiner D.P., 2003, ApJS, 146, 407\\
Frew D.J., Parker Q.A., Russeil D., 2005, MNRAS, submitted\\
Gaustad J.E., McCullough P.R., Rosing W., \& Van Buren D., 2001, PASP, 113, 1326\\
Georgelin Y.P., Georgelin Y.M., 1970, A\&AS, 3, 1\\
Georgelin Y.M., Russeil D., Amram P. et al., 2000, A\&A, 357,c308\\
Gerola H., Seiden P., 1978, ApJ, 223, 129\\
Green A.J., Cram L.E., Large M.I., Ye T., 1999, ApJS, 122, 207\\
Gum C.S., 1952, Observatory, 72, 151\\
Gum C.S., 1955, Mem.RAS, 67, 155\\
Haffner L.M., Reynolds R.J., Tufte S.L., Madsen G.J., Jaehnig K.P., Percival J.W., 2003, ApJS, 149, 405\\ 
Hambly N.C., Miller L., MacGillivray H.T., Herd J.T., Cormack W.A., 1998, MNRAS, 298, 897\\
Hambly N.C., et al., 2001a, MNRAS, 326, 1279\\
Hambly N.C., Irwin M.J., MacGillivray H.T., 2001b, MNRAS, 326, 1295\\
Hambly N.C., Davenhall A.C., Irwin M.J., MacGillivray H.T., 2001c, MNRAS, 326, 1315\\
Hase V.F., Shajn G.A.,  1955, Isv. Krym. Astrofiz. Obs., 15, 11\\
Hog E., Fabricius C., Makarov V.V., Urban S., Corbin T., Wycoff G., Bastian U., Schwekendiek P., Wicenec A., 2000, A\&A, 355, L27\\
Hopewell E. C., Barlow M. J., Drew J. E., Unruh Y. C., Pierce M. J.,
Parker Q. A., Knigge C., Phillipps S., Zijlstra A. A., 2005, MNRAS, submitted.\\
Jarrett T, Chester T., Cutri R., Schneider S., Skrutskie M., Huchra J.P., 2000, AJ, 119, 2498\\ 
Johnson H.M., 1955, ApJ, 121, 604\\
Johnson H.M., 1956, ApJ, 124, 90\\
Keller S.C., Grebel E.K., Miller G.J., Yoss K.M., 2001, AJ, 122, 248\\ 
Kennicutt R.C., 1992, ApJ, 388, 310\\
Kodak publication P-315, 1987, Scientific imaging with Kodak films and plates\\
Lasker B.M., et al., 1988, ApJS,68,1\\
Mader S.L., Zealey W.J., Parker Q.A., Masheder M.R.W., 1999, MNRAS, 310,331\\
Meaburn J., 1978, Aplied Optics, 17, 1271\\
Meaburn J., 1980, MNRAS, 192, 365\\
Meaburn J., White N.J., 1982, MNRAS, 200, 771\\
Mel'Nik A.M., Efremov Y.N., 1995, AstL, 21, 10\\
Miller L., Cormack W.A.., Paterson M.G., Beard S.M., Lawrence L., 1992, 
In MacGillivray H.T., Thomson E.B., eds, Digitised
Optical Sky Surveys, Kluwer, Dordrecht, p.133\\
Morgan D.H., Parker Q.A., \& Russeil D., 2001, MNRAS, 322, 877.\\
Morgan D.H., Parker Q.A., Cohen M., 2003, MNRAS, 346, 729\\
Morgan D.H., Parker Q.A., 2005, MNRAS, in press\\
Nossal S., et al., 2001, J.Geophys.Res. 5605\\
Osterbrock D.E., 1989, Astrophysics of Gaseous Nebulae, University Science Books\\
Parker Q.A., Bland-Hawthorn J., 1998, PASA, 15, 33\\
Parker Q.A., Malin D.F., 1999, PASA, 16, 288\\
Parker Q.A., Phillipps S., 1997, PASA, 15, 28\\
Parker Q.A., Phillipps S., 1998, A\&G, 39, 4.10\\
Parker Q.A., Phillipps S., 2003, in ASP Conf.Ser. 289: Proceedings of the IAU 8th Asian-Pacific
Regional Meeting, Volume I, 165\\
Parker Q.A., Hartley M., Russeil D., Acker A., Ochsenbein F., Morgan D.H.,
Beaulieu S., Morris R., Phillipps S., Cohen,M., 2003, ASP Conf.Ser. eds M.Dopita,
S.Kwok and R Sutherland, p.41\\ 
Parker Q.A., Morgan D.H., 2003, MNRAS, 341, 961\\
Parker Q.A., Frew D.J., Stupar M., 2004, AAO Newsletter, 104, 9\\
Peyaud A.E.J., Parker Q.A., Acker A., 2003, in SF2A-2003: Semaine de l'Astrophysique Francaise, 311\\
Phillipps S., Parker Q.A., 1993, MNRAS, 265, 385\\
Pickering W.H., 1890, Sidereal Messenger, 9, 2\\
Pierce M.J., Frew D.J., Parker Q.A., Koppen J., 2004, PASA, 21, 334\\
Pierce, M, 2005, PhD thesis, University of Bristol\\
Price S.D., Egan M.P., Carey S.J., Mizuno D., Kuchar T. 2001, AJ, 121, 2819\\
Rizzo J.R., Arnal E.M., 1998, A\&A 332, 1025\\
Rodgers A.W., Campbell C.T., Whiteoak J.B., 1960, MNRAS, 121, 103\\
Russeil D., et al., 1997, A\&A, 319, 788\\
Russeil D., et al., 1998, PASA, 15, 9\\
Russeil D., Adami C., Amram P., Coarer E., Georgelin T.M., Marcelin M., Parker Q.A., 2005, A\&A, 429, 497\\
Schlegel D.J., Finkbeiner D.P., Davis M., 1998, ApJ, 500, 525\\
Sharpless S., 1953, ApJ, 118, 362\\
Sharpless S., 1959, ApJS, 4, 257\\
Sivan, 1974, A\&A,16, 163\\
Stephenson C.B., \& Sanduleak N., 1977, ApJS, 33, 459\\
Storkey A.J., Hambly N.C., Williams C.K.I., Mann R.G., 2004, MNRAS, 347, 36\\
Sung H., Chun M., Bessell M.S., 2000, AJ, 120, 333\\
Tenorio-Tagle G., Palous J., 1987, A\&A, 186, 287\\
Tritton S.B., 1993, UKSTU Handboook, publication of the Royal Observatory Edinburgh\\
Walker A., Zealey W.J., Parker Q.A., 2001, PASA, 18, 259\\
Watson F.G. 1984, MNRAS, 206, 661\\
Zacharias N., Urban S.E., Zacharias M.I., Wycoff G.L., Hall D.M., Monet D.G., Raffert T.J., 
2004, AJ, 127,3043
 \end{document}